# The Gravitational Spacecraft

**Fran De Aquino**
Maranhao State University, Physics Department, S.Luis/MA, Brazil.

There is an electromagnetic factor of correlation between gravitational mass and inertial mass, which in specific electromagnetic conditions, can be reduced, made negative and increased in numerical value. This means that gravitational forces can be reduced, inverted and intensified by means of electromagnetic fields. Such control of the gravitational interaction can have a lot of practical applications. For example, a new concept of spacecraft and aerospace flight arises from the possibility of the electromagnetic control of the gravitational mass. The novel spacecraft called Gravitational Spacecraft possibly will change the paradigm of space flight and transportation in general. Here, its operation principles and flight possibilities, it will be described. Also it will be shown that other devices based on gravity control, such as the Gravitational Motor and the Quantum Transceivers, can be used in the spacecraft, respectively, for Energy Generation and Telecommunications.

**CONTENTS**

## 1. Introduction

The discovery of the correlation between gravitational mass and inertial mass [1] has shown that the gravity can be *reduced*, *nullified* and *inverted*. Starting from this discovery several ways were proposed in order to obtain experimentally the local gravity control [2]. Consequently, new concepts of spacecraft and aerospace flight have arisen. This novel spacecraft, called Gravitational Spacecraft, can be equipped with other devices also based on gravity control, such as the Gravitational Motor and the Quantum Transceiver that can be used, respectively, for energy generation and telecommunications. Based on the theoretical background which led to the gravity control, the operation principles of the Gravitational Spacecraft and of the devices above mentioned, will be described in this work.

## 2. Gravitational Shielding

The contemporary greatest challenge of the Theoretical Physics was to prove that, Gravity is a *quantum* phenomenon. Since the General Relativity describes gravity as related to the curvature of the space-time then, the quantization of the gravity implies the quantization of the proper space-time. Until the end of the century XX, several attempts to quantify gravity were accomplished. However, all of them resulted fruitless [3, 4].

In the beginning of this century, it has been clearly noticed that there was something unsatisfactory about the whole notion of quantization and that the quantization process had many ambiguities. Then, a new approach has been proposed starting from the generalization of the *action function**. The result has been the derivation of a theoretical background, which finally led to the so-sought quantization of the gravity and of the space-time. Published under the title: "*Mathematical Foundations of the Relativistic Theory of Quantum Gravity*"†, this theory predicts a consistent *unification* of Gravity with Electromagnetism. It shows that the *strong* equivalence principle is reaffirmed and, consequently Einstein's equations are preserved. In fact, Einstein's equations can be deduced directly from the *Relativistic Theory of Quantum Gravity*. This shows, therefore, that the General Relativity is a particularization of this new theory, just as the Newton's theory is a particular case from the General Relativity. Besides, it was deduced from the new theory an important correlation between the *gravitational mass* and the *inertial mass*, which shows that the gravitational mass of a particle can be *decreased* and even made *negative*, independently of its inertial mass, i.e., while the gravitational mass is

* The formulation of the *action* in Classical Mechanics extends to the Quantum Mechanics and it has been the basis for the development of the *Strings Theory*.

† http://arxiv.org/abs/physics/0212033

progressively reduced, the inertial mass does not vary. This is highly relevant because it means that the weight of a body can also be reduced and even inverted in certain circumstances, since Newton's gravity law defines the weight $P$ of a body as the product of its *gravitational mass* $m_g$ by the local gravity acceleration $g$, i.e.,

$$P = m_g g \qquad (1)$$

It arises from the mentioned law that the gravity acceleration (or simply the gravity) produced by a body with gravitational mass $M_g$ is given by

$$g = \frac{GM_g}{r^2} \qquad (2)$$

The physical property of mass has two distinct aspects: *gravitational mass* $m_g$ and *inertial mass* $m_i$. The gravitational mass produces and responds to gravitational fields. It supplies the mass factors in Newton's famous inverse-square law of gravity $\left(F = GM_g m_g / r^2\right)$. The inertial mass is the mass factor in *Newton's 2nd Law of Motion* $\left(F = m_i a\right)$. These two masses are not equivalent but correlated by means of the following factor [1]:

$$\frac{m_g}{m_{i0}} = \left\{1 - 2\left[\sqrt{1 + \left(\frac{\Delta p}{m_{i0} c}\right)^2} - 1\right]\right\} \qquad (3)$$

Where $m_{i0}$ is the *rest* inertial mass and $\Delta p$ is the variation in the particle's *kinetic momentum*; $c$ is the speed of light.

This equation shows that only for $\Delta p = 0$ the gravitational mass is equal to the inertial mass. Instances in which $\Delta p$ is produced by *electromagnetic radiation*, Eq. (3) can be rewritten as follows:

$$\frac{m_g}{m_{i0}} = \left\{1 - 2\left[\sqrt{1 + \left(\frac{n_r^2 D}{\rho c^3}\right)^2} - 1\right]\right\} \qquad (4)$$

Where $n_r$ is the *refraction index* of the particle; $D$ is the power density of the electromagnetic radiation absorbed by the particle; and $\rho$ its density of inertial mass.

It was shown [1] that there is an additional effect of *gravitational shielding* produced by a substance whose gravitational mass was reduced or made negative. This effect shows that just *above the substance* the gravity acceleration $g_1$ will be reduced at the same proportion $\chi = m_g / m_{i0}$, i.e., $g_1 = \chi g$, ($g$ is the gravity acceleration *bellow* the substance).

Equation (4) shows, for example, that, in the case of a gas at ultra-low pressure (*very low density of inertial mass*), the *gravitational mass* of the gas can be strongly reduced or made negative by means of the incidence of electromagnetic radiation with power density relatively low.

Thus, it is possible to use this effect in order to produce gravitational shieldings and, thus, *to control the local gravity*.

The *Gravity Control Cells* (GCC) shown in the article "*Gravity Control by means of Electromagnetic*

*Field through Gas or Plasma at Ultra-Low Pressure*"‡, are devices designed on the basis, of this effect, and usually are chambers containing gas or plasma at ultra-low pressure. Therefore, when an oscillating electromagnetic field is applied upon the gas its gravitational mass will be reduced and, consequently, the gravity *above* the mentioned GCC will also be reduced at the same proportion.

It was also shown that it is possible to make a gravitational shielding even with the chamber filled with *Air* at one atmosphere. In this case, the *electric conductivity of the air* must be strongly increased in order to reduce the intensity of the electromagnetic field or the power density of the applied radiation.

This is easily obtained by *ionizing the air* in the local where we want to build the gravitational shielding. There are several manners of ionizing the air. One of them is by means of ionizing radiation produced by a radioactive source of low intensity, for example, by using the radioactive element *Americium* (Am-241). The Americium is widely used as air ionizer in smoke detectors. Inside the detectors, there is just a little amount of americium 241 (about of 1/5000 grams) in the form of $AmO_2$. Its cost is very low (about of US$ 1500 per gram). The dominant radiation is composed of alpha particles. Alpha particles cannot cross a paper sheet and are also blocked by some centimeters of air. The Americium used in the smoke detectors can only be dangerous if inhaled.

The Relativistic Theory of Quantum Gravity also shows the existence of a *generalized equation for the inertial forces* which has the following form

$$F_i = M_g a \quad (5)$$

This expression means a *new law for the Inertia*. Further on, it will be shown that it *incorporates the Mach's principle* to Gravitation theory [5].

Equation (3) tell us that the gravitational mass is only equal to the inertial mass when $\Delta p = 0$. Therefore, we can easily conclude that only in this particular situation the new expression of $F_i$ reduces to $F_i = m_i a$, which is the expression for Newton's 2nd Law of Motion. Consequently, this Newton's law is just a particular case from the new law expressed by the Eq. (5), which clearly shows how *the local inertial forces are correlated to the gravitational interaction of the local system with the distribution of cosmic masses* (via $m_g$) and thus, *incorporates* definitively *the Mach's principle* to the Gravity theory.

The Mach's principle postulates that: "*The local inertial forces would be produced by the gravitational interaction of the local system with the distribution of cosmic masses*". However, in spite of the several attempts carried out, this principle had not yet been incorporated to the Gravitation theory. Also Einstein had carried out several attempts. The *ad hoc* introduction of the cosmological

‡ http://arxiv.org/abs/physics/0701091

term in his gravitation equations has been one of these attempts.

With the advent of equation (5), *the origin of the inertia* - that was considered the most obscure point of the particles' theory and field theory – becomes now evident.

In addition, this equation also reveals that, if the gravitational mass of a body is very close to zero or if there is around the body a *gravitational shielding* which reduces closely down to zero the *gravity accelerations due to the rest of the Universe*, then the intensities of the inertial forces that act on the body become also very close to zero.

This conclusion is highly relevant because it shows that, under these conditions, the spacecraft could describe, with great velocities, unusual trajectories (such as curves in right angles, abrupt inversion of direction, etc.) without inertial impacts on the occupants of the spacecraft. Obviously, out of the above-mentioned condition, the spacecraft and the crew would be destroyed due to the strong presence of the inertia.

When we make a sharp curve with our car we are pushed towards a direction contrary to that of the motion of the car. This happens due to existence of *the inertial forces*. However, if our car is involved by a *gravitational shielding*, which reduces strongly the gravitational interaction of the car (and everything that is inside the car) with the rest of the Universe, then in accordance with the Mach's principle, the local inertial forces would also be strongly reduced and, consequently, we would not feel anything during the maneuvers of the car.

## 3. Gravitational Motor: Free Energy

It is known that the energy of the gravitational field of the Earth can be converted into rotational kinetic energy and electric energy. In fact, this is exactly what takes place in hydroelectric plants. However, the construction these hydroelectric plants have a high cost of construction and can only be built, obviously, where there are rivers.

The gravity control by means of any of the processes mentioned in the article: "*Gravity Control by means of Electromagnetic Field through Gas or Plasma at Ultra-Low Pressure*" allows the inversion of the weight of any body, practically at any place. Consequently, the conversion of the gravitational energy into rotational mechanical energy can also be carried out at any place.

In Fig. (1), we show a schematic diagram of a *Gravitational Motor*. The first *Gravity Control Cell* (GCC1) changes the local gravity from $g$ to $g' = -ng$, propelling the left side of the rotor in a direction contrary to the motion of the right side. The second GCC changes the gravity back again to $g$ i.e., from $g' = -ng$ to $g$, in such a way that the gravitational change occurs just on the region indicated in Fig.1. Thus, a *torque* $T$ given by

$$T = (-F' + F)r = \left[-(m_g/2)g' + (m_g/2)g\right]r = \\ = (n+1)\tfrac{1}{2}m_g gr$$

Is applied on the rotor of gravitational mass $m_g$, making the rotor spin with angular velocity $\omega$.

The average *power*, $P$, of the motor is $P = T\omega$. However, $-g' + g = \omega^2 r$. Thus, we have

$$P = \tfrac{1}{2} m_i \sqrt{(n+1)^3 g^3 r} \qquad (6)$$

Consider a cylindrical rotor of iron $(\rho = 7800 Kg.m^{-3})$ with height $h = 0.5m$, radius $r = R/3 = 0.0545m$ and inertial mass $m_i = \rho \pi R^2 h = 327.05kg$. By adjusting the GCC 1 in order to obtain $\chi_{air(1)} = m_{g(air)}/m_{i(air)} = -n = -19$ and, since $g = 9.81 m.s^{-2}$, then Eq. (6) gives

$$P \cong 2.19 \times 10^5 \, watts \cong 219 \ KW \cong 294 HP$$

This shows that this small motor can be used, for example, to substitute the conventional motors used in the cars. It can also be coupled to an electric generator in order to produce *electric energy*. The conversion of the rotational mechanical energy into electric energy is not a problem since it is a problem technologically resolved several decades ago. Electric generators are usually produced by the industries and they are commercially available, so that it is enough to couple a gravitational motor to an electric generator for we obtaining electric energy. In this case, just a gravitational motor with the power above mentioned it would be enough to supply the need of electric energy of, for example, at least 20 residences. Finally, it can substitute the conventional motors of the same power, with the great advantage of *not needing of fuel for its opera*tion. What means that the gravitational motors can produce energy practically free.

It is easy to see that gravitational motors of this kind can be designed for powers needs of just some watts up to millions of kilowatts.

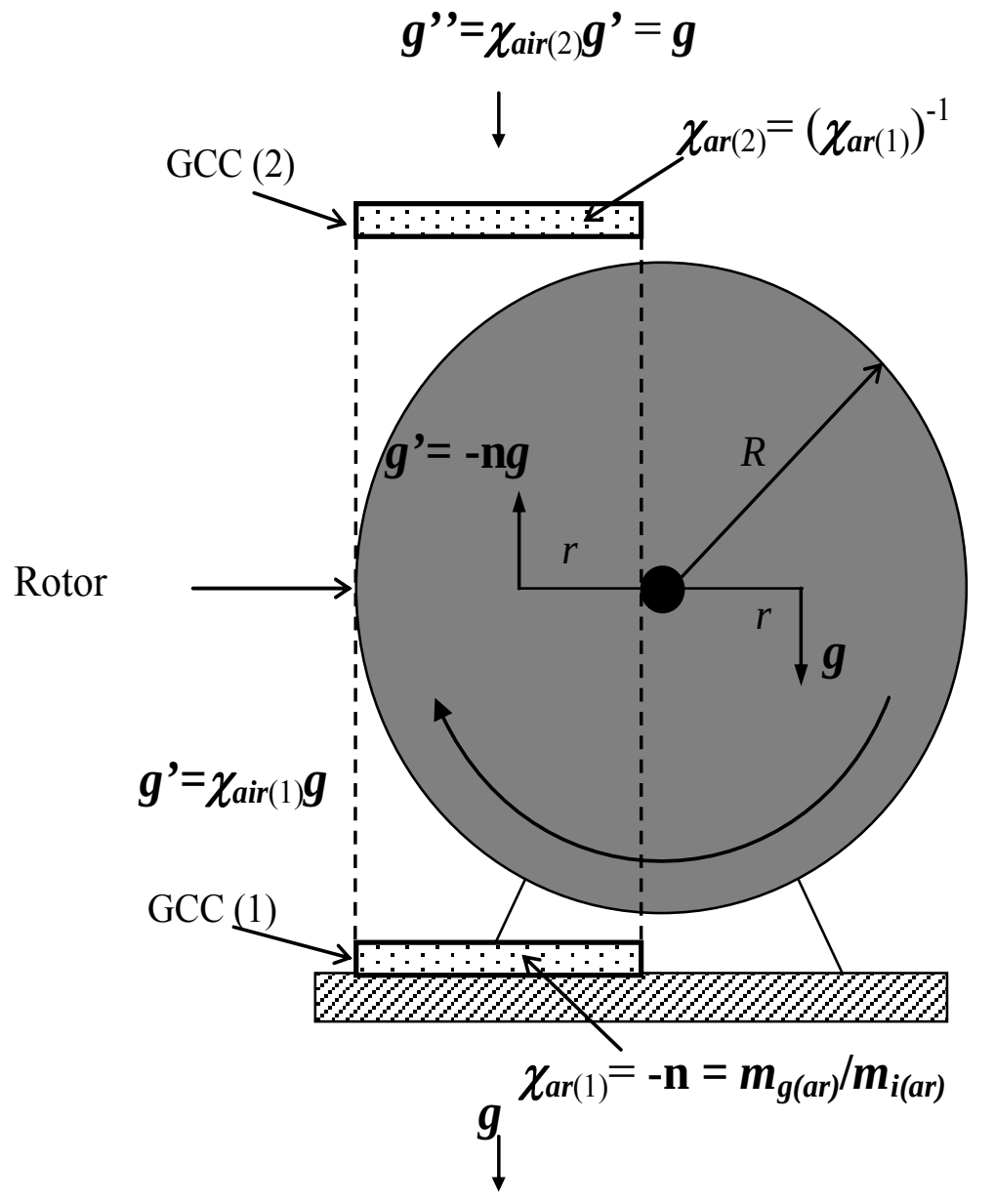

Fig. 1 – Gravitational Motor - The first *Gravity Control Cell* (GCC1) changes the local gravity from $g$ to $g' = -ng$, propelling the left side of the rotor in contrary direction to the motion of the right side. The second GCC changes the gravity back again to $g$ i.e., from $g' = -ng$ to $g$, in such a way that the gravitational change occurs just on the region shown in figure above.

## 4. The Gravitational Spacecraft

Consider a metallic sphere with radius $r_s$ in the terrestrial atmosphere. If the external surface of the sphere is recovered with a radioactive substance (for example, containing Americium 241) then the air in the space close to the surface of the sphere will be strongly ionized by the radiation emitted from the radioactive element and, consequently, the electric conductivity of the air close to sphere will become strongly increased.

By applying to the sphere an electric potential of low frequency $V_{rms}$, in order to produce an electric field $E_{rms}$ starting from the surface of the sphere, then very close to the surface, the intensity of the electric field will be $E_{rms} = V_{rms}/r_s$ and, in agreement with Eq. (4), the *gravitational mass* of the *Air* in this region will be expressed by

$$m_{g(air)} = \left\{1 - 2\left[\sqrt{1 + \frac{\mu_0}{4c^2}\left(\frac{\sigma_{air}}{4\pi f}\right)^3 \frac{V_{rms}^4}{r_s^4 \rho_{air}^2}} - 1\right]\right\} m_{i0(air)} \quad (7)$$

Therefore we will have

$$\chi_{air} = \frac{m_{g(air)}}{m_{i0(air)}} = \left\{1 - 2\left[\sqrt{1 + \frac{\mu_0}{4c^2}\left(\frac{\sigma_{air}}{4\pi f}\right)^3 \frac{V_{rms}^4}{r_s^4 \rho_{air}^2}} - 1\right]\right\} \quad (8)$$

The gravity accelerations acting on the sphere, due to the rest of the Universe (See Fig. 2), will be given by

$$g_i' = \chi_{air} g_i \qquad i = 1,2,...,n$$

Note that by varying $V_{rms}$ or the frequency $f$, we can easily to *reduce and control* $\chi_{air}$. Consequently, we can also control the intensities of the gravity accelerations $g_i'$ in order to produce a *controllable gravitational shielding* around the sphere.

Thus, the *gravitational forces* acting on the sphere, due to the rest of the Universe, will be given by

$$F_{gi} = M_g g_i' = M_g (\chi_{air} g_i)$$

where $M_g$ is the gravitational mass of the sphere.

The gravitational shielding around of the sphere reduces both the gravity accelerations acting on the sphere, due to the rest of the Universe, and the gravity acceleration produced by the gravitational mass $M_g$ of the own sphere. That is, if inside the shielding the gravity produced by the sphere is $g = -GM_g/r^2$, then, *out of* the shielding it becomes $g' = \chi_{air} g$. Thus, $g' = \chi_{air}\left(-GM_g/r^2\right) = -G\left(\chi_{air}M_g\right)/r^2 = -Gm_g/r^2$, where

$$m_g = \chi_{air} M_g$$

Therefore, *for the Universe out of the shielding* the gravitational mass of the sphere is $m_g$ and not $M_g$. In these circumstances, the *inertial forces* acting on the sphere, in agreement with the *new law for inertia*, expressed by Eq. (5), will be given by

$$F_{ii} = m_g a_i \quad (9)$$

Thus, these forces will be almost null when $m_g$ becomes almost null by means of the action of the gravitational shielding. This means that, in these circumstances, the sphere practically loses its *inertial properties*. This effect leads to *a new concept of spacecraft and aerospatial flight*. The spherical form of the spacecraft is just *one* form that the Gravitational Spacecraft can have, since the gravitational shielding can also be obtained with other formats.

An important aspect to be observed is that it is possible to control the gravitational mass of the spacecraft, $M_{g(spacecraf)}$, simply by controlling the gravitational mass of a body *inside* the spacecraft. For instance, consider a parallel plate capacitor inside the spacecraft. The gravitational mass of the *dielectric* between the plates of the capacitor can be controlled by means of the ELF electromagnetic field through it. Under these circumstances, the *total* gravitational mass of the spacecraft will be given by

$$M_{g(spacecraf)}^{total} = M_{g(spacecraf)} + m_g = \\ = M_{i0} + \chi_{dielectric} m_{i0} \qquad (10)$$

where $M_{i0}$ is the rest inertial mass of the spacecraft(without the dielectric) and $m_{i0}$ is the rest inertial mass of the dielectric; $\chi_{dielectric} = m_g / m_{i0}$, where $m_g$ is the gravitational mass of the dielectric. By decreasing the value of $\chi_{dielectric}$, the gravitational mass of the spacecraft decreases. It was shown, that the value of $\chi$ can be negative. Thus, when $\chi_{dielectric} \cong -M_{i0}/m_{i0}$, the *gravitational mass of the spacecraft gets very close to zero*. When $\chi_{dielectric} < -M_{i0}/m_{i0}$, the gravitational mass of the spacecraft becomes negative.

Therefore, *for an observer out of the spacecraft*, the gravitational mass of the spacecraft is $M_{g(spacecraf)} = M_{i0} + \chi_{dielectric} m_{i0}$, and not $M_{i0} + m_{i0}$.

Another important aspect to be observed is that we can *control the gravity inside the spacecraft*, in order to produce, for example, a gravity acceleration equal to the Earth's gravity $\left(g = 9.81 m.s^{-2}\right)$. This will be very useful in the case of space flight, and can be easily obtained by putting in the ceiling of the spacecraft the system shown in Fig. 3. This system has three GCC with nuclei of ionized air (or air at low pressure). Above these GCC there is a massive block with mass $M_g$.

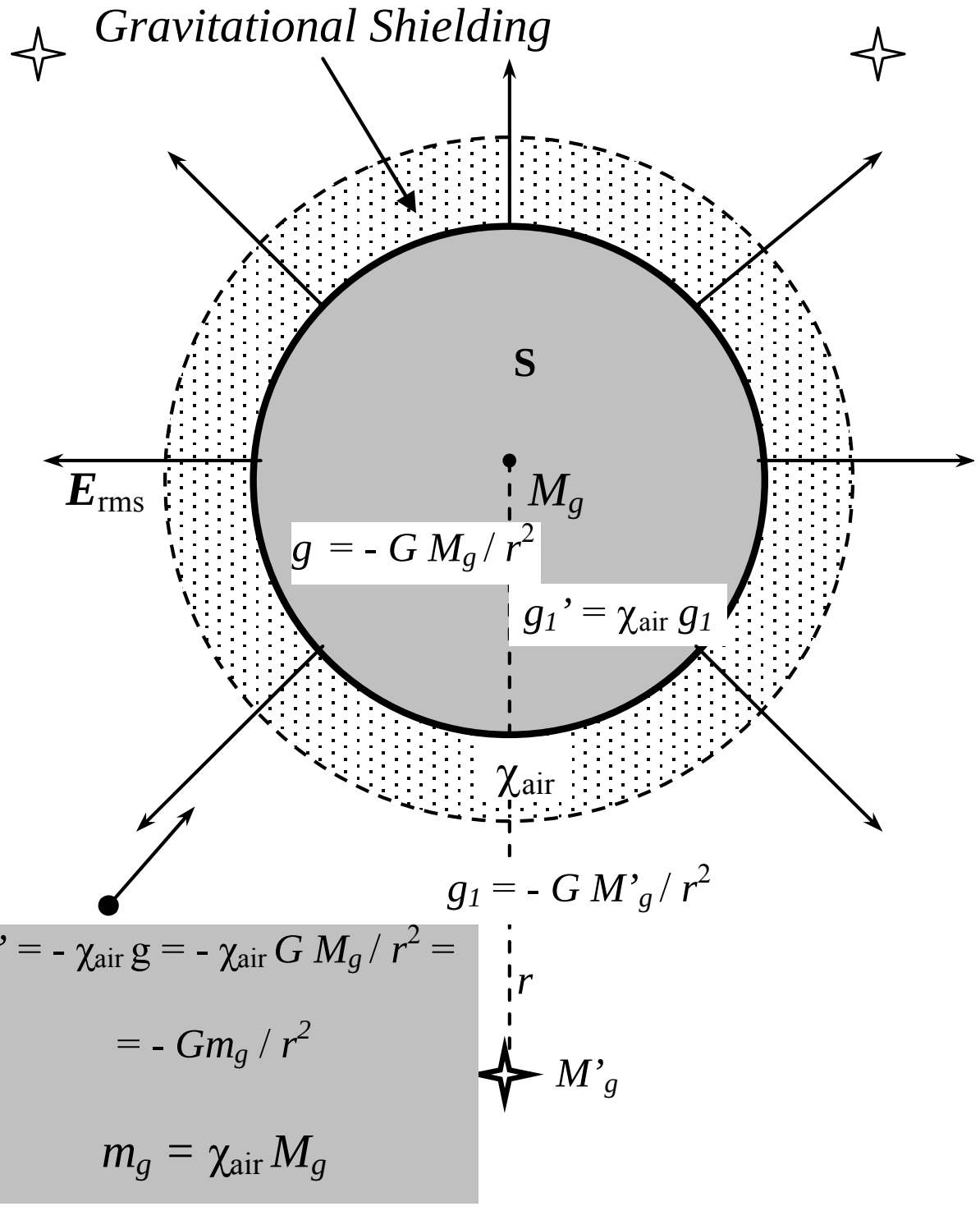

Fig.2- The gravitational shielding reduces the *gravity accelerations* ( $g_1'$) acting on the sphere (due to the rest of the Universe) and also reduces the gravity acceleration that the sphere produces upon all the particles of the Universe ($g'$). For the Universe, the gravitational mass of the sphere will be $m_g = \chi_{air} M_g$.

As we have shown [2], a gravitational *repulsion* is established between the mass $M_g$ and any positive gravitational mass below the mentioned system. This means that the particles in this region will stay subjected to a gravity acceleration $a_b$, given by

$$\vec{a}_b \cong (\chi_{air})^3 \vec{g}_M \cong -(\chi_{air})^3 G \frac{M_g}{r_0^2} \hat{\mu} \qquad (11)$$

If the Air inside the GCCs is sufficiently ionized, in such way that $\sigma_{air} \cong 10^3 S.m^{-1}$, and if $f = 1\ Hz$, $\rho_{air} \cong 1\ kg.m^{-3}$, $V_{rms} \cong 10\ KV$ and $d = 1\ cm$ then the Eq.8 shows that inside the GCCs we will have

$$\chi_{air} = \frac{m_{g(air)}}{m_{i0(air)}} = \left\{1 - 2\left[\sqrt{1 + \frac{\mu_0}{4c^2}\left(\frac{\sigma_{air}}{4\pi f}\right)^3 \frac{V_{rms}^4}{d^4 \rho_{air}^2}} - 1\right]\right\} \cong -10^3$$

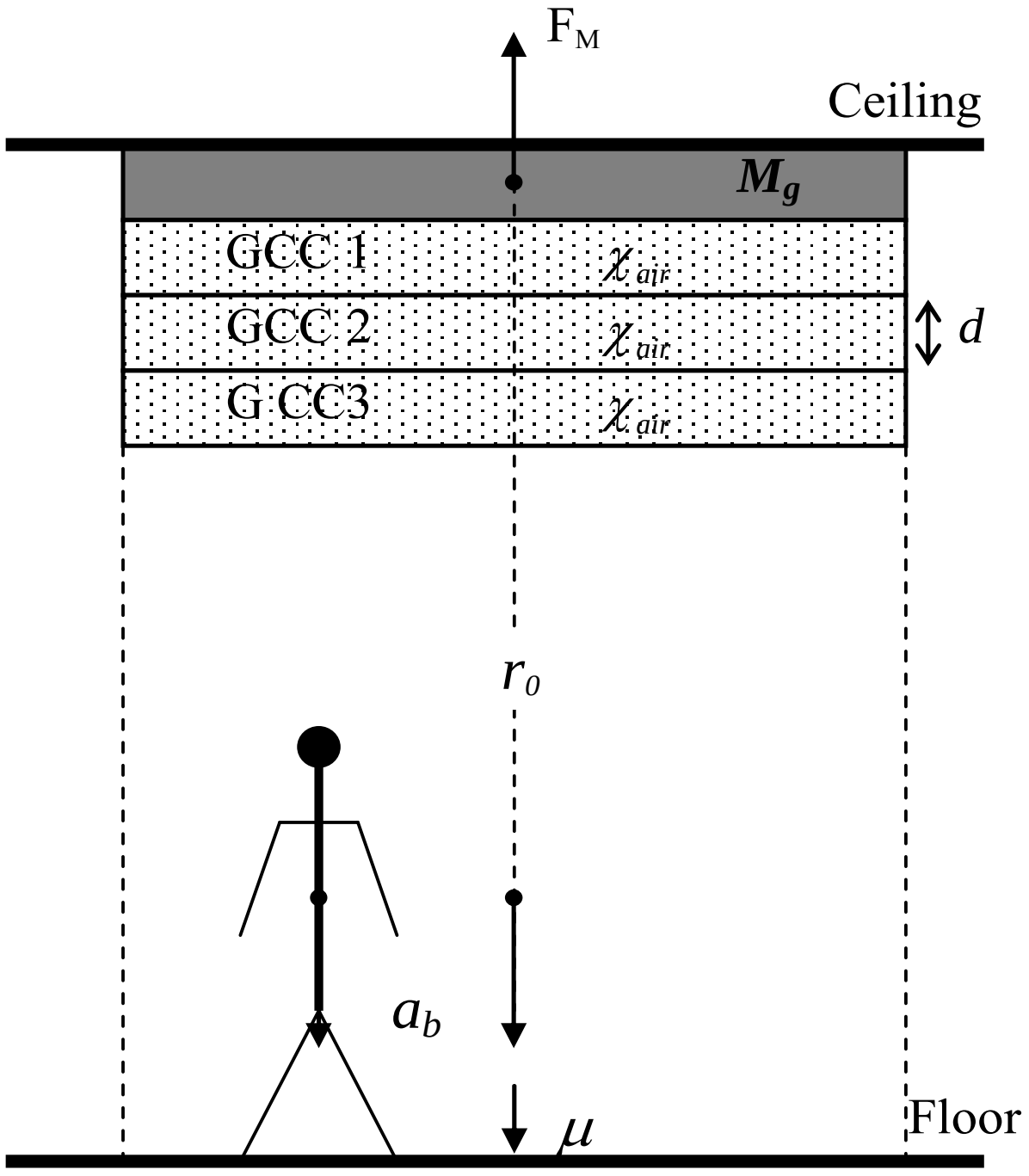

$$\vec{a}_b \cong (\chi_{air})^3 \vec{g}_M \cong -(\chi_{air})^3 G \frac{M_g}{r_0^2} \hat{\mu}$$

Fig.3 – If the Air inside the GCC is sufficiently ionized, in such way that $\sigma_{air} \cong 10^3 S.m^{-1}$ and if $f = 1\ Hz$; $d = 1cm$; $\rho_{air} \cong 1\ kg.m^{-3}$ and $V_{rms} \cong 10\ KV$ then Eq. 8 shows that inside the CCGs we will have $\chi_{air} \cong -10^3$. Therefore, for $M_g \cong M_i \cong 100kg$ and $r_o \cong 1m$ the gravity acceleration inside the spacecraft will be directed from the ceiling to the floor of the spacecraft and its intensity will be $a_b \approx 10 m.s^{-2}$.

Therefore the equation (11) gives

$$a_b \approx +10^9 G \frac{M_g}{r_0^2} \qquad (12)$$

For $M_g \cong M_i \cong 100kg$ and $r_0 \cong 1m$ (See Fig.3), the gravity inside the spacecraft will be directed from the ceiling to the floor and its intensity will have the following value

$$a_b \approx 10 m.s^{-2} \qquad (13)$$

Therefore, an interstellar travel in a gravitational spacecraft will be particularly comfortable, since we can travel during all the time subjected to the gravity which we are accustomed to here in the Earth.

We can also use the system shown in Fig. 3 as a thruster in order to propel the spacecraft. Note that the gravitational repulsion that occurs between the block with mass $M_g$ and any particle after the GCCs *does not depend on* of the place where the system is working. Thus, this *Gravitational Thruster* can propel the gravitational spacecraft in *any direction*. Moreover, it can work in the terrestrial atmosphere as well as in the cosmic space. In this case, the energy that produces the propulsion is obviously the *gravitational energy*, which is always present in any point of the Universe.

The schematic diagram in Fig. 4 shows in details the operation of the Gravitational Thruster. A gas of any type injected into the chamber beyond the GCCs acquires an acceleration $a_{gas}$, as shown in Fig.4, the intensity of which, as we have seen, is given by

$$a_{gas} = (\chi_{gas})^3 g_M \cong -(\chi_{gas})^3 G \frac{M_g}{r_0^2} \qquad (14)$$

Thus, if inside of the GCCs, $\chi_{gas} \cong -10^9$ then the equation above gives

$$a_{gas} \cong +10^{27} G \frac{M_g}{r_0^2} \qquad (15)$$

For $M_g \cong M_i \cong 10kg$, $r_0 \cong 1m$ we have $a_{gas} \cong 6.6 \times 10^{17} m.s^{-2}$. With this enormous acceleration the particles of the gas reach velocities close to the speed of the light in just a few nanoseconds. Thus, if the emission rate of the gas is $dm_{gas}/dt \cong 10^{-3} kg/s \cong 4000 litres/hour$, then the trust produced by the gravitational thruster will be

$$F = v_{gas}\frac{dm_{gas}}{dt} \cong c\frac{dm_{gas}}{dt} \cong 10^5 N \qquad (16)$$

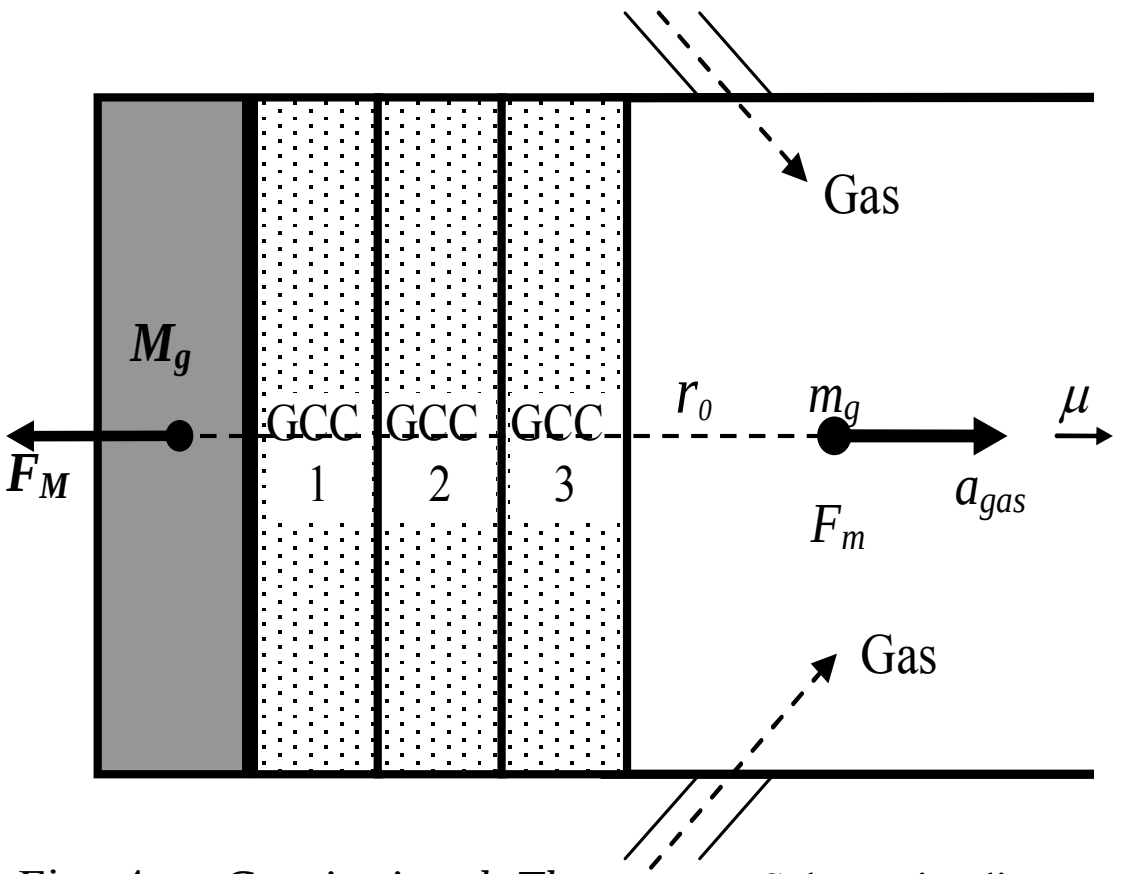

Fig. 4 – *Gravitational Thruster* – Schematic diagram showing the operation of the Gravitational Thruster. Note that in the case of very strong $\chi_{air}$, for example $\chi_{air} \cong -10^9$, the gravity accelerations upon the boxes of the second and third GCCs become very strong. Obviously, the walls of the mentioned boxes cannot to stand the enormous pressures. However, it is possible to build a similar system [2] with 3 or more GCCs, *without material boxes*. Consider for example, a surface with several radioactive sources (Am-241, for example). The *alpha* particles emitted from the Am-241 cannot reach besides 10cm of air. Due to the trajectory of the alpha particles, three or more successive layers of air, with different electrical conductivities $\sigma_1$, $\sigma_2$ and $\sigma_3$, will be established in the ionized region. It is easy to see that the gravitational shielding effect produced by these three layers is similar to the effect produced by the 3 GCCs above.

It is easy to see that the gravitational thrusters are able to produce strong trusts (similarly to the produced by the powerful thrusters of the modern aircrafts) *just by consuming the injected gas for its operation.*

It is important to note that, if $F$ is the thrust produced by the gravitational thruster then, in agreement with Eq. (5), the spacecraft acquires an acceleration $a_{spacecraft}$, expressed by the following equation

$$a_{spacecraft} = \frac{F}{M_{g(spacecraft)}} = \frac{F}{\chi_{out} M_{i(spacecraft)}} \qquad (17)$$

Where $\chi_{out}$, given by Eq. (8), is the factor of gravitational shielding which depends on the external medium where the spacecraft is placed. By adjusting the shielding for $\chi_{out} = 0.01$ and if $M_{spacecraft} = 10^4 Kg$ then for a thrust $F \cong 10^5 N$, the acceleration of the spacecraft will be

$$a_{spacecraft} = 1000 m.s^{-2} \qquad (18)$$

With this acceleration, in just at 1(one) day, the velocity of the spacecraft will be close to the speed of light. However it is easy to see that $\chi_{out}$ can still be much more reduced and, consequently, the thrust much more increased so that it is possible to increase up to 1 million times the acceleration of the spacecraft.

It is important to note that, the inertial effects upon the spacecraft will be reduced by $\chi_{out} = M_g/M_i \cong 0.01$. Then, in spite of its effective acceleration to be $a = 1000 m.s^{-2}$, the effects for the crew of the spacecraft will be equivalents to an acceleration of only

$$a' = \frac{M_g}{M_i} a \approx 10 m.s^{-1}$$

This is the magnitude of the acceleration on the passengers in a contemporary commercial jet.

Then, it is noticed that the gravitational spacecrafts can be subjected to enormous *accelerations* (or *decelerations*) without imposing any harmful impacts whatsoever on the spacecrafts or its crew.

We can also use the system shown in Fig. 3, as a *lifter*, inclusively within the spacecraft, in order to lift peoples or things into the spacecraft as shown in Fig. 5. Just using two GCCs, the gravitational acceleration produced below the GCCs will be

$$\vec{a}_g = (\chi_{air})^2 g_M \cong -(\chi_{air})^2 G M_g / r_0^2 \,\hat{\mu} \qquad (19)$$

Note that, in this case, if $\chi_{air}$ is *negative*, the acceleration $\vec{a}_g$ will have a direction *contrary* to the versor $\hat{\mu}$, i.e., the body will be *attracted* in the direction of the GCCs, as shown in Fig.5. In practice, this will occur when the air inside the GCCs is sufficiently ionized, in such a way that $\sigma_{air} \cong 10^3 S.m^{-1}$. Thus, if the internal thickness of the GCCs is now $d = 1\ mm$ and if $f = 1\ Hz$; $\rho_{air} \cong 1\ kg.m^{-3}$ and $V_{rms} \cong 10\ KV$, we will then have $\chi_{air} \cong -10^5$. Therefore, for $M_g \cong M_i \cong 100kg$ and, for example, $r_0 \cong 10m$ the gravitational acceleration acting on the body will be $a_b \approx 0.6 m.s^{-2}$. It is obvious that this value can be easily increased or decreased, simply by varying the voltage $V_{rms}$. Thus, by means of this *Gravitational Lifter,* we can lift or lower persons or materials with great versatility of operation.

It was shown [1] that, when the gravitational mass of a particle is reduced into the range, $+0.159M_i$ to $-0.159M_i$, it becomes imaginary, i.e., its masses (gravitational and inertial) becomes imaginary. Consequently, the particle disappears from our ordinary Universe, i.e., it becomes *invisible* for us. This is therefore a manner of to obtain the transitory invisibility of persons, animals, spacecraft, etc. However, the factor $\chi = M_{g(imaginary)} / M_{i(imaginary)}$ remains real because

$$\chi = \frac{M_{g(imaginary)}}{M_{i(imaginary)}} = \frac{M_g i}{M_i i} = \frac{M_g}{M_i} = real$$

Thus, if the gravitational mass of the particle is reduced by means of the absorption of an amount of electromagnetic energy $U$, for example, then we have

$$\chi = \frac{M_g}{M_i} = \left\{1 - 2\left[\sqrt{1 + \left(U/m_{i0}c^2\right)^2} - 1\right]\right\}$$

This shows that the energy $U$ *continues acting* on the particle turned imaginary. In practice this means that *electromagnetic fields act on imaginary particles*. Therefore, the internal electromagnetic field of a GCC remains acting upon the particles inside the GCC even when their gravitational masses are in the range $+0.159M_i$ to $-0.159M_i$, turning them *imaginaries.* This is very important because it means that the GCCs of a gravitational spacecraft remain working even when the spacecraft becomes imaginary.

Under these conditions, the gravity accelerations acting on the imaginary spacecraft, due to the rest of the Universe will be, as we have see, given by

$$g_i' = \chi\ g_i \qquad i = 1,2,...,n$$

Where $\chi = M_{g(imaginary)} / M_{i(imaginary)}$ and $g_i = -G m_{gi(imaginary)} / r_i^2$. Thus, the gravitational forces acting on the spacecraft will be given by

$$F_{gi} = M_{g(imaginary)} g_i' =$$
$$= M_{g(imaginary)}\left(-\chi G m_{gj(imaginary)} / r_j^2\right) =$$
$$= M_g i\left(-\chi G m_{gi} i / r_i^2\right) = +\chi G M_g m_{gi} / r_i^2 . \quad (20)$$

Note that these forces are *real*. By calling that, the *Mach's principle* says that the *inertial effects* upon a particle are consequence of the gravitational interaction of the particle with the rest of the Universe. Then we can conclude that the inertial forces acting on the spacecraft in imaginary state are also *real*. Therefore, it can travel in the imaginary space-time using the gravitational thrusters.

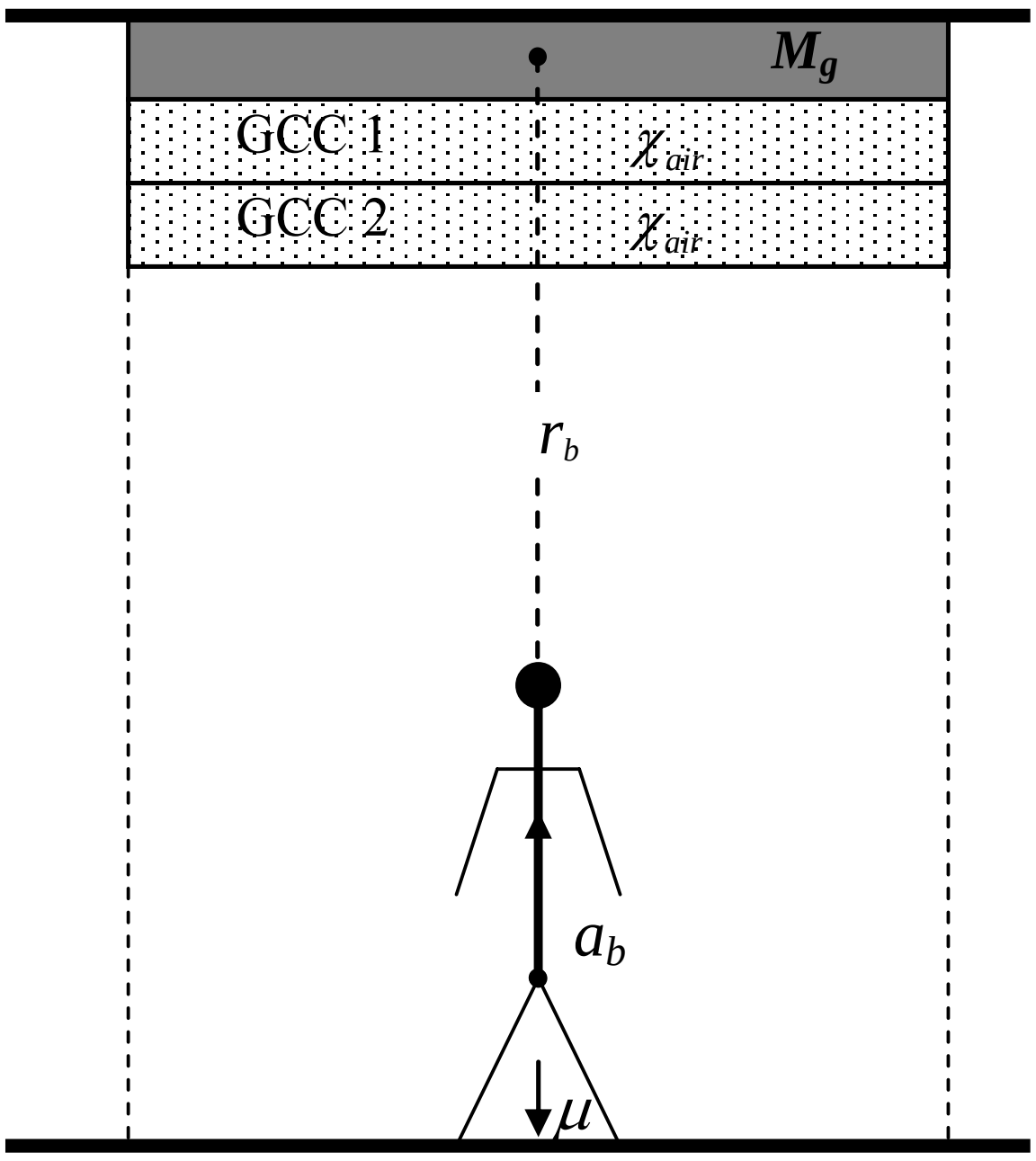

$$\vec{a}_b \cong (\chi_{air})^2 \vec{g}_M \cong -(\chi_{air})^2 G \frac{M_g}{r_b^2} \hat{\mu}$$

Fig.5 – *The Gravitational Lifter* – If the air inside the GCCs is sufficiently ionized, in such way that $\sigma_{air} \cong 10^3 S.m^{-1}$ and the internal thickness of the GCCs is now $d = 1\ mm$ then, if $f = 1\ Hz$; $\rho_{air} \cong 1\ kg.m^{-3}$ and $V_{rms} \cong 10\ KV$, we have $\chi_{air} \cong -10^5$. Therefore, for $M_g \cong M_i \cong 100kg$ and $r_0 \cong 10m$ the gravity acceleration acting on the body will be $a_b \approx 0.6 m.s^{-2}$.

It was also shown [1] that *imaginary particles* can have *infinity velocity* in the imaginary space-time. Therefore, this is also the upper limit of velocity for the gravitational spacecrafts traveling in the imaginary space-time. On the other hand, the travel in the imaginary space-time can be very safe, because there will not be any material body in the trajectory of the spacecraft.

It is easy to show that the gravitational forces between two thin layers of air (with masses $m_{g1}$ and $m_{g2}$ ) around the spacecraft , are expressed by

$$\vec{F}_{12} = -\vec{F}_{21} = -(\chi_{air})^2 G \frac{m_{i1} m_{i2}}{r^2} \hat{\mu} \qquad (21)$$

Note that these forces can be strongly increased by increasing the value of $\chi_{air}$. In these circumstances, the air around the spacecraft would be strongly compressed upon the external surface of the spacecraft creating an atmosphere around it. This can be particularly useful in order to minimize the *friction* between the spacecraft and the atmosphere of the planet in the case of very high speed movements of the spacecraft. With the atmosphere around the spacecraft the friction will occur between the atmosphere of the spacecraft and the atmosphere of the planet. In this way, the friction will be minimum and the spacecraft could travel at very high speeds without overheating.

However, in order for this to occur, it is necessary to put the gravitational shielding in another position as shown in Fig.2. Thus, the values of $\chi_{airB}$ and $\chi_{airA}$ will be independent (See Fig.6). Thus, while inside the gravitational shielding, the value of $\chi_{airB}$ is put close to zero, in order to strongly reduce the gravitational mass of the spacecraft (inner part of the shielding), the value of $\chi_{airA}$ must be reduced to about $-10^8$ in order to strongly increase the gravitational attraction between the air molecules around the spacecraft. Thus, by substituting $\chi_{airA} \cong -10^8$ intoEq.21, we get

$$\vec{F}_{12} = -\vec{F}_{21} = -10^{16} G \frac{m_{i1} m_{i2}}{r^2} \hat{\mu} \qquad (22)$$

If, $m_{i1} \cong m_{i2} = \rho_{air} V_1 \cong \rho_{air} V_2 \cong 10^{-8} kg$ and $r = 10^{-3} m$ then Eq. 22 gives

$$\vec{F}_{12} = -\vec{F}_{21} \cong -10^{-4} N \qquad (23)$$

These forces are much more intense than the *inter-atomic forces* (the forces that unite the atoms and molecules) the intensities of which are of the order of $1 - 1000 \times 10^{-8} N$. Consequently, the air around the spacecraft will be strongly compressed upon the surface of the spacecraft and thus will produce a crust of air which will accompany the spacecraft during its displacement and will protect it from the friction with the atmosphere of the planet.

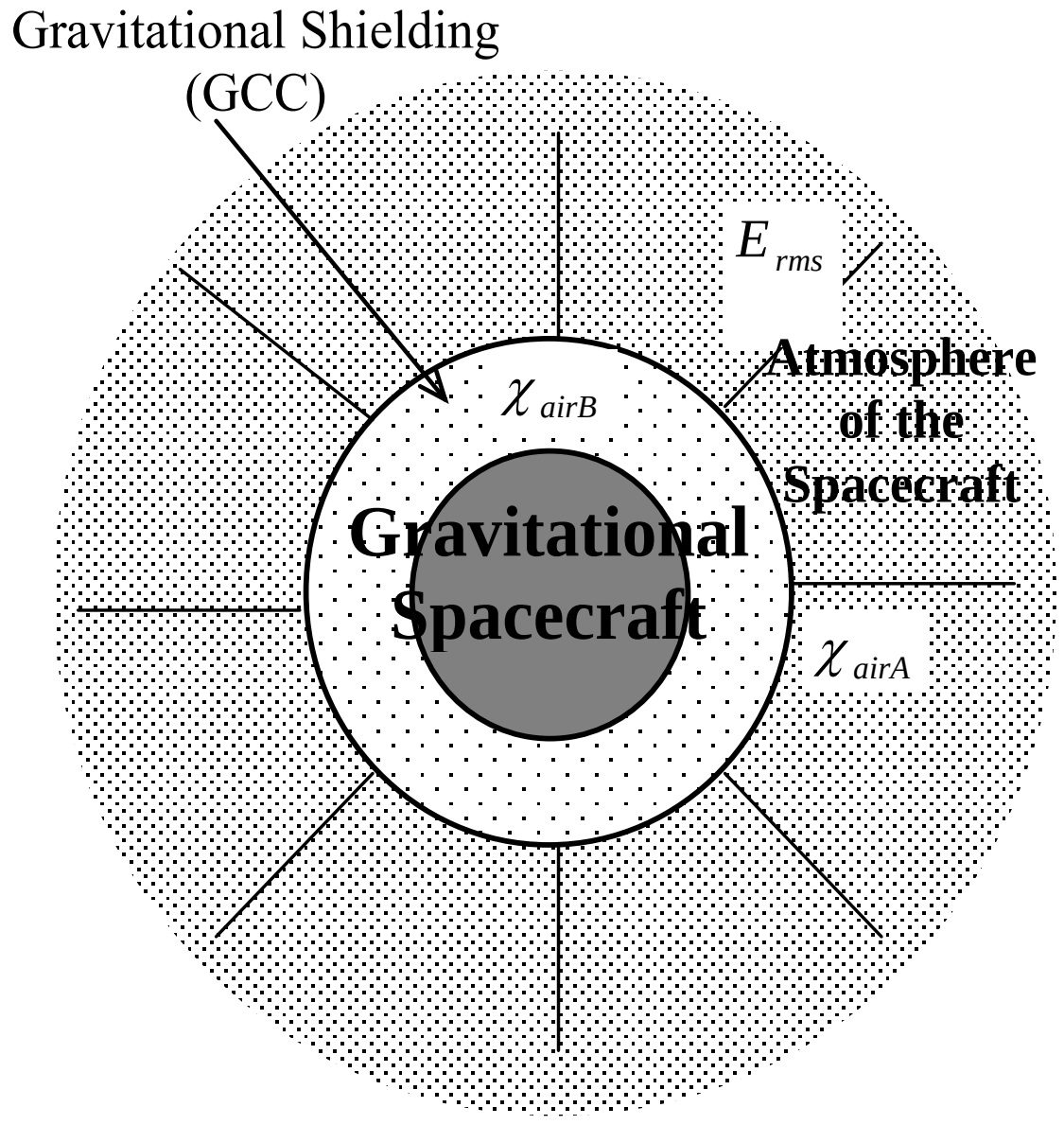

Fig. 6 – *Artificial atmosphere* around the gravitational spacecraft - while inside the gravitational shielding the value of $\chi_{airB}$ is putted close to zero, in order to strongly reduces the gravitational mass of the spacecraft (inner part of the shielding), the value of $\chi_{airA}$ must be reduced for about $-10^8$ in order to strongly increase the gravitational attraction between the air molecules around the spacecraft.

## 5. The Imaginary Space-time

The speed of light in free space is, as we know, about of 300.000 km/s. The speeds of the fastest modern airplanes of the present time do not reach 2 km/s and the speed of rockets do not surpass 20 km/s. This shows how much our aircraft and rockets are slow when compared with the speed of light.

The star nearest to the Earth (excluding the Sun obviously) is the Alpha of Centaur, which is about of *4 light-years* distant from the Earth (Approximately 37.8 trillions of kilometers). Traveling at a speed about 100 times greater than the maximum speed of our faster spacecrafts, we would take about 600 years to reach Alpha of Centaur. Then imagine how many years we would take to leave our own galaxy. In fact, it is not difficult to see that our spacecrafts are very slow, even for travels in our own solar system.

One of the fundamental characteristics of the gravitational spacecraft, as we already saw, is its capability to acquire enormous accelerations without submitting the crew to any discomfort.

Impelled by gravitational thrusters gravitational spacecrafts can acquire accelerations until $10^8 m.s^{-2}$ or more. This means that these spacecrafts can reach speeds very close to the speed of light in just a few seconds. These gigantic accelerations can be unconceivable for a layman, however they are common in our Universe. For example, when we submit an electron to an electric field of just $1\,Volt/m$ it acquires an acceleration $a$, given by

$$a = \frac{eE}{m_e} = \frac{\left(1.6\times10^{-19}C\right)\left(1\ V/m\right)}{9.11\times10^{-31}} \cong 10^{11}m.s^{-2}$$

As we see, this acceleration is about 100 times greater than that acquired by the gravitational spacecraft previously mentioned.

By using the gravitational shieldings it is possible to reduce the inertial effects upon the spacecraft. As we have shown, they are reduced by the factor $\chi_{out} = M_g/M_i$. Thus, if the inertial mass of the spacecraft is $M_i = 10.000kg$ and, by means of the gravitational shielding effect the gravitational mass of the spacecraft is reduced to $M_g \approx 10^{-8}M_i$ then , in spite of the effective acceleration to be gigantic, for example, $a \approx 10^{9}m.s^{-2}$, the effects for the crew of the spacecraft would be *equivalents to* an acceleration $a'$ of only

$$a' = \frac{M_g}{M_i}a = \left(10^{-8}\right)\left(10^{9}\right) \approx 10m.s^{-2}$$

This acceleration is similar to that which the passengers of a contemporary commercial jet are subjected.

Therefore the crew of the gravitational spacecraft would be comfortable while the spacecraft would reach speeds close to the speed of light in few seconds. However to travel at such velocities in the Universe may note be practical. Take for example, Alpha of Centaur (4 light-years far from the Earth): a round trip to it would last about eight years. Trips beyond that star could take then several decades, and this obviously is impracticable. Besides, to travel at such a speed would be very dangerous, because a shock with other celestial bodies would be inevitable. However, as we showed [1] there is a possibility of a spacecraft travel *quickly* far beyond our galaxy without the risk of being destroyed by a sudden shock with some celestial body. The solution is the gravitational spacecraft travel through the *Imaginary* or *Complex Space-time.*

It was shown [1] that it is possible to carry out a transition to the *Imaginary space-time* or *Imaginary Universe*. It is enough that the body has its *gravitational mass* reduced to a value in the range of $+0.159M_i$ to $-0.159M_i$. In these circumstances, the masses of the body (gravitational and inertial) become *imaginaries* and, so does the body. (Fig.7). Consequently, the body disappears from our ordinary space-time and appears in the imaginary space-time. In other words, it becomes *invisible* for an observer at the real Universe. Therefore, this is a way to get temporary *invisibility* of human beings, animals, spacecrafts, etc.

Thus, a spacecraft can leave our Universe and appear in the Imaginary Universe, where it can travel at any speed since in the Imaginary Universe *there is no speed limit for the gravitational spacecraft*, as it occurs in our Universe, where the particles cannot surpass the light speed. In this way, as the gravitational spacecraft is propelled by the gravitational thrusters, it can attain accelerations up to $10^{9}m.s^{-2}$, then after one day of trip with this acceleration, it can reach velocities $V \approx 10^{14} m.s^{-1}$(about 1 million times the speed of light). With this velocity, after 1 month of trip the spacecraft would have traveled about$10^{21} m$. In order to have idea of this distance, it is enough to remind that the diameter of our Universe (visible Universe) is of the order of$10^{26} m$.

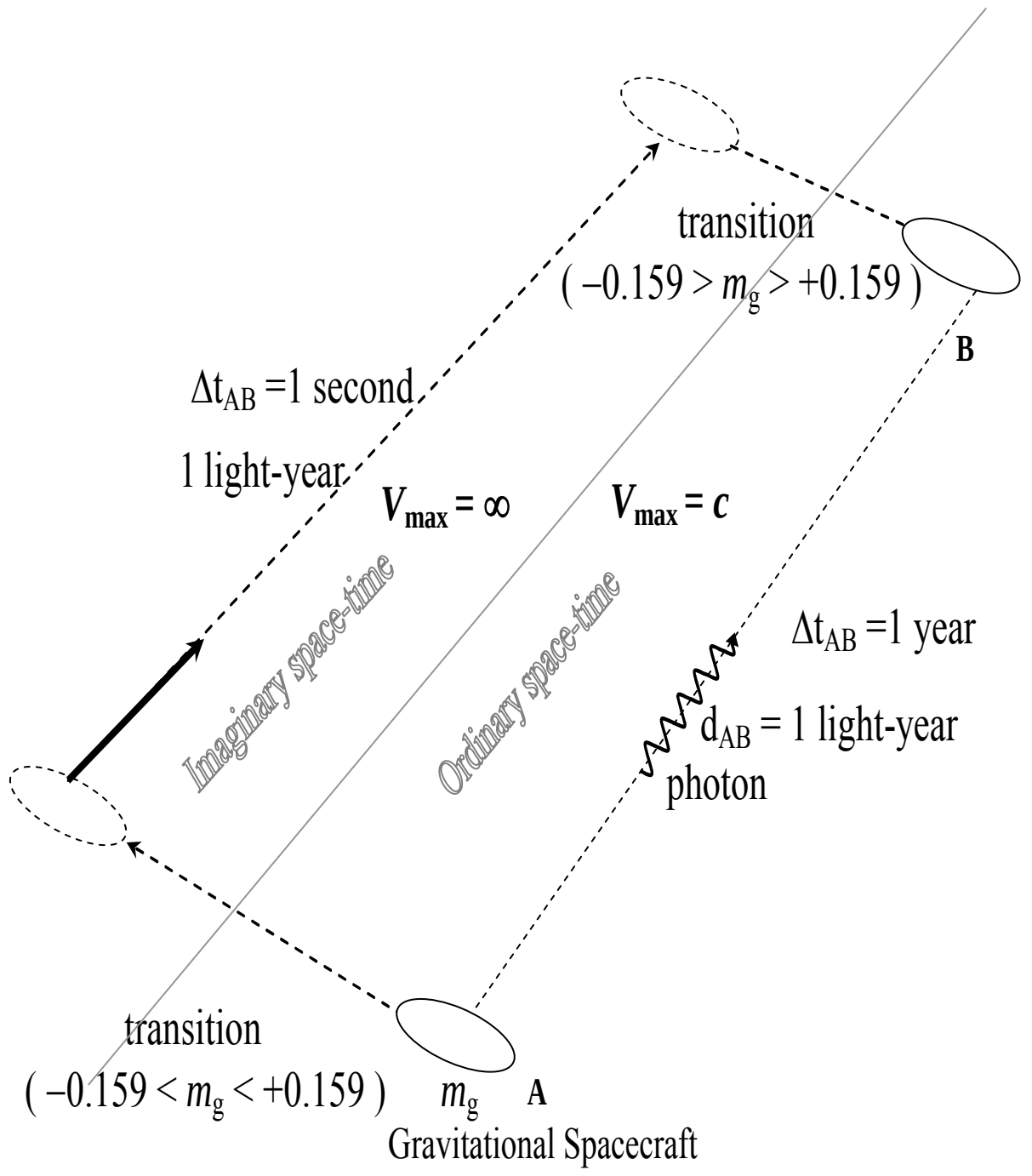

Fig. 7 – Travel in the *Imaginary* Space-time.

Due to the extremely low density of the imaginary bodies, the collision between them cannot have the same consequences of the collision between the dense real bodies.

Thus for a gravitational spacecraft in imaginary state the problem of the collision doesn't exist in high-speed. Consequently, the gravitational spacecraft can transit freely in the imaginary Universe and, in this way reach easily any point of our real Universe once they can make the transition back to our Universe by only increasing the gravitational mass of the spacecraft in such way that it leaves the range of $+0.159M_i$ to$-0.159M_i$. Thus the spacecraft can reappear in our Universe near its target.

The return trip would be done in similar way. That is to say, the spacecraft would transit in the imaginary Universe back to the departure place where would reappear in our Universe and it would make the approach flight to the wanted point. Thus, trips through our Universe that would delay millions of years, at speeds close to the speed of light, could be done in just a few *months* in the imaginary Universe.

What will an observer see when in the imaginary space-time? It will see light, bodies, planets, stars, etc., everything formed by imaginary photons, imaginary atoms, imaginary protons, imaginary neutrons and imaginary electrons. That is to say, the observer will find an Universe similar to ours, just formed by particles with imaginary masses. The term *imaginary* adopted from the Mathematics, as we already saw, gives the false impression that these masses do not exist. In order to avoid this misunderstanding we researched the true nature of that new mass type and matter.

The existence of imaginary mass associated to the *neutrino* is well-known. Although its imaginary mass is not physically observable, its square is. This amount is found experimentally to be negative. Recently, it was shown [1] that *quanta* of imaginary mass exist associated to the *photons, electrons, neutrons,* and *protons*, and that these imaginary masses would have psychic properties (elementary capability of "choice"). Thus, the true nature of this new kind of mass and matter shall be psychic and, therefore we should not use the term *imaginary* any longer. Consequently from the above exposed we can conclude that the gravitational spacecraft penetrates in the *Psychic Universe* and not in an "imaginary" Universe.

In this Universe, the matter would be, obviously composed by psychic molecules and psychic atoms formed by psychic neutrons, psychic protons and psychic electrons. i.e., the matter would have psychic mass and consequently it would be *subtle*, much less dense than the matter of our *real* Universe.

Thus, from a quantum viewpoint, the psychic particles are similar to the material particles, so that we can use the Quantum Mechanics to describe the psychic particles. In this case, by analogy to the material particles, a particle with psychic mass $m_\Psi$ will be described by the following expressions:

$$\vec{p}_\psi = \hbar\vec{k}_\psi$$
$$E_\psi = \hbar\omega_\psi$$

Where $\vec{p}_\psi = m_\Psi\vec{V}$ is the *momentum* carried by the wave and $E_\psi$ its energy; $\left|\vec{k}_\psi\right| = 2\pi/\lambda_\psi$ is the *propagation number* and $\lambda_\psi = h/m_\Psi V$ the *wavelength* and $\omega_\psi = 2\pi f_\psi$ its cyclic *frequency*.

The variable quantity that characterizes DeBroglie's waves is called *Wave Function*, usually indicated by $\Psi$. The wave function associated to a material particle describes the dynamic state of the particle: its value at a particular point x, y, z, t is related to the probability of finding the particle in that place and instant. Although $\Psi$ does not have a physical interpretation, its square $\Psi^2$ (or $\Psi\ \Psi^*$) calculated for a particular point x, y, z, t is *proportional to the probability of experimentally finding the particle in that place and instant*.

Since $\Psi^2$ is proportional to the probability $P$ of finding the particle described by $\Psi$, the integral of $\Psi^2$ on the *whole space* must be finite – inasmuch as the particle is someplace. Therefore, if

$$\int_{-\infty}^{+\infty}\Psi^2 dV = 0$$

The interpretation is that the particle does not exist. Conversely, if

$$\int_{-\infty}^{+\infty}\Psi^2 dV = \infty$$

*the particle will be everywhere simultaneously*.

The wave function $\Psi$ corresponds, as we know, to the displacement $y$ of the undulatory motion of a rope. However, $\Psi$ as opposed to $y$, is not a measurable quantity and can, hence, being a *complex* quantity. For this reason, it is admitted that $\Psi$ is described in the $x$-direction by

$$\Psi = Be^{-(2\pi i/h)(Et-px)}$$

This equation is the mathematical description of the wave associated with a free material particle, with total energy $E$ and *momentum* $p$, moving in the direction $+x$.

As concerns the psychic particle, the variable quantity characterizing psyche waves will also be called wave function, denoted by $\Psi_\Psi$ ( to distinguish it from the material particle wave function), and, by analogy with equation of $\Psi$, expressed by:

$$\Psi_\Psi = \Psi_0 e^{-(2\pi\, i/h)(E_\Psi t - p_\Psi x)}$$

If an experiment involves a large number of identical particles, all described by the same wave function $\Psi$ , the *real* density of mass $\rho$ of these particles in x, y, z, t is proportional to the corresponding value $\Psi^2$ ($\Psi^2$ is known as *density of probability*. If $\Psi$ is *complex* then $\Psi^2 = \Psi\Psi^*$. Thus, $\rho \propto \Psi^2 = \Psi.\Psi^*$). Similarly, in the case of psychic particles, the *density of psychic mass*, $\rho_\Psi$, in x, y, z, will be expressed by $\rho_\Psi \propto \Psi_\Psi^2 = \Psi_\Psi \Psi_\Psi^*$. It is known that $\Psi_\Psi^2$ is always *real* and *positive* while $\rho_\Psi = m_\Psi / V$ is an *imaginary* quantity. Thus, as the *modulus* of an imaginary number is always real and positive, we can transform the proportion $\rho_\Psi \propto \Psi_\Psi^2$, in equality in the following form:

$$\Psi_\Psi^2 = k|\rho_\Psi|$$

Where $k$ is a *proportionality constant* (real and positive) to be determined.

In Quantum Mechanics we have studied the *Superposition Principle*, which affirms that, if a particle (or system of particles) is in a *dynamic state* represented by a wave function $\Psi_1$ and may also be in another dynamic state described by $\Psi_2$ then, the general dynamic state of the particle may be described by $\Psi$, where $\Psi$ is a linear combination (superposition) of $\Psi_1$ and $\Psi_2$, i.e.,

$$\Psi = c_1\Psi_1 + c_2\Psi_2$$

The C*omplex constants* $c_1$ e $c_2$ respectively express the percentage of dynamic state, represented by $\Psi_1$ e $\Psi_2$ in the formation of the general dynamic state described by $\Psi$.

In the case of psychic particles (psychic bodies, consciousness, etc.), by analogy, if $\Psi_{\Psi 1}$, $\Psi_{\Psi 2}$,..., $\Psi_{\Psi n}$ refer to the different dynamic states the psychic particle takes, then its general dynamic state may be described by the wave function $\Psi_\Psi$ , given by:

$$\Psi_\Psi = c_1\Psi_{\Psi 1} + c_2\Psi_{\Psi 2} + \ldots + c_n\Psi_{\Psi n}$$

The state of superposition of wave functions is, therefore, common for both psychic and material particles. In the case of material particles, it can be verified, for instance, when an electron changes from one orbit to another. Before effecting the transition to another energy level, the electron carries out "virtual transitions" [6]. A kind of *relationship* with other electrons before performing the real transition. During this relationship period, its wave function remains "*scattered*" by *a wide region of the space* [7] thus superposing the wave functions of the other electrons. In this relationship the electrons *mutually* influence each other, with the possibility of *intertwining* their wave functions[§]. When this happens, there occurs the so-called *Phase Relationship* according to quantum-mechanics concept.

In the electrons "virtual" transition mentioned before, the "listing" of all the possibilities of the electrons is described, as we know, by *Schrödinger's wave equation.*

[§] Since the electrons are simultaneously waves and particles, their wave aspects will interfere with each other; besides superposition, there is also the possibility of occurrence of *intertwining* of their wave functions.

Otherwise, it is general for material particles. By analogy, in the case of psychic particles, we may say that the "listing" of all the possibilities of the psyches involved in the relationship will be described by *Schrödinger's equation* – for psychic case, i.e.,

$$\nabla^2\Psi_\Psi + \frac{p_\Psi^2}{\hbar^2}\Psi_\Psi = 0$$

Because the wave functions are capable of intertwining themselves, the quantum systems may "penetrate" each other, thus establishing an internal relationship where all of them are affected by the relationship, no longer being isolated systems but becoming an integrated part of a larger system. This type of internal relationship, which exists only in quantum systems, was called *Relational Holism* [8].

We have used the Quantum Mechanics in order to describe the foundations of the Psychic Universe which the Gravitational Spacecrafts will find, and that influences us daily. These foundations recently discovered – particularly the *Psychic Interaction*, show us that a rigorous description of the Universe cannot to exclude the psychic energy and the psychic particles. This verification makes evident the need of to redefine the Psychology with basis on the quantum foundations recently discovered. This has been made in the article: "*Physical Foundations of Quantum Psychology*"** [9], recently published, where it is shown that the Psychic Interaction leads us to understand the Psychic Universe and the extraordinary relationship that the human consciousnesses establish among themselves and with the Ordinary Universe. Besides, we have shown that the Psychic Interaction postulates a new model for the evolution theory, in which the evolution is interpreted not only as a biological fact, but mainly as *psychic* fact. Therefore, is not only the mankind that evolves in the Earth's planet, but all the ecosystem of the Earth.

** http://htpprints.yorku.ca/archive/00000297

## 6. Past and Future

It was shown [1,9] that the *collapse* of the *psychic* wave function must suddenly also express in reality (*real* space-time) all the possibilities described by it. This is, therefore, *a point of decision* in which there occurs the compelling need of *realization* of the *psychic form*. We have seen that the *materialization* of the psychic form, in the real space-time, occurs when it contains enough *psychic mass* for the total materialization†† of the psychic form (*Materialization Condition*). When this happens, all the psychic energy contained in the psychic form is transformed in real energy in the real space-time. Thus, in the psychic space-time just the *holographic* register of the psychic form, which gives origin to that fact, survives, since the psychic energy deforms the *metric* of the psychic space-time‡‡, producing the holographic register. Thus, the past survive in the psychic space-time just in the form of holographic register. That is to say, all that have occurred in the past is holographically registered in the *psychic space-time*. Further ahead, it will be seen that this register can be accessed by an observer in the *psychic* space-time as well as by an observer in the *real* space-time.

†† By this we mean not only materialization proper but also the movement of matter to realize its psychic content (including radiation).

‡‡ As shown in *General Theory of Relativity* the energy modifies the metric of the space-time (deforming the space-time).

A psychic form is intensified by means of a continuous addition of psychic mass. Thus, when it acquires sufficiently psychic mass, its realization occurs in the real space-time. Thus the future is going being built in the present. By means of our current thoughts we shape the psychic forms that will go (or will not) take place in the future. Consequently, those psychic forms are continually being holographically registered in the *psychic space-time* and, just as the holographic registrations of the *past* these future registration can also be accessed by the *psychic* space-time as well as by the *real* space-time.

The access to the holographic registration of the past doesn't allow, obviously, the modification of the past. This is not possible because there would be a clear violation of the *principle of causality* that says that the causes should precede the effects. However, the psychic forms that are being shaped now in order to manifest themselves in the future, can be modified before they manifest themselves. Thus, the access to the registration of those psychic forms becomes highly relevant for our present life, since we can avoid the manifestation of many unpleasant facts in the future.

Since both registrations are in the *psychic* space-time, then the access to their information only occur by means of the interaction with another psychic body, for example, our *consciousness* or a *psychic observer* (body totally formed by psychic mass). We have seen that, if the gravitational mass of a body is reduced to within the range $+0.159M_i$ to $-0.159M_i$, its gravitational and inertial masses become *imaginaries* (*psychics*) and, therefore, the body becomes a *psychic body*. Thus, a *real observer* can also become in a *psychic observer*. In this way, a gravitational spacecraft can transform all its inertial mass into psychic mass, and thus carry out a transition to the psychic space-time and become a psychic spacecraft. In these circumstances, an observer inside the spacecraft also will have its mass transformed into psychic mass, and, therefore, the observer also will be transformed into a psychic observer. What will this observer see when it penetrates the psychic Universe? According to the *Correspondence principle*, all that exists in the real Universe must have the correspondent in the psychic Universe and vice-versa. This principle reminds us that we live in more than one world. At the present time, we live in the real Universe, but we can also live in the psychic Universe. Therefore, the psychic observer will see the psychic bodies and their correspondents in the real Universe. Thus, a pilot of a gravitational spacecraft, in travel through the psychic space-time, won't have difficulty to spot the spacecraft in its trips through the Universe.

The fact of the psychic forms manifest themselves in the real space-time exactly at its images and likeness, it indicates that real forms (forms in the real space-time) are prior to all reflective *images* of psychic forms of the past. Thus, the real space-time is a mirror of the psychic space-time. Consequently, any register in the psychic space-time will have a correspondent image in the real space-time. This means that it is possible that we find in the real space-time the *image* of the holographic register existing in the psychic space-time, corresponding to our *past*. Similarly, every psychic form that is being shaped in the psychic space-time will have reflective image in the real space-time. Thus, the *image* of the holographic register of our future (existing in the psychic space-time) can also be found in the *real* space-time.

Each image of the holographic register of our future will be obviously correlated to a future epoch in the temporal coordinate of the space-time. In the same way, each image of the holographic registration of our past will be correlated to a passed time in the temporal coordinate of the referred space-time. Thus, in order to access the mentioned registrations we should accomplish trips to the past or future in the real space-time. This is possible now, with the advent of the gravitational spacecrafts because they allow us to reach speeds close to the speed of light. Thus, by varying the gravitational mass of the spacecraft for *negative* or *positive* we can go respectively to the *past* or *future* [1].

If the gravitational mass of a particle is *positive*, then $t$ is always *positive* and given by

$$t=+t_0\Big/\sqrt{1-V^2/c^2}$$

This leads to the well-known relativistic prediction that the particle goes to the *future* if $V\rightarrow c$. However, if the gravitational mass of the particle is negative, then $t$ is also *negative* and, therefore, given by

$$t=-t_0\Big/\sqrt{1-V^2/c^2}$$

In this case, the prevision is that the particle goes to the *past* if $V\rightarrow c$. In this way, *negative gravitational mass* is the necessary condition to the particle to go to the *past*.

Since the acceleration of a spacecraft with gravitational mass $m_g$, is given by $a=F/m_g$, where $F$ is the thrust of its thrusters, then the more we reduce the value of $m_g$ the bigger the acceleration of the spacecraft will be. However, since the value of $m_g$ cannot be reduced to the range $+0.159M_i$ to $-0.159M_i$ because the spacecraft would become a psychic body, and it needs to remain in the real space-time in order to access the past or the future in the real space-time, then, the ideal values for the spacecraft to operate with safety would be $\pm 0.2m_i$. Let us consider a gravitational spacecraft whose inertial mass is $m_i=10.000kg$. If its gravitational mass was made *negative* and equal to $m_g=-0.2m_i=-2000kg$ and, at this instant the thrust produced by the thrusters of the spacecraft was $F = 10^5 N$ then, the spacecraft would acquire acceleration $a = F/m_g = 50 m.s^{-2}$ and, after $t = 30 days = 2.5 \times 10^6 s$, the speed of the spacecraft would be $v = 1.2 \times 10^8 m.s^{-1} = 0.4c$. Therefore, right after that the spacecraft returned to the Earth, its crew would find the Earth in the *past* (due to the *negative* gravitational mass of the spacecraft) at a time $t = -t_0 / \sqrt{1 - V^2/c^2}$ ; $t_0$ is the time measured by an observer at rest on the Earth. Thus, if $t_0 = 2009$ AD, the time interval $\Delta t = t - t_0$ would be expressed by

$$\Delta t = t - t_0 = -t_0 \left( \frac{1}{\sqrt{1 - V^2/c^2}} - 1 \right) = -t_0 \left( \frac{1}{\sqrt{1 - 0.16}} - 1 \right) \cong$$

$$\cong -0.091 t_0 \cong -183 years$$

That is, the spacecraft would be in the year 1826 AD. On the other hand, if the gravitational mass of the spacecraft would have become positive $m_g = +0.2 m_i = +2000 kg$, instead of negative, then the spacecraft would be in the future at $\Delta t = +183$ *years* from 2009. That is, it would be in the year 2192 AD.

## 7. Instantaneous Interestelar Communications

Consider a cylindrical GCC (GCC antenna) as shown in Fig.8. The *gravitational mass* of the *air* inside the GCC is

$$m_{g(air)} = \left\{ 1 - 2 \left[ \sqrt{1 + \frac{\sigma_{(air)} B^4}{4\pi f \mu \rho^2_{(air)} c^2}} - 1 \right] \right\} m_{i(air)} \quad (24)$$

Where $\sigma_{(ar)}$ is the electric conductivity of the ionized air inside the GCC and $\rho_{(ar)}$ is its density; $f$ is the frequency of the magnetic field.

By varying $B$ one can vary $m_{g(air)}$ and consequently to vary the gravitational field generated by $m_{g(air)}$, producing then *Gravitational Radiation*. Then a GCC can work as a *Gravitational Antenna*.

Apparently, Newton's theory of gravity had no gravitational waves because, if a gravitational field changed in some way, that change would have taken place *instantaneously* everywhere in space, and one can think that there is not a wave in this case. However, we have already seen that the gravitational interaction can be repulsive, besides attractive. Thus, as with electromagnetic interaction, the gravitational interaction must be produced by the exchange of "virtual" *quanta of* spin 1 and mass null, i.e., the *gravitational* "virtual" *quanta* (*graviphoton*) must have spin 1 and not 2. Consequently, the fact that a change in a gravitational field reaches *instantaneously* every point in space occurs simply due to the speed of the *graviphoton* to be *infinite*. It is known that there is no speed limit for "*virtual*" photons. On the other hand, the *electromagnetic quanta* ("virtual" photons) can not communicate the *electromagnetic interaction* to an infinite distance.

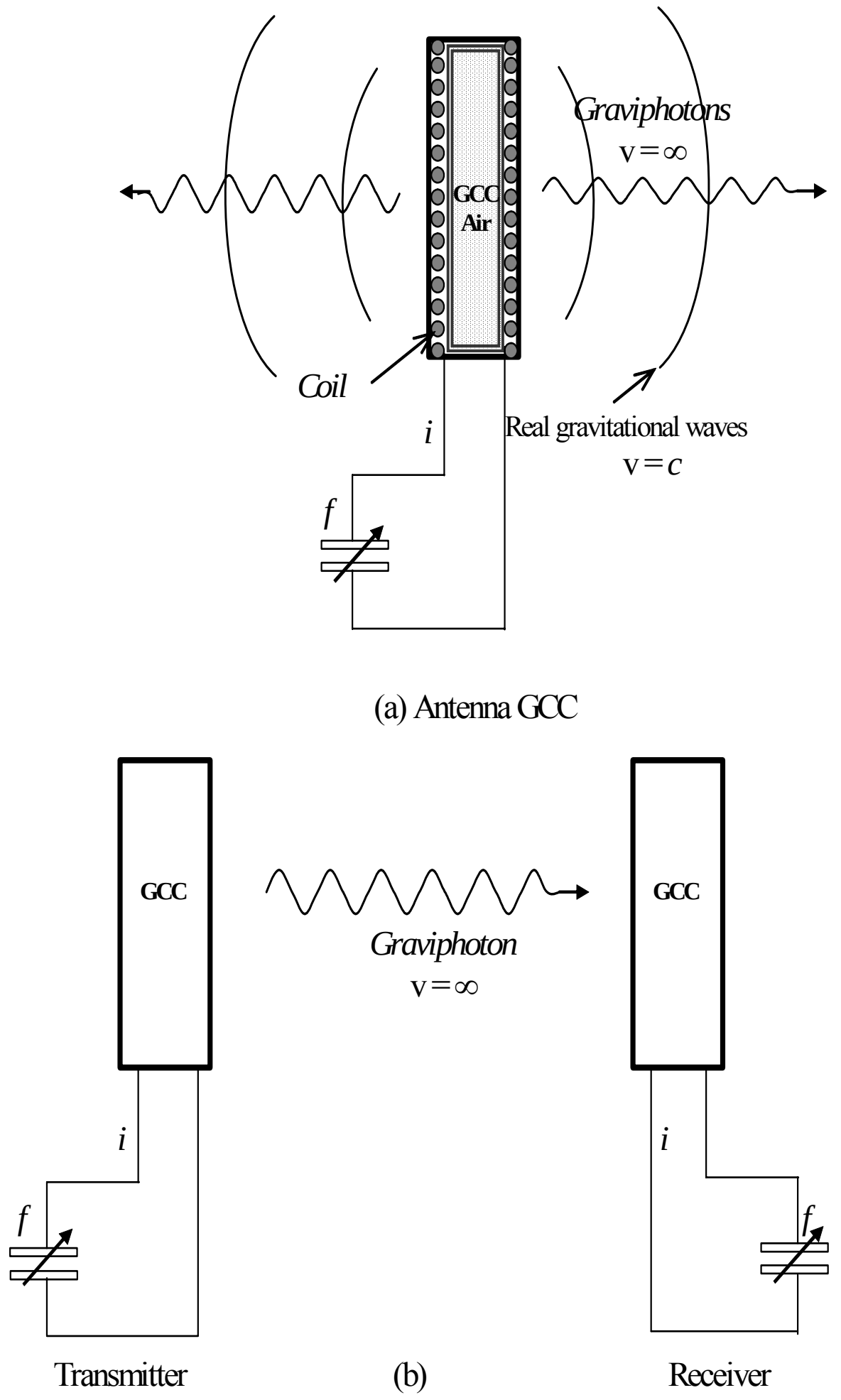

Fig. 8 – Transmitter and Receiver of *Virtual* Gravitational Radiation.

Thus, there are *two types* of gravitational radiation: the *real* and *virtual*, which is constituted of graviphotons; the *real* gravitational waves are ripples in the space-time generated by *gravitational field* changes. According to Einstein's theory of gravity the velocity of propagation of these waves is equal to the speed of light [10].

Unlike the electromagnetic waves the *real* gravitational waves have low interaction with matter and consequently low scattering. Therefore *real* gravitational waves are suitable as a means of transmitting information. However, when the distance between transmitter and receiver is too large, for example of the order of magnitude of several light-years, the transmission of information by means of gravitational waves becomes impracticable due to the long time necessary to receive the information. On the other hand, there is no delay during the transmissions by means of *virtual* gravitational radiation. In addition, the scattering of this radiation is null. Therefore the *virtual* gravitational radiation is very suitable as a means of transmitting information at any distances, including astronomical distances.

As concerns detection of the *virtual* gravitational radiation from GCC antenna, there are many options. Due to *Resonance Principle* a similar GCC antenna (receiver) *tuned at the same frequency* can absorb energy from an incident *virtual* gravitational radiation (See Fig.8 (b)). Consequently, the gravitational mass of the air inside the GCC receiver will vary such as the gravitational mass of the air inside the GCC transmitter. This will induce a magnetic field similar to the magnetic field of the GCC transmitter and therefore the current through the coil inside the GCC receiver will have the same characteristics of the current through the coil inside the GCC transmitter. However, the *volume* and *pressure* of the air inside the two GCCs must be exactly the same; also the *type* and the *quantity of atoms* in the air inside the two GCCs must be exactly the same. Thus, the GCC antennas are simple but they are not easy to build.

Note that a GCC antenna radiates *graviphotons* and *gravitational waves* simultaneously (Fig. 8 (a)). Thus, it is not only a gravitational antenna: it is a *Quantum Gravitational Antenna* because it can also emit and detect gravitational "virtual" *quanta* (graviphotons), which, in turn, can transmit information *instantaneously* from any distance in the Universe *without* scattering.

Due to the difficulty to build two similar GCC antennas and, considering that the electric current in the receiver antenna can be detectable even if the gravitational mass of the nuclei of the antennas are not *strongly* reduced, then we propose to replace the gas at the nuclei of the antennas by a thin *dielectric lamina*. When the *virtual* gravitational radiation strikes upon the dielectric lamina, its gravitational mass varies similarly to the gravitational mass of the dielectric lamina of the transmitter antenna, inducing an electromagnetic field $(E, B)$ similar to the transmitter antenna. Thus, the electric current in the receiver antenna will have the same characteristics of the current in the transmitter antenna. In this way, it is then possible to build two similar antennas whose nuclei have the same volumes and the same types and quantities of atoms.

Note that the Quantum Gravitational Antennas can also be used to transmit *electric power*. It is easy to see that the Transmitter and Receiver can work with strong voltages and electric currents. This means that strong electric power can be transmitted among Quantum Gravitational Antennas. This obviously solves the problem of *wireless* electric power transmission. Thus, we can conclude that the spacecrafts *do not necessarily need* to have a system for generation of electric energy inside them. Since the electric energy to be used in the spacecraft can be *instantaneously transmitted* from *any point of the Universe*, by means of the above mentioned systems of transmission and reception of "virtual" gravitational waves.

## 8. Origin of Gravity and Genesis of the Gravitational Energy

It was shown [1] that the "virtual" *quanta* of the *gravitational interaction* must have spin 1 and not 2, and that they are "virtual" photons (*graviphotons*) with *zero mass* outside the *coherent* matter. Inside the coherent matter the graviphotons mass is *non-zero*. Therefore, the gravitational forces are also *gauge* forces, because they are yielded by the exchange of "virtual" *quanta* of spin 1, such as the electromagnetic forces and the weak and strong nuclear forces.

Thus, the gravitational forces are produced by the exchanging of "virtual" photons (Fig.9). Consequently, this is precisely the *origin of the gravity*.

Newton's theory of gravity does not explain *why* objects attract one another; it simply models this observation. Also Einstein's theory does not explain the origin of gravity. Einstein's theory of gravity only describes gravity with more precision than Newton's theory does.

Besides, there is nothing in both theories explaining the *origin of the energy* that produces the gravitational forces. Earth's gravity attracts all objects on the surface of our planet. This has been going on for over 4.5 billions years, yet no known energy source is being converted to support this tremendous ongoing energy expenditure. Also is the enormous continuous energy expended by Earth's gravitational field for maintaining the Moon in its orbit - millennium after millennium. In spite of the ongoing energy expended by Earth's gravitational field to hold objects down on surface and the Moon in orbit, why the energy of the field never diminishes in strength or drains its energy source? Is this energy expenditure balanced by a conversion of energy from an unknown energy source?

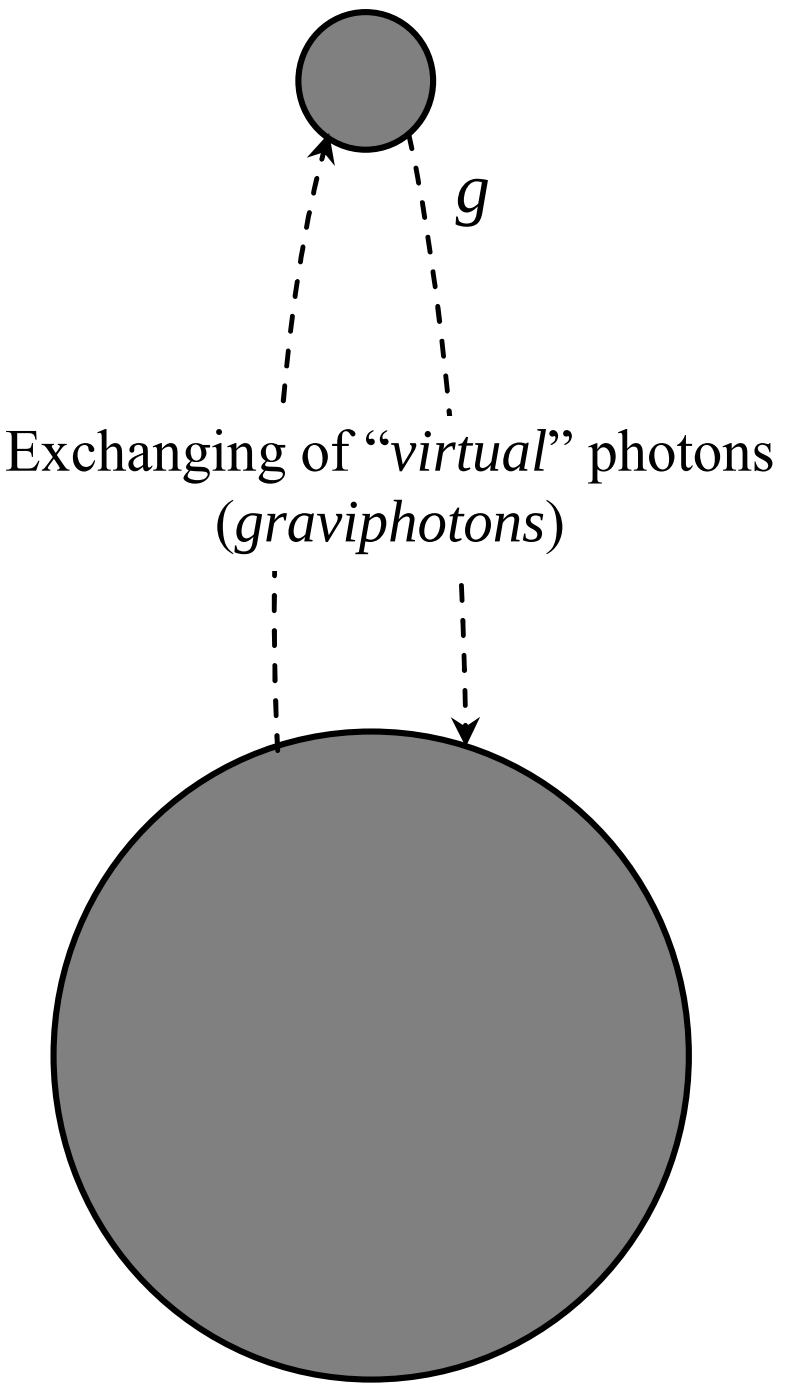

Fig. 9 – *Origin of Gravity*: The gravitational forces are produced by the exchanging of "virtual" photons (*graviphotons*).

The energy $W$ necessary to support the effort expended by the gravitational forces $F$ is well-known and given by

$$W = \int_{\infty}^{r} F dr = -G \frac{M_g m_g}{r}$$

According to the *Principle of Energy Conservation*, the spending of this energy must be compensated by a conversion of another type of energy.

The Uncertainty Principle tells us that, due to the occurrence of exchange of *graviphotons* in a time interval $\Delta t < \hbar / \Delta E$ (where $\Delta E$ is the energy of the graviphoton), the energy variation $\Delta E$ cannot be detected in the system $M_g - m_g$. Since the total energy $W$ is the sum of the energy of the $n$ graviphotons, i.e., $W = \Delta E_1 + \Delta E_2 + ... + \Delta E_n$, then the energy $W$ *cannot be detected as well*. However, as we know it can be converted into another type of energy, for example, in rotational kinetic energy, as in the hydroelectric plants, or in the *Gravitational Motor*, as shown in this work.

It is known that a *quantum* of energy $\Delta E = hf$, which varies during a time interval $\Delta t = 1/f = \lambda/c < \hbar/\Delta E$ (wave period) cannot be experimentally detected. This is an *imaginary* photon or a "*virtual*" photon. Thus, the graviphotons are *imaginary* photons, i.e., the energies $\Delta E_i$ of the graviphotons are imaginaries energies and therefore the energy $W = \Delta E_1 + \Delta E_2 + ... + \Delta E_n$ is also an *imaginary* energy. Consequently, it belongs to the *imaginary space-time*.

It was shown [1] that, imaginary energy is equal to *psychic energy*. Consequently, the *imaginary space-time* is, in fact, the *psychic space-time*, which contains the Supreme Consciousness. Since the Supreme Consciousness has *infinite* psychic mass [1], then the *psychic space-time* contains *infinite psychic energy*. This is highly relevant, because it confers to the *Psychic Universe* the characteristic of *unlimited source of energy*. Thus, as the origin of the gravitational energy it is correlated to the psychic energy, then the spending of gravitational energy can be supplied *indefinitely* by the Psychic Universe.

This can be easily confirmed by the fact that, in spite of the enormous amount of energy expended by Earth's gravitational field to hold objects down on the surface of the planet and maintain the Moon in its orbit, the energy of Earth's gravitational field never diminishes in strength or drains its energy source.

***Acknowledgement***

The author would like to thank Dr. *Getúlio Marques Martins* (COPPE – UFRJ, Rio de Janeiro-Brasil) for revising the manuscript.

# APPENDIX A: The Simplest Method to Control the Gravity

In this Appendix we show the simplest method to control the gravity.

Consider a body with mass density $\rho$ and the following electric characteristics: $\mu_r, \varepsilon_r, \sigma$ (relative permeability, relative permittivity and electric conductivity, respectively). Through this body, passes an electric current $I$, which is the sum of a sinusoidal current $i_{osc} = i_0 \sin \omega t$ and the DC current $I_{DC}$, i.e., $I = I_{DC} + i_0 \sin \omega t$ ; $\omega = 2\pi f$. If $i_0 << I_{DC}$ then $I \cong I_{DC}$. Thus, the current $I$ varies with the frequency $f$, but the variation of its intensity is quite small in comparison with $I_{DC}$, i.e., $I$ will be practically constant (Fig. A1). This is of fundamental importance for maintaining the value of the gravitational mass of the body, $m_g$, sufficiently stable during all the time.

The *gravitational mass* of the *body* is given by [1]

$$m_g = \left\{1 - 2\left[\sqrt{1 + \left(\frac{n_r U}{m_{i0} c^2}\right)^2} - 1\right]\right\} m_{i0} \qquad (A1)$$

where $U$, is the electromagnetic energy absorbed by the body and $n_r$ is the index of refraction of the body.

Equation (A1) can also be rewritten in the following form

$$\frac{m_g}{m_{i0}} = \left\{1 - 2\left[\sqrt{1 + \left(\frac{n_r W}{\rho c^2}\right)^2} - 1\right]\right\} \qquad (A2)$$

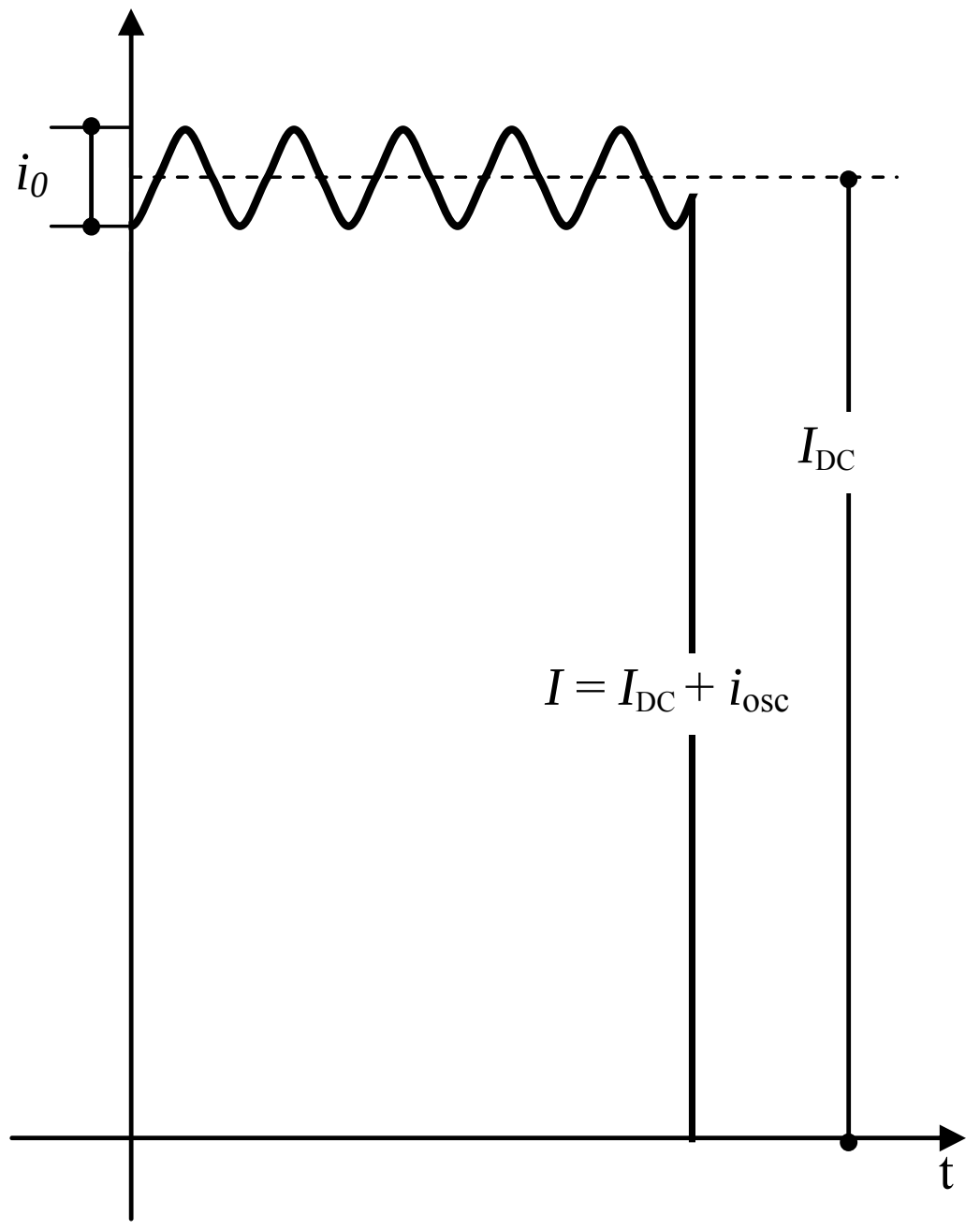

Fig. A1 - The electric current $I$ varies with frequency $f$. But the variation of $I$ is quite small in comparison with $I_{DC}$ due to $i_o << I_{DC}$. In this way, we can consider $I \cong I_{DC}$.

where, $W = U/V$ is the *density of electromagnetic energy* and $\rho = m_{i0}/V$ is the density of inertial mass.

The *instantaneous values* of the density of electromagnetic energy in an *electromagnetic* field can be deduced from Maxwell's equations and has the following expression

$$W = \tfrac{1}{2}\varepsilon E^2 + \tfrac{1}{2}\mu H^2 \qquad (A3)$$

where $E = E_m \sin \omega t$ and $H = H \sin \omega t$ are the *instantaneous values* of the electric field and the magnetic field respectively.

It is known that $B = \mu H$, $E/B = \omega/k_r$ [11] and

$$v=\frac{dz}{dt}=\frac{\omega}{\kappa_r}=\frac{c}{\sqrt{\frac{\varepsilon_r\mu_r}{2}\left(\sqrt{1+(\sigma/\omega\varepsilon)^2}+1\right)}} \qquad (A4)$$

where $k_r$ is the real part of the *propagation vector* $\vec{k}$ (also called *phase constant* ); $k=\left|\vec{k}\right|=k_r+ik_i$ ; $\varepsilon$ , $\mu$ and $\sigma$, are the electromagnetic characteristics of the medium in which the incident (or emitted) radiation is propagating( $\varepsilon=\varepsilon_r\varepsilon_0$ ; $\varepsilon_0=8.854\times10^{-12}F/m$ ; $\mu=\mu_r\mu_0$ where $\mu_0=4\pi\times10^{-7}H/m$). It is known that for *free-space* $\sigma=0$ and $\varepsilon_r=\mu_r=1$. Then Eq. (A4) gives

$$v=c$$

From (A4), we see that the *index of refraction* $n_r=c/v$ is given by

$$n_r=\frac{c}{v}=\sqrt{\frac{\varepsilon_r\mu_r}{2}\left(\sqrt{1+(\sigma/\omega\varepsilon)^2}+1\right)} \qquad (A5)$$

Equation (A4) shows that $\omega/\kappa_r=v$. Thus, $E/B=\omega/k_r=v$, i.e.,

$$E=vB=v\mu H \qquad (A6)$$

Then, Eq. (A3) can be rewritten in the following form:

$$W=\tfrac{1}{2}\left(\varepsilon v^2\mu\right)\mu H^2+\tfrac{1}{2}\mu H^2 \qquad (A7)$$

For $\sigma<<\omega\varepsilon$, Eq. (A4) reduces to

$$v=\frac{c}{\sqrt{\varepsilon_r\mu_r}}$$

Then, Eq. (A7) gives

$$W=\tfrac{1}{2}\left[\varepsilon\left(\frac{c^2}{\varepsilon_r\mu_r}\right)\mu\right]\mu H^2+\tfrac{1}{2}\mu H^2=\mu H^2$$

This equation can be rewritten in the following forms:

$$W=\frac{B^2}{\mu} \qquad (A8)$$

or

$$W=\varepsilon\, E^2 \qquad (A9)$$

For $\sigma>>\omega\varepsilon$, Eq. (A4) gives

$$v=\sqrt{\frac{2\omega}{\mu\sigma}} \qquad (A10)$$

Then, from Eq. (A7) we get

$$W=\tfrac{1}{2}\left[\varepsilon\left(\frac{2\omega}{\mu\sigma}\right)\mu\right]\mu H^2+\tfrac{1}{2}\mu H^2=\left(\frac{\omega\varepsilon}{\sigma}\right)\mu H^2+\tfrac{1}{2}\mu H^2\cong$$

$$\cong\tfrac{1}{2}\mu H^2 \qquad (A11)$$

Since $E=vB=v\mu H$, we can rewrite (A11) in the following forms:

$$W\cong\frac{B^2}{2\mu} \qquad (A12)$$

or

$$W\cong\left(\frac{\sigma}{4\omega}\right)E^2 \qquad (A13)$$

By comparing equations (A8) (A9) (A12) and (A13), we can see that Eq. (A13) shows that the best way to obtain a strong value of $W$ *in practice* is by applying an *Extra Low-Frequency* (ELF) *electric field* $(w=2\pi f<<1Hz)$ through a *medium with high electrical conductivity.*

Substitution of Eq. (A13) into Eq. (A2), gives

$$m_g=\left\{1-2\left[\sqrt{1+\frac{\mu}{4c^2}\left(\frac{\sigma}{4\pi f}\right)^3\frac{E^4}{\rho^2}}-1\right]\right\}m_{i0}=$$

$$=\left\{1-2\left[\sqrt{1+\left(\frac{\mu_0}{256\pi^3c^2}\right)\left(\frac{\mu_r\sigma^3}{\rho^2 f^3}\right)E^4}-1\right]\right\}m_{i0}=$$

$$=\left\{1-2\left[\sqrt{1+1.758\times10^{-27}\left(\frac{\mu_r\sigma^3}{\rho^2 f^3}\right)E^4}-1\right]\right\}m_{i0}$$

$$(A14)$$

Note that $E=E_m\sin\omega t$. The average value for $E^2$ is equal to $\tfrac{1}{2}E_m^2$ because $E$ varies sinusoidaly ($E_m$ is the maximum value for $E$). On the other hand, $E_{rms} = E_m/\sqrt{2}$. Consequently we can change $E^4$ by $E_{rms}^4$, and the equation above can be rewritten as follows

$$m_g = \left\{1-2\left[\sqrt{1+1.758\times10^{-27}\left(\frac{\mu_r\sigma^3}{\rho^2 f^3}\right)E_{rms}^4}-1\right]\right\}m_{i0}$$

Substitution of the well-known equation of the *Ohm's vectorial Law*: $j = \sigma E$ into (A14), we get

$$m_g = \left\{1-2\left[\sqrt{1+1.758\times10^{-27}\frac{\mu_r j_{rms}^4}{\sigma\rho^2 f^3}}-1\right]\right\}m_{i0} \qquad (A15)$$

where $j_{rms} = j/\sqrt{2}$.

Consider a 15 cm square *Aluminum thin foil* of *10.5 microns thickness* with the following characteristics: $\mu_r = 1$ ; $\sigma = 3.82\times10^7 S.m^{-1}$; $\rho = 2700\,Kg.m^{-3}$. Then, (A15) gives

$$m_g = \left\{1-2\left[\sqrt{1+6.313\times10^{-42}\frac{j_{rms}^4}{f^3}}-1\right]\right\}m_{i0} \qquad (A16)$$

Now, consider that the ELF electric current $I = I_{DC} + i_0 \sin \omega t$, $(i_0 << I_{DC})$ passes through that Aluminum foil. Then, the current density is

$$j_{rms} = \frac{I_{rms}}{S} \cong \frac{I_{DC}}{S} \qquad (A17)$$

where

$$S = 0.15m\left(10.5\times10^{-6}m\right) = 1.57\times10^{-6}m^2$$

If the ELF electric current has frequency $f = 2\mu Hz = 2\times10^{-6}Hz$, then, the gravitational mass of the aluminum foil, given by (A16), is expressed by

$$m_g = \left\{1-2\left[\sqrt{1+7.89\times10^{-25}\frac{I_{DC}^4}{S^4}}-1\right]\right\}m_{i0} =$$

$$= \left\{1-2\left[\sqrt{1+0.13 I_{DC}^4}-1\right]\right\}m_{i0} \qquad (A18)$$

Then,

$$\chi = \frac{m_g}{m_{i0}} \cong \left\{1-2\left[\sqrt{1+0.13 I_{DC}^4}-1\right]\right\} \qquad (A19)$$

For $I_{DC} = 2.2A$, the equation above gives

$$\chi = \left(\frac{m_g}{m_{i0}}\right) \cong -1 \qquad (A20)$$

This means that *the gravitational shielding* produced by the aluminum foil can change the gravity acceleration *above* the foil down to

$$g' = \chi\, g \cong -1g \qquad (A21)$$

Under these conditions, the Aluminum foil works basically as a Gravity Control Cell (GCC).

In order to check these theoretical predictions, we suggest an experimental set-up shown in Fig.A2.

A 15cm square Aluminum foil of *10.5 microns thickness* with the following composition: Al 98.02%; Fe 0.80%; Si 0.70%; Mn 0.10%; Cu 0.10%; Zn 0.10%; Ti 0.08%; Mg 0.05%; Cr 0.05%, and with the following characteristics: $\mu_r = 1$; $\sigma = 3.82\times10^7 S.m^{-1}$; $\rho = 2700 Kg.m^{-3}$, is fixed on a 17 cm square *Foam Board* §§ plate of 6mm thickness as shown in Fig.A3. This device (the simplest

§§ *Foam board* is a very strong, *lightweigh*t (density: 24.03 kg.m$^{-3}$) and easily cut material used for the mounting of photographic prints, as backing in picture framing, in 3D design, and in painting. It consists of three layers — an inner layer of polystyrene clad with outer facing of either white clay coated paper or brown Kraft paper.

Gravity Control Cell GCC) is placed on a pan balance shown in Fig.A2.

Above the Aluminum foil, a *sample* (any type of material, any mass) connected to a dynamometer will check the decrease of the *local gravity acceleration* upon the sample $(g' = \chi\, g)$, due to the gravitational shielding produced by the decreasing of gravitational mass of the Aluminum foil $(\chi = m_g / m_{i0})$. Initially, the sample lies 5 cm above the Aluminum foil. As shown in Fig.A2, the board with the dynamometer can be displaced up to few meters in height. Thus, the initial distance between the Aluminum foil and the sample can be increased in order to check the reach of the gravitational shielding produced by the Aluminum foil.

In order to generate the ELF electric current of $f = 2\mu Hz$, we can use the widely-known Function Generator HP3325A (Op.002 High Voltage Output) that can generate sinusoidal voltages with *extremely-low* frequencies down to $f = 1\times 10^{-6}\,Hz$ and amplitude up to 20V ($40V_{pp}$ into $500\Omega$ load). The maximum output current is $0.08A_{pp}$; output impedance <2Ω at ELF .

Figure A4 (a) shows the equivalent electric circuit for the experimental set-up. The electromotive forces are: $\varepsilon_1$ (HP3325A) and $\varepsilon_2$ (*12V* DC Battery).The values of the *resistors* are : $R_1 = 500\Omega - 2W$ ; $r_{i1} < 2\Omega$ ; $R_2 = 4\Omega - 40W$ ; $r_{i2} < 0.1\Omega$ ; $R_p = 2.5\times 10^{-3}\Omega$ ; *Rheostat* (0≤ $R$ ≤10Ω - 90W). The *coupling transformer* has the following characteristics: air core with diameter $\phi = 10mm$ ; area $S = \pi\phi^2/4 = 7.8\times 10^{-5}m^2$ ; wire#12AWG; $N_1 = N_2 = N = 20$; $l = 42mm$; $L_1 = L_2 = L = \mu_0 N^2 (S/l) = 9.3\times 10^{-7}H$ .Thus, we get

$$Z_1 = \sqrt{(R_1 + r_{i1})^2 + (\omega L)^2} \cong 501\Omega$$

and

$$Z_2 = \sqrt{(R_2 + r_{i2} + R_p + R)^2 + (\omega L)^2}$$

For $R = 0$ we get $Z_2 = Z_2^{min} \cong 4\Omega$; for $R = 10\Omega$ the result is $Z_2 = Z_2^{max} \cong 14\Omega$. Thus,

$$Z_{1,total}^{min} = Z_1 + Z_{1,reflected}^{min} = Z_1 + Z_2^{min}\left(\frac{N_1}{N_2}\right)^2 \cong 505\Omega$$

$$Z_{1,total}^{max} = Z_1 + Z_{1,reflected}^{max} = Z_1 + Z_2^{max}\left(\frac{N_1}{N_2}\right)^2 \cong 515\Omega$$

The maxima *rms* currents have the following values:

$$I_1^{max} = \tfrac{1}{\sqrt{2}} 40V_{pp} \big/ Z_{1,total}^{min} = 56mA$$

(The maximum output current of the Function Generator HP3325A (Op.002 High Voltage Output) is $80mA_{pp} \cong 56.5mA_{rms}$);

$$I_2^{max} = \frac{\varepsilon_2}{Z_2^{min}} = 3A$$

and

$$I_3^{max} = I_2^{max} + I_1^{max} \cong 3A$$

The new expression for the *inertial forces*, (Eq.5) $\vec{F}_i = M_g \vec{a}$ , shows that the inertial forces are proportional to *gravitational mass*. Only in the particular case of $m_g = m_{i0}$, the expression above reduces to the well-known Newtonian expression $\vec{F}_i = m_{i0}\vec{a}$ . The equivalence between gravitational and inertial forces $(\vec{F}_i \equiv \vec{F}_g)$ [1] shows then that a balance measures the *gravitational mass* subjected to acceleration $a = g$. Here, the decrease in the *gravitational mass* of the Aluminum foil will be measured by a pan balance with the following characteristics: range 0-200g; readability 0.01g.

The mass of the Foam Board plate is: $\cong 4.17g$, the mass of the Aluminum foil is: $\cong 0.64g$, the total mass of the ends and the electric wires of connection is $\cong 5g$. Thus, *initially* the balance will show $\cong 9.81g$. According to (A18), when the electric current through the Aluminum foil (resistance $r_p^* = l/\sigma S = 2.5\times10^{-3}\Omega$) reaches the value $I_3 \cong 2.2A$, we will get $m_{g(Al)} \cong -m_{i0(Al)}$. Under these circumstances, the balance will show:

$$9.81g - 0.64g - 0.64g \cong 8.53g$$

and the gravity acceleration $g'$ *above* the Aluminum foil, becomes $g' = \chi\, g \cong -1g$.

It was shown [1] that, when the gravitational mass of a particle is reduced to the gravitational mass ranging between $+0.159M_i$ to $-0.159M_i$, it becomes *imaginary*, i.e., the gravitational and the inertial masses of the particle become *imaginary*. Consequently, the particle *disappears* from our ordinary space-time. This phenomenon can be observed in the proposed experiment, i.e., *the Aluminum foil will disappear* when its gravitational mass becomes smaller than $+0.159M_i$. It will become visible again, only when its gravitational mass becomes smaller than $-0.159M_i$, or when it becomes greater than $+0.159M_i$.

Equation (A18) shows that the gravitational mass of the Aluminum foil, $m_{g(Al)}$, goes *close to zero* when $I_3 \cong 1.76A$. Consequently, the gravity acceleration *above* the Aluminum foil also goes close to zero since $g' = \chi\, g = m_{g(Al)}/m_{i0(Al)}$. Under these circumstances, the Aluminum foil remains *invisible*.

Now consider a rigid Aluminum wire # 14 AWG. The area of its cross section is

$$S = \pi\left(1.628\times10^{-3}m\right)^2/4 = 2.08\times10^{-6}m^2$$

If an ELF electric current with frequency $f = 2\mu Hz = 2\times10^{-6}Hz$ passes through this wire, its gravitational mass, given by (A16), will be expressed by

$$m_g = \left\{1-2\left[\sqrt{1+6.313\times10^{-42}\frac{j_{rms}^4}{f^3}}-1\right]\right\}m_{i0} =$$
$$= \left\{1-2\left[\sqrt{1+7.89\times10^{-25}\frac{I_{DC}^4}{S^4}}-1\right]\right\}m_{i0} =$$
$$= \left\{1-2\left[\sqrt{1+0.13I_{DC}^4}-1\right]\right\}m_{i0} \qquad (A22)$$

For $I_{DC} \cong 3A$ the equation above gives

$$m_g \cong -3.8m_{i0}$$

Note that we can replace the Aluminum foil for this wire in the experimental set-up shown in Fig.A2. It is important also to note that an ELF electric current that passes through a wire - which makes a spherical form, as shown in Fig A5 - reduces the gravitational mass of the wire (Eq.

A22), and the gravity *inside sphere* at the same proportion, $\chi = m_g/m_{i0}$, (Gravitational Shielding Effect). In this case, that effect can be checked by means of the Experimental set-up 2 (Fig.A6). Note that the spherical form can be transformed into an ellipsoidal form or a disc in order to coat, for example, a Gravitational Spacecraft. It is also possible to coat with a wire several forms, such as cylinders, cones, cubes, etc.

The circuit shown in Fig.A4 (a) can be modified in order to produce a new type of Gravitational Shielding, as shown in Fig.A4 (b). In this case, the Gravitational Shielding will be produced in the Aluminum plate, with thickness $h$, of the parallel plate capacitor connected in the point $P$ of the circuit (See Fig.A4 (b)). Note that, in this circuit, the Aluminum foil (resistance $R_p$) (Fig.A4(a)) has been replaced by a Copper wire # 14 AWG with *1cm* length ($l = 1cm$) in order to produce a resistance $R_\phi = 5.21 \times 10^{-5}\Omega$. Thus, the voltage in the point $P$ of the circuit will have the maximum value $V_p^{max} = 1.1 \times 10^{-4} V$ when the resistance of the rheostat is null $(R = 0)$ and the minimum value $V_p^{min} = 4.03 \times 10^{-5} V$ when $R = 10\Omega$. In this way, the voltage $V_p$ (with frequency $f = 2\mu Hz$) applied on the capacitor will produce an electric field $E_p$ with intensity $E_p = V_p/h$ through the Aluminum plate of thickness $h = 3mm$. It is important to note that *this plate cannot be connected to ground* (earth), in other words, cannot be grounded, because, in this case, the electric field through it will be *null* ***.

According to Eq. A14, when $E_p^{max} = V_p^{max}/h = 0.036\,V/m$, $f = 2\mu Hz$ and $\sigma_{Al} = 3.82 \times 10^7 S/m$, $\rho_{Al} = 2700\,kg/m^3$ (Aluminum), we get

$$\chi = \frac{m_{(Al)}}{m_{i(Al)}} \cong -0.9$$

Under these conditions, the maximum *current density* through the plate with thickness $h$ will be given by $j^{max} = \sigma_{Al} E_p^{max} = 1.4 \times 10^6 A/m^2$ (It is well-known that the maximum current density supported by the Aluminum is $\approx 10^8 A/m^2$).

Since the area of the plate is $A = (0.2)^2 = 4 \times 10^{-2} m^2$, then the maximum current is $i^{max} = j^{max} A = 56kA$. Despite this enormous current, the maximum dissipated power will be just $P^{max} = (i^{max})^2 R_{plate} = 6.2W$, because the resistance of the plate is very small, i.e., $R_{plate} = h/\sigma_{Al} A \cong 2 \times 10^{-9}\Omega$.

Note that the area $A$ of the plate (where the Gravitational Shielding takes place) can have several geometrical configurations. For example, it can be the area of the external surface of an ellipsoid, sphere, etc. Thus, it can be the area of the external surface of a Gravitational Spacecraft. In this case, if $A \cong 100m^2$, for example, the maximum dissipated

*** When the voltage $V_p$ is applied on the capacitor, the charge distribution in the dielectric induces positive and negative charges, respectively on opposite sides of the Aluminum plate with thickness $h$. If the plate is not connected to the ground (Earth) this charge distribution produces an electric field $E_p = V_p/h$ through the plate. However, if the plate is connected to the ground, the negative charges (electrons) escapes for the ground and the positive charges are redistributed along the entire surface of the Aluminum plate making *null* the electric field through it.

power will be $P^{max} \cong 15.4kW$, i.e., approximately $154W/m^2$.

All of these systems work with Extra-Low Frequencies $(f << 10^{-3} Hz)$. Now, we show that, by simply changing *the geometry of the surface of the Aluminum foil*, it is possible to increase the working frequency $f$ up to more than *1Hz*.

Consider the Aluminum foil, now with several semi-spheres stamped on its surface, as shown in Fig. A7 . The semi-spheres have radius $r_0 = 0.9\ mm$, and are joined one to another. The Aluminum foil is now coated by an insulation layer with relative permittivity $\varepsilon_r$ and dielectric strength $k$. A voltage source is connected to the Aluminum foil in order to provide a voltage $V_0$ (rms) with frequency $f$. Thus, the electric potential $V$ at a distance $r$, in the interval from $r_0$ to $a$, is given by

$$V = \frac{1}{4\pi\varepsilon_r\varepsilon_0}\frac{q}{r} \qquad (A23)$$

In the interval $a < r \leq b$ the electric potential is

$$V = \frac{1}{4\pi\varepsilon_0}\frac{q}{r} \qquad (A24)$$

since for the air we have $\varepsilon_r \cong 1$.

Thus, on the surface of the metallic spheres $(r = r_0)$ we get

$$V_0 = \frac{1}{4\pi\varepsilon_r\varepsilon_0}\frac{q}{r_0} \qquad (A25)$$

Consequently, the electric field is

$$E_0 = \frac{1}{4\pi\varepsilon_r\varepsilon_0}\frac{q}{r_0^2} \qquad (A26)$$

By comparing (A26) with (A25), we obtain

$$E_0 = \frac{V_0}{r_0} \qquad (A27)$$

The electric potential $V_b$ at $r = b$ is

$$V_b = \frac{1}{4\pi\varepsilon_0}\frac{q}{b} = \frac{\varepsilon_r V_0 r_0}{b} \qquad (A28)$$

Consequently, the electric field $E_b$ is given by

$$E_b = \frac{1}{4\pi\varepsilon_0}\frac{q}{b^2} = \frac{\varepsilon_r V_0 r_0}{b^2} \qquad (A29)$$

From $r = r_0$ up to $r = b = a + d$ *the electric field is approximately constant* (See Fig. A7). Along the distance $d$ it will be called $E_{air}$. For $r > a + d$, the electric field stops being constant. Thus, the intensity of the electric field at $r = b = a + d$ is approximately equal to $E_0$, i.e., $E_b \cong E_0$. Then, we can write that

$$\frac{\varepsilon_r V_0 r_0}{b^2} \cong \frac{V_0}{r_0} \qquad (A30)$$

whence we get

$$b \cong r_0\sqrt{\varepsilon_r} \qquad (A31)$$

Since the intensity of the electric field through the air, $E_{air}$, is $E_{air} \cong E_b \cong E_0$, then, we can write that

$$E_{air} = \frac{1}{4\pi\varepsilon_0}\frac{q}{b^2} = \frac{\varepsilon_r V_0 r_0}{b^2} \qquad (A32)$$

Note that, $\varepsilon_r$ refers to the *relative permittivity of the insulation layer, which is covering the Aluminum foil.*

If the intensity of this field is greater than the dielectric strength of the air $(3\times10^6 V/m)$ there will occur the well-known *Corona effect*. Here, this effect is necessary in order to increase the electric conductivity of the air at this region (layer with thickness *d*). Thus, we will assume

$$E_{air}^{min} = \frac{\varepsilon_r V_0^{min} r_0}{b^2} = \frac{V_0^{min}}{r_0} = 3\times10^6 V/m$$

and

$$E_{air}^{max} = \frac{\varepsilon_r V_0^{max} r_0}{b^2} = \frac{V_0^{max}}{r_0} = 1\times10^7 V/m \quad (A33)$$

The electric field $E_{air}^{min} \leq E_{air} \leq E_{air}^{max}$ will produce an *electrons flux* in a direction and an *ions flux* in an opposite direction. From the viewpoint of electric current, the ions flux can be considered as an "electrons" flux at the same direction of the real electrons flux. Thus, the current density through the air, $j_{air}$, will be the *double* of the current density expressed by the well-known equation of Langmuir-Child

$$j=\frac{4}{9}\varepsilon_r\varepsilon_0\sqrt{\frac{2e}{m_e}}\frac{V^{\frac{3}{2}}}{d^2}=\alpha\frac{V^{\frac{3}{2}}}{d^2}=2.33\times10^{-6}\frac{V^{\frac{3}{2}}}{d^2} \qquad (A34)$$

where $\varepsilon_r \cong 1$ for the *air*; $\alpha = 2.33\times10^{-6}$ is the called *Child's constant*.

Thus, we have

$$j_{air}=2\alpha\frac{V^{\frac{3}{2}}}{d^2} \qquad (A35)$$

where $d$, in this case, is the thickness of the air layer where the electric field is approximately constant and $V$ is the voltage drop given by

$$V=V_a-V_b=\frac{1}{4\pi\varepsilon_0}\frac{q}{a}-\frac{1}{4\pi\varepsilon_0}\frac{q}{b}=$$
$$=V_0r_0\varepsilon_r\left(\frac{b-a}{ab}\right)=\left(\frac{\varepsilon_r r_0 d}{ab}\right)V_0 \qquad (A36)$$

By substituting (A36) into (A35), we get

$$j_{air}=\frac{2\alpha}{d^2}\left(\frac{\varepsilon_r r_0 dV_0}{ab}\right)^{\frac{3}{2}}=\frac{2\alpha}{d^{\frac{1}{2}}}\left(\frac{\varepsilon_r r_0 V_0}{b^2}\right)^{\frac{3}{2}}\left(\frac{b}{a}\right)^{\frac{3}{2}}=$$
$$=\frac{2\alpha}{d^{\frac{1}{2}}}E_{air}^{\frac{3}{2}}\left(\frac{b}{a}\right)^{\frac{3}{2}} \qquad (A37)$$

According to the equation of the *Ohm's vectorial Law*: $j=\sigma E$, we can write that

$$\sigma_{air}=\frac{j_{air}}{E_{air}} \qquad (A38)$$

Substitution of (A37) into (A38) yields

$$\sigma_{air}=2\alpha\left(\frac{E_{air}}{d}\right)^{\frac{1}{2}}\left(\frac{b}{a}\right)^{\frac{3}{2}} \qquad (A39)$$

If the insulation layer has thickness $\Delta = 0.6\ mm$, $\varepsilon_r \cong 3.5$ (1-60Hz), $k=17kV/mm$ (Acrylic sheet 1.5mm thickness), and the semi-spheres stamped on the metallic surface have $r_0 = 0.9\ mm$ (See Fig.A7) then $a=r_0+\Delta=1.5\ mm$. Thus, we obtain from Eq. (A33) that

$$V_0^{min}=2.7kV$$
$$V_0^{max}=9kV \qquad (A40)$$

From equation (A31), we obtain the following value for $b$:

$$b=r_0\sqrt{\varepsilon_r}=1.68\times10^{-3}m \qquad (A41)$$

Since $b=a+d$ we get

$$d=1.8\times10^{-4}m$$

Substitution of $a$, $b$, $d$ and A(32) into (A39) produces

$$\sigma_{air}=4.117\times10^{-4}E_{air}^{\frac{1}{2}}=1.375\times10^{-2}V_0^{\frac{1}{2}}$$

Substitution of $\sigma_{air}$, $E_{air}(rms)$ and $\rho_{air}=1.2\ kg.m^{-3}$ into (A14) gives

$$\frac{m_{g(air)}}{m_{i0(air)}}=\left\{1-2\left[\sqrt{1+1.758\times10^{-27}\frac{\sigma_{air}^3E_{air}^4}{\rho_{air}^2f^3}}-1\right]\right\}=$$
$$=\left\{1-2\left[\sqrt{1+4.923\times10^{-21}\frac{V_0^{5.5}}{f^3}}-1\right]\right\} \qquad (A42)$$

For $V_0=V_0^{max}=9kV$ and $f=2Hz$, the result is

$$\frac{m_{g(air)}}{m_{i0(air)}}\cong-1.2$$

Note that, by increasing $V_0$ the values of $E_{air}$ and $\sigma_{air}$ are increased. Thus, as show (A42), there are two ways for decrease the value of $m_{g(air)}$: increasing the value of $V_0$ or decreasing the value of $f$.

Since $E_0^{max} = 10^7 V/m = 10kV/mm$ and $\Delta = 0.6\ mm$ then the dielectric strength of the insulation must be $\geq 16.7kV/mm$. As mentioned above, the dielectric strength of the acrylic is $17kV/mm$.

It is important to note that, due to the strong value of $E_{air}$ (Eq. A37) the *drift velocity* $v_d$, $(v_d = j_{air}/ne = \sigma_{air}E_{air}/ne)$ of the free charges inside the ionized air put them at a distance $x = v_d/t = 2fv_d \cong 0.4m$, which is much greater than the distance $d = 1.8\times10^{-4}m$. Consequently, the number $n$ of free charges decreases strongly inside the air layer of thickness $d$ †††, except, obviously, in a thin layer, very close to the dielectric, where the number of free charges remains sufficiently increased, to maintain the air conductivity with $\sigma_{air} \cong 1.1S/m$ (Eq. A39).

The thickness $h$ of this thin air layer close to the dielectric can be easily evaluated starting from the charge distribution in the neighborhood of the dielectric, and of the repulsion forces established among them. The result is $h = \sqrt{0.06e/4\pi\varepsilon_0 E} \cong 4\times10^{-9}m$. This is, therefore, the thickness of the *Air* Gravitational Shielding. If the area of this Gravitational Shielding is equal to the area of a format A4 sheet of paper, i.e., $A = 0.20\times0.291 = 0.0582m^2$, we obtain the following value for the resistance $R_{air}$ of the Gravitational Shielding: $R_{air} = h/\sigma_{air}A \cong 6\times10^{-8}\Omega$. Since the maximum electrical current through this air layer is $i^{max} = j^{max}A \cong 400kA$, then the maximum power radiated from the Gravitational Shielding is $P_{air}^{max} = R_{air}\left(i_{air}^{max}\right)^2 \cong 10kW$. This means that a very strong light will be radiated from this type of Gravitational Shielding. Note that this device can also be used as a lamp, which will be much more efficient than conventional lamps.

Coating a ceiling with this lighting system enables the entire area of ceiling to produce light. This is a form of lighting very different from those usually known.

Note that the value $P_{air}^{max} \cong 10kW$, defines the power of the transformer shown in Fig.A10. Thus, the maximum current in the secondary is $i_s^{max} = 9kV/10kW = 0.9A$.

Above the Gravitational Shielding, $\sigma_{air}$ is reduced to the normal value of conductivity of the atmospheric air $\left(\approx 10^{-14}S/m\right)$. Thus, the power radiated from this region is

$$P_{air}^{max} = (d-h)\left(i_{air}^{max}\right)^2/\sigma_{air}A = $$
$$= (d-h)A\sigma_{air}\left(E_{air}^{max}\right)^2 \cong 10^{-4}W$$

Now, we will describe a method to coat the Aluminum semi-spheres with acrylic in the necessary dimension $(\Delta = a - r_0)$. First, take an Aluminum plate with $21cm\times29.1cm$ (A4 format). By means of a convenient process, several semi-spheres can be stamped on its surface. The semi-spheres have radius $r_0 = 0.9\ mm$, and are joined one to another. Next, take an acrylic sheet (A4 format) with 1.5mm thickness (See Fig.A8 (a)). Put a heater below the Aluminum plate in order to heat the Aluminum (Fig.A8 (b)). When the Aluminum is sufficiently heated up, the acrylic sheet and the Aluminum plate are pressed, one against the other, as shown in Fig. A8 (c). The two D devices shown in this figure are used in order to impede that the press compresses the acrylic and the aluminum to a distance shorter than $y+a$. After some seconds, remove the press and the heater. The device is ready to be subjected to a voltage $V_0$ with frequency $f$, as shown in Fig.A9. Note that, in this case, the balance is not necessary, because *the substance that produces the gravitational shielding* is an *air layer* with thickness $d$ *above* the acrylic sheet. This is, therefore, more a type of Gravity Control Cell (GCC) with *external gravitational shielding*.

††† Reducing therefore the conductivity, $\sigma_{air}$, to the normal value of the conductivity of atmospheric air.

It is important to note that this GCC can be made very thin and as flexible as a fabric. Thus, it can be used to produce *anti- gravity clothes*. These clothes can be extremely useful, for example, to walk on the surface of high gravity planets.

Figure A11 shows some geometrical forms that can be stamped on a metallic surface in order to produce a Gravitational Shielding effect, similar to the produced by the *semi-spherical form*.

An obvious evolution from the semi-spherical form is the *semi-cylindrical* form shown in Fig. A11 (b); Fig.A11(c) shows *concentric metallic rings* stamped on the metallic surface, an evolution from Fig.A11 (b). These geometrical forms produce the same effect as the semi-spherical form, shown in Fig.A11 (a). By using concentric metallic rings, it is possible to build *Gravitational Shieldings* around bodies or spacecrafts with several formats (spheres, ellipsoids, etc); Fig. A11 (d) shows a Gravitational Shielding around a Spacecraft with *ellipsoidal form*.

The previously mentioned Gravitational Shielding, produced on a thin layer of ionized air, has a *behavior different from* the Gravitational Shielding produced on a *rigid substance*. When the gravitational masses of the air molecules, inside the shielding, are reduced to within the range $+0.159m_i < m_g < -0.159m_i$, they go to the *imaginary space-time*, as previously shown in this article. However, the electric field $E_{air}$ stays at the real space-time. Consequently, the molecules return immediately to the real space-time in order to return soon after to the *imaginary* space-time, due to the action of the electric field $E_{air}$.

In the case of the Gravitational Shielding produced on a *solid substance*, when the molecules of the substance go to the *imaginary* space-time, *the electric field that produces the effect, also goes to the imaginary space-time together with them*, since in this case, the substance of the Gravitational Shielding is rigidly connected to the metal that produces the electric field. (See Fig. A12 (b)). This is the fundamental difference between the *non-solid* and *solid* Gravitational Shieldings.

Now, consider a Gravitational Spacecraft that is able to produce an *Air* Gravitational Shielding and also a *Solid* Gravitational Shielding, as shown in Fig. A13 (a) ‡‡‡. Assuming that the intensity of the electric field, $E_{air}$, necessary to reduce the gravitational mass of the *air molecules* to within the range $+0.159m_i < m_g < -0.159m_i$, *is much smaller* than the intensity of the electric field, $E_{rs}$, necessary to reduce the gravitational mass of the *solid substance* to within the range $+0.159m_i < m_g < -0.159m_i$, then we conclude that the Gravitational Shielding made of ionized air goes to the imaginary space-time *before* the Gravitational Shielding made of *solid substance*. When this occurs the spacecraft does not go to the imaginary space-time together with the Gravitational Shielding of air, because the air molecules are not rigidly connected to the spacecraft. Thus, while the air molecules go into the imaginary space-time, the spacecraft stays in the *real space-time*, and remains subjected to the effects of the Gravitational Shielding around it, since the shielding does not stop to work, due to its extremely short permanence at the imaginary space-time. Under these circumstances, the gravitational mass of the Gravitational Shielding can be reduced to $m_g \cong 0$. For example, $m_g \cong 10^{-4} kg$. Thus, if the *inertial mass* of the Gravitational Shielding is $m_{i0} \cong 1kg$, then $\chi = m_g / m_{i0} \cong 10^{-4}$. As we have seen, this means that *the inertial effects on the spacecraft* will be reduced by $\chi \cong 10^{-4}$. Then, in spite of the effective acceleration of the spacecraft be, for example, $a = 10^5 m.s^{-2}$, the effects on the crew of the spacecraft will be equivalent to an acceleration of only

$$a' = \frac{m_g}{m_{i0}} a = \chi \; a \approx 10 m.s^{-1}$$

This is the magnitude of the acceleration upon the passengers in a contemporary commercial jet.

Then, it is noticed that Gravitational Spacecrafts can be subjected to enormous *accelerations* (or *decelerations*) without imposing any harmful impacts whatsoever on the spacecrafts or its crew.

Now, imagine that the intensity of the electric field that produces the Gravitational Shielding around the spacecraft is *increased* up to reaching the value $E_{rs}$ that reduces the gravitational mass of the *solid* Gravitational Shielding to within the range $+0.159m_i < m_g < -0.159m_i$. Under these circumstances, the *solid* Gravitational Shielding goes to the imaginary space-time and, since it is rigidly connected to the spacecraft, also the spacecraft goes to the imaginary space-time together with the Gravitational Shielding. Thus, the spacecraft can travel within the imaginary space-time and make use of the Gravitational Shielding around it.

‡‡‡ The *solid* Gravitational Shielding can also be obtained by means of *an ELF electric current through a metallic lamina* placed *between the semi-spheres and the Gravitational Shielding of Air* (See Fig.A13 (a)). The gravitational mass of the solid Gravitational Shielding will be controlled just by means of the intensity of the ELF electric current. Recently, it was discovered that Carbon nanotubes (CNTs) can be added to *Alumina* ($Al_2O_3$) to convert it into a good electrical conductor. It was found that the electrical conductivity increased up to 3375 S/m at 77°C in samples that were 15% nanotubes by volume [12]. It is known that the density of α-Alumina is $3.98\times10^3 kg.m^{-3}$ and that it can withstand 10-20 KV/mm. Thus, these values show that the Alumina-CNT can be used to make a *solid* Gravitational Shielding.

As we have already seen, the maximum velocity of propagation of the interactions in the imaginary space-time is *infinite* (in the real space-time this limit is equal to the light velocity $c$). This means that *there are no limits for the velocity of the spacecraft in the imaginary space-time*. Thus, the acceleration of the spacecraft can reach, for example, $a = 10^9 m.s^{-2}$, which leads the spacecraft to attain velocities $V \approx 10^{14} m.s^{-1}$ (about 1 million times the speed of light) after one day of trip. With this velocity, after 1 month of trip the spacecraft would have traveled about $10^{21} m$. In order to have idea of this distance, it is enough to remind that the diameter of our Universe (visible Universe) is of the order of $10^{26} m$.

Due to the extremely low density of the *imaginary* bodies, the collision between them cannot have the same consequences of the collision between the real bodies.

Thus, *for a Gravitational Spacecraft in imaginary state, the problem of the collision in high-speed doesn't exist.* Consequently, the Gravitational Spacecraft can transit freely in the imaginary Universe and, in this way, reach easily any point of our real Universe once they can make the transition back to our Universe by only increasing the gravitational mass of the Gravitational Shielding of the spacecraft in such way that it leaves the range of $+0.159 M_i$ to $-0.159 M_i$.

The return trip would be done in similar way. That is to say, the spacecraft would transit in the imaginary Universe back to the departure place where would reappear in our Universe. Thus, trips through our Universe that would delay millions of years, at speeds close to the speed of light, could be done in just a few *months* in the imaginary Universe.

In order to produce the acceleration of $a \approx 10^9 m.s^{-2}$ upon the spacecraft we propose a Gravitational Thruster with 10 GCCs (10 Gravitational Shieldings) of the type with several semi-spheres stamped on the metallic surface, as previously shown, or with the *semi-cylindrical* form shown in Figs. A11 (b) and (c). The 10 GCCs are filled with air at 1 atm and 300K. If the insulation layer is made with *Mica* $(\varepsilon_r \cong 5.4)$ and has thickness $\Delta = 0.1\ mm$, and the semi-spheres stamped on the metallic surface have $r_0 = 0.4\ mm$ (See Fig.A7) then $a = r_0 + \Delta = 0.5\ mm$. Thus, we get

$$b = r_0\sqrt{\varepsilon_r} = 9.295\times10^{-4} m$$

and

$$d = b - a = 4.295\times10^{-4} m$$

Then, from Eq. A42 we obtain

$$\chi_{air} = \frac{m_{g(air)}}{m_{i0(air)}} = \left\{1 - 2\left[\sqrt{1 + 1.758\times10^{-27}\frac{\sigma_{air}^3 E_{air}^4}{\rho_{air}^2 f^3}} - 1\right]\right\} =$$

$$= \left\{1 - 2\left[\sqrt{1 + 1.0\times10^{-18}\frac{V_0^{5.5}}{f^3}} - 1\right]\right\}$$

For $V_0 = V_0^{max} = 15.6kV$ and $f = 0.12Hz$, the result is

$$\chi_{air} = \frac{m_{g(air)}}{m_{i0(air)}} \cong -1.6\times10^4$$

Since $E_0^{max} = V_0^{max}/r_0$ is now given by $E_0^{max} = 15.6kV/0.9mm = 17.3kV/mm$ and $\Delta = 0.1\ mm$ then the dielectric strength of the insulation must be $\geq 173kV/mm$. As shown in the table below[§§§], *0.1mm - thickness of* Mica *can withstand 17.6 kV* (that is greater than $V_0^{max} = 15.6kV$), in such way that the dielectric strength is *176 kV/mm.*

The Gravitational Thrusters are positioned at the spacecraft, as shown in Fig. A13 (b). Then, when the spacecraft is in the *intergalactic space*, the gravity acceleration upon the gravitational mass $m_{gt}$ of the bottom of the thruster (See Fig.A13 (c)), is given by [2]

$$\vec{a} \cong (\chi_{air})^{10} \vec{g}_M \cong -(\chi_{air})^{10} G \frac{M_g}{r^2} \hat{\mu}$$

where $M_g$ is the gravitational mass in front of the spacecraft.

For simplicity, let us consider just the effect of a hypothetical volume $V = 10 \times 10^3 \times 10^3 = 10^7 m^3$ of intergalactic matter in front of the spacecraft $(r \cong 30m)$. The average density of matter in the *intergalactic medium (IGM)* is $\rho_{ig} \approx 10^{-26} kg.m^{-3})$[****]. Thus, for $\chi_{air} \cong -1.6 \times 10^4$ we get

$$a = -(-1.6 \times 10^4)^{10} (6.67 \times 10^{-11}) \left( \frac{10^{-19}}{30^2} \right) =$$

$$= -10^9 m.s^{-2}$$

In spite of this gigantic acceleration, the inertial effects for the crew of the spacecraft can be strongly reduced if, for example, the gravitational mass of the Gravitational Shielding is reduced down to $m_g \cong 10^{-6} kg$ and its inertial mass is $m_{i0} \cong 100kg$. Then, we get $\chi = m_g / m_{i0} \cong 10^{-8}$. Therefore, *the inertial effects on the spacecraft* will be reduced by $\chi \cong 10^{-8}$, and consequently, the inertial effects on the crew of the spacecraft would be *equivalent to* an acceleration $a'$ of only

$$a' = \frac{m_g}{m_{i0}} a = (10^{-8})(10^9) \approx 10 m.s^{-2}$$

Note that the Gravitational Thrusters in the spacecraft must have a very small diameter (of the order of *millimeters*) since, obviously, the hole through the Gravitational Shielding cannot be large. Thus, these thrusters are in fact, *Micro-Gravitational Thrusters*. As shown in Fig. A13 (b), it is possible to place several micro-gravitational thrusters in the spacecraft. This gives to the Gravitational Spacecraft, several degrees of freedom and shows the enormous superiority of this spacecraft in relation to the contemporaries spacecrafts.

The density of matter in the *intergalactic medium (IGM)* is about *10 ⁻²⁶ kg.m⁻³* , which is very less than the density of matter in the *interstellar medium* (~$10^{-21}$ $kg.m^{-3}$) that is less than the density of matter in the *interplanetary medium* (~$10^{-20}$ $kg.m^{-3}$). The density of matter is enormously

---

[§§§] The *dielectric strength* of some dielectrics can have different values in lower thicknesses. This is, for example, the case of the *Mica*.

| *Dielectric* | *Thickness* (mm) | *Dielectric Strength* (kV/mm) |
|---|---|---|
| Mica | 0.01 mm | 200 |
| **Mica** | **0.1 mm** | **176** |
| Mica | 1 mm | 61 |

[****] Some theories put the average density of the Universe as the equivalent of *one hydrogen atom per cubic meter* [13,14]. The density of the universe, however, is clearly not uniform. Surrounding and stretching between galaxies, there is rarefied plasma [15] that is thought to possess a cosmic filamentary structure [16] and that is slightly denser than the average density in the universe. This material is called the *intergalactic medium (IGM)* and is mostly ionized hydrogen; i.e. a plasma consisting of equal numbers of electrons and protons. The IGM is thought to exist at a density of 10 to 100 times the average density of the Universe (10 to 100 hydrogen atoms per cubic meter, i.e., $\approx 10^{-26} kg.m^{-3}$).

increased inside the Earth's atmosphere ($1.2kg.m^{-3}$ near to Earth's surface). Figure A14 shows the gravitational acceleration acquired by a Gravitational Spacecraft, in these media, using Micro-Gravitational thrusters.

In relation to the *Interstellar* and *Interplanetary medium*, the *Intergalactic medium* requires the greatest value of $\chi_{air}$ ($\chi$ inside the *Micro-Gravitational Thrusters*), i.e., $\chi_{air} \cong -1.6\times10^{4}$. This value strongly decreases when the spacecraft is within the Earth's atmosphere. In this case, it is sufficient only[††††] $\chi_{air} \cong -10$ in order to obtain:

$$a = -(\chi_{air})^{10} G \frac{\rho_{atm} V}{r^2} \cong$$

$$\cong -(-10)^{10}\left(6.67\times10^{-11}\right)\frac{1.2\left(10^{7}\right)}{(20)^2} \cong 10^4\, m.s^{-2}$$

With this acceleration the Gravitational Spacecraft can reach about *50000 km/h* in a few seconds. Obviously, the Gravitational Shielding of the spacecraft will reduce strongly *the inertial effects upon the crew* of the spacecraft, in such way that the inertial effects of this strong acceleration will not be felt. In addition, the *artificial atmosphere*, which is possible to build around the spacecraft, by means of gravity control technologies shown in this article (See Fig.6) and [2], will protect it from the *heating* produced by the friction with the Earth's atmosphere. Also, the gravity can be controlled inside of the Gravitational Spacecraft in order to maintain a value close to the Earth's gravity as shown in Fig.3.

Finally, it is important to note that a Micro-Gravitational Thruster does not work *outside* a Gravitational Shielding, because, in this case, *the resultant upon the thruster is null* due to the symmetry (See Fig. A15 (a)). Figure A15 (b) shows a micro-gravitational thruster inside a Gravitational Shielding. This thruster has 10 Gravitational Shieldings, in such way that the gravitational acceleration upon the *bottom* of the thruster, due to a gravitational mass $M_g$ *in front* of the thruster, is $a_{10} = \chi_{air}^{10} a_0$ where $a_0 = -GM_g/r^2$ is the gravitational acceleration acting on the front of the micro-gravitational thruster. *In the opposite direction*, the gravitational acceleration upon the bottom of the thruster, produced by a gravitational mass $M_g$, is

$$a_0' = \chi_s\left(-GM_g/r'^2\right) \cong 0$$

since $\chi_s \cong 0$ due to the Gravitational Shielding around the micro-thruster (See Fig. A15 (b)). Similarly, the acceleration in front of the thruster is

$$a_{10}' = \chi_{air}^{10} a_0' = \left[\chi_{air}^{10}\left(-GM_g/r'^2\right)\right]\chi_s$$

where $\left[\chi_{air}^{10}\left(-GM_g/r'^2\right)\right] < a_{10}$, since $r' > r$. Thus, for $a_{10} \cong 10^9 m.s^{-2}$ and $\chi_s \approx 10^{-8}$ we conclude that $a_{10}' < 10 m.s^{-2}$. This means that $a_{10}' << a_{10}$. Therefore, we can write that the resultant on the micro-thruster can be expressed by means of the following relation

$$R \cong F_{10} = \chi_{air}^{10} F_0$$

---

[††††] This value is within the range of values of $\chi$ ($\chi < -10^{3}$. *See Eq.A15*), which can be produced by means of *ELF electric currents* through metals as *Aluminum*, etc. This means that, in this case, if convenient, we can replace *air* inside the GCCs of the Gravitational Micro-thrusters by metal laminas with *ELF electric currents* through them.

Figure A15 (c) shows a Micro-Gravitational Thruster with *10 Air Gravitational Shieldings* (10 GCCs). Thin Metallic laminas are placed after each *Air* Gravitational Shielding in order to retain the electric field $E_b = V_0/x$, produced by metallic *surface behind* the semi-spheres. The laminas with semi-spheres stamped on its surfaces are connected to the ELF voltage source $V_0$ and the thin laminas in front of the Air Gravitational Shieldings are grounded. The air inside this Micro-Gravitational Thruster is at 300K, 1atm.

We have seen that the insulation layer of a GCC can be made up of Acrylic, Mica, etc. Now, we will design a GCC using *Water* (*distilled water*, $\varepsilon_{r(H_2O)} = 80$) and Aluminum *semi-cylinders* with radius $r_0 = 1.3mm$ . Thus, for $\Delta = 0.6mm$, the new value of $a$ is $a = 1.9mm$. Then, we get

$$b = r_0\sqrt{\varepsilon_{r(H_2O)}} = 11.63\times10^{-3}m \qquad (A43)$$

$$d = b - a = 9.73\times10^{-3}m \qquad (A44)$$

and

$$E_{air} = \frac{1}{4\pi\varepsilon_{r(air)}\varepsilon_0}\frac{q}{b^2} =$$

$$= \varepsilon_{r(H_2O)}\frac{V_0 r_0}{\varepsilon_{r(air)}b^2} =$$

$$= \frac{V_0/r_0}{\varepsilon_{r(air)}} \cong \frac{V_0}{r_0} = 1111.1\ V_0 \qquad (A45)$$

Note that

$$E_{(H_2O)} = \frac{V_0/r_0}{\varepsilon_{r(H_2O)}}$$

and

$$E_{(acrylic)} = \frac{V_0/r_0}{\varepsilon_{r(acrylic)}}$$

Therefore, $E_{(H_2O)}$ and $E_{(acrylic)}$ are much smaller than $E_{air}$. Note that for $V_0 \le 9kV$ the intensities of $E_{(H_2O)}$ and $E_{(acrylic)}$ are not sufficient to produce the ionization effect, which increases the electrical conductivity. Consequently, the conductivities of the water and the acrylic remain $<< 1\ S.m^{-1}$. In this way, with $E_{(H_2O)}$ and $E_{(acrylic)}$ much smaller than $E_{air}$, and $\sigma_{(H2O)} << 1$, $\sigma_{(acrylic)} << 1$, the decrease in both the gravitational mass of the acrylic and the gravitational mass of water, according to Eq.A14, is negligible. This means that only in the air layer the decrease in the gravitational mass will be relevant.

Equation A39 gives the electrical conductivity of the air layer, i.e.,

$$\sigma_{air} = 2\alpha\left(\frac{E_{air}}{d}\right)^{\frac{1}{2}}\left(\frac{b}{a}\right)^{\frac{3}{2}} = 0.029V_0^{\frac{1}{2}} \qquad (A46)$$

Note that $b = r_0\sqrt{\varepsilon_{r(H_2O)}}$. Therefore, here the value of $b$ is larger than in the case of the acrylic. Consequently, *the electrical conductivity of the air layer will be larger here than in the case of acrylic.*

Substitution of $\sigma_{(air)}$, $E_{air}$ (*rms*) and $\rho_{air} = 1.2kg.m^{-3}$ into Eq. A14, gives

$$\frac{m_{g(air)}}{m_{i0(air)}} = \left\{1 - 2\left[\sqrt{1 + 4.54\times10^{-20}\frac{V_0^{5.5}}{f^3}} - 1\right]\right\} \qquad (A47)$$

For $V_0 = V_0^{\max} = 9kV$ and $f = 2Hz$, the result is

$$\frac{m_{g(air)}}{m_{i0(air)}} \cong -8.4$$

This shows that, by using *water* instead of acrylic, the result is much better.

In order to build the GCC based on the calculations above (See Fig. A16), take an Acrylic plate with *885mm* x *885m* and *2mm* thickness, then paste on it an Aluminum sheet with *895.2mm* x *885mm* and *0.5mm* thickness(note that two edges of the Aluminum sheet are bent as shown in Figure A16 (b)). Next, take *342* Aluminum yarns with *884mm* length and *2.588mm* diameter (wire # 10 AWG) and insert them side by side on the Aluminum sheet. See in Fig. A16 (b) the detail of fixing of the yarns on the Aluminum sheet. Now, paste acrylic strips (with *13.43mm* height and *2mm* thickness) around the Aluminum/Acrylic, making a box. Put *distilled water* (approximately *1 litter*) inside this box, up to a height of exactly *3.7mm* from the edge of the acrylic base. Afterwards, paste an Acrylic lid (*889mm* x *889mm* and *2mm* thickness) on the box. Note that above the water there is an *air* layer with *885mm* x *885mm* and *7.73mm* thickness (See Fig. A16). This thickness plus the acrylic lid thickness (*2mm*) is equal to $d = b - a = 9.73mm$ where $b = r_0\sqrt{\varepsilon_{r(H_2O)}} = 11.63mm$ and $a = r_0 + \Delta = 1.99mm$, since $r_0 = 1.3mm$, $\varepsilon_{r(H_2O)} = 80$ and $\Delta = 0.6mm$.

Note that the gravitational action of the electric field $E_{air}$, extends itself only up to the distance $d$, which, in this GCC, is given by the sum of the Air layer thickness (*7.73mm*) plus the thickness of the Acrylic lid (*2mm*).

Thus, it is ensured the gravitational effect on the air layer while it is practically nullified in the acrylic sheet above the air layer, since $E_{(acrylic)} << E_{air}$ and $\sigma_{(acrylic)} << 1$.

With this GCC, we can carry out an experiment where the *gravitational mass of the air layer* is progressively reduced when the voltage applied to the GCC is increased (or when the frequency is decreased). A precision balance is placed below the GCC in order to measure the mentioned mass decrease for comparison with the values predicted by Eq. A(47). In total, this GCC weighs about *6kg*; the *air layer 7.3grams*. The balance has the following characteristics: *range 0-6kg; readability 0.1g*. Also, in order to prove the *Gravitational Shielding Effect*, we can put a *sample* (connected to a dynamometer) above the GCC in order to check the gravity acceleration in this region.

In order to prove *the exponential effect* produced by the superposition of the Gravitational Shieldings, we can take three similar GCCs and put them one above the other, in such way that above the GCC 1 the gravity acceleration will be $g' = \chi g$; above the GCC2 $g'' = \chi^2 g$, and above the GCC3 $g''' = \chi^3 g$. Where $\chi$ is given by Eq. (A47).

It is important to note that the intensity of the electric field through the air *below* the GCC is *much smaller* than the intensity of the electric field through the air layer inside the GCC. In addition, the electrical conductivity of the air below the GCC is much smaller than the conductivity of the air layer inside the GCC. Consequently, the decrease of the gravitational mass of the air below the GCC, according to Eq.A14, is negligible. This means that the GCC1, GCC2 and GCC3 can be simply overlaid, on the experiment proposed above. However, since it is necessary to put samples among them in order to measure the gravity above each GCC, we suggest a spacing of 30cm or more among them.

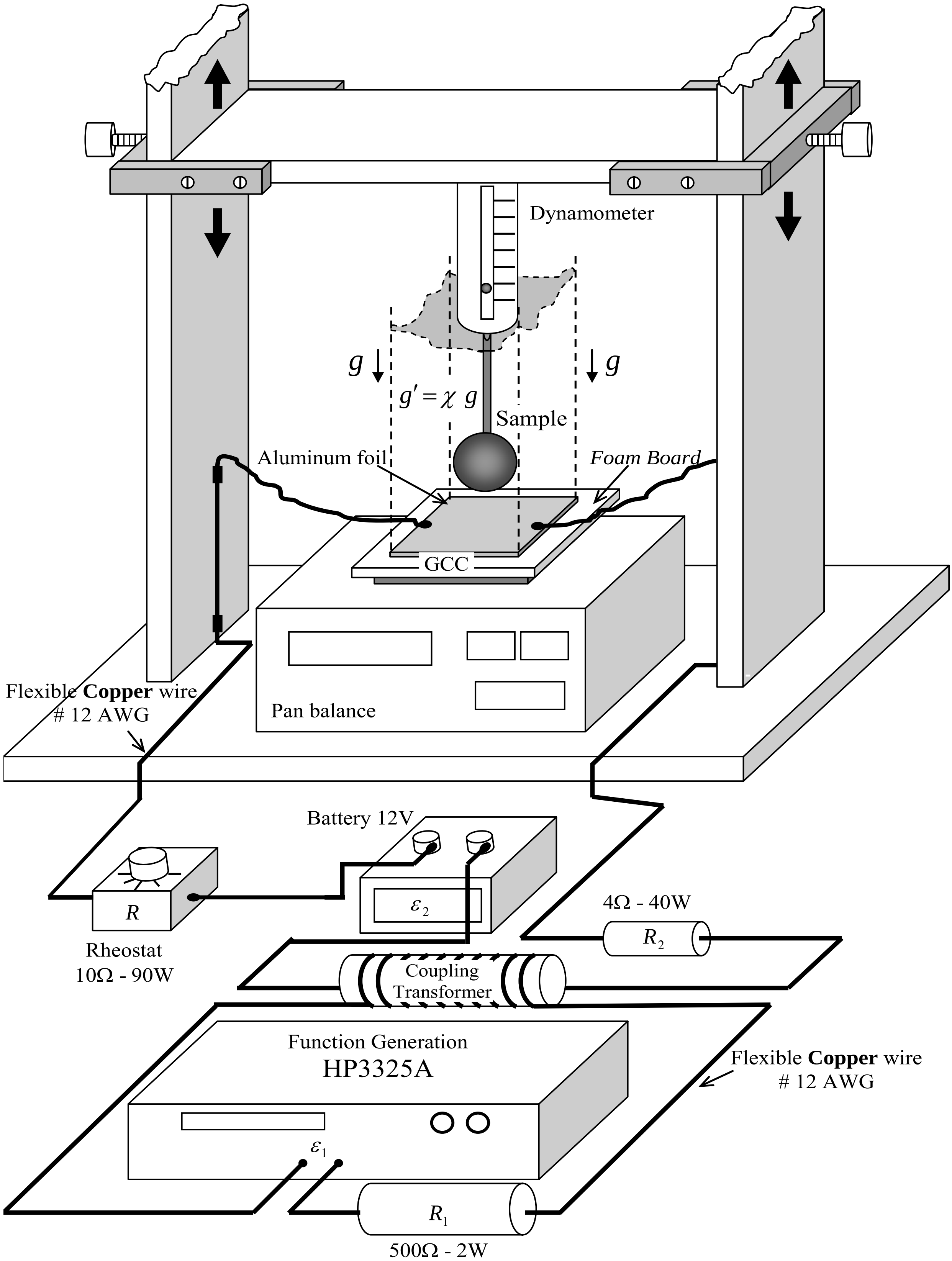

Fig. A2 – Experimental Set-up 1.

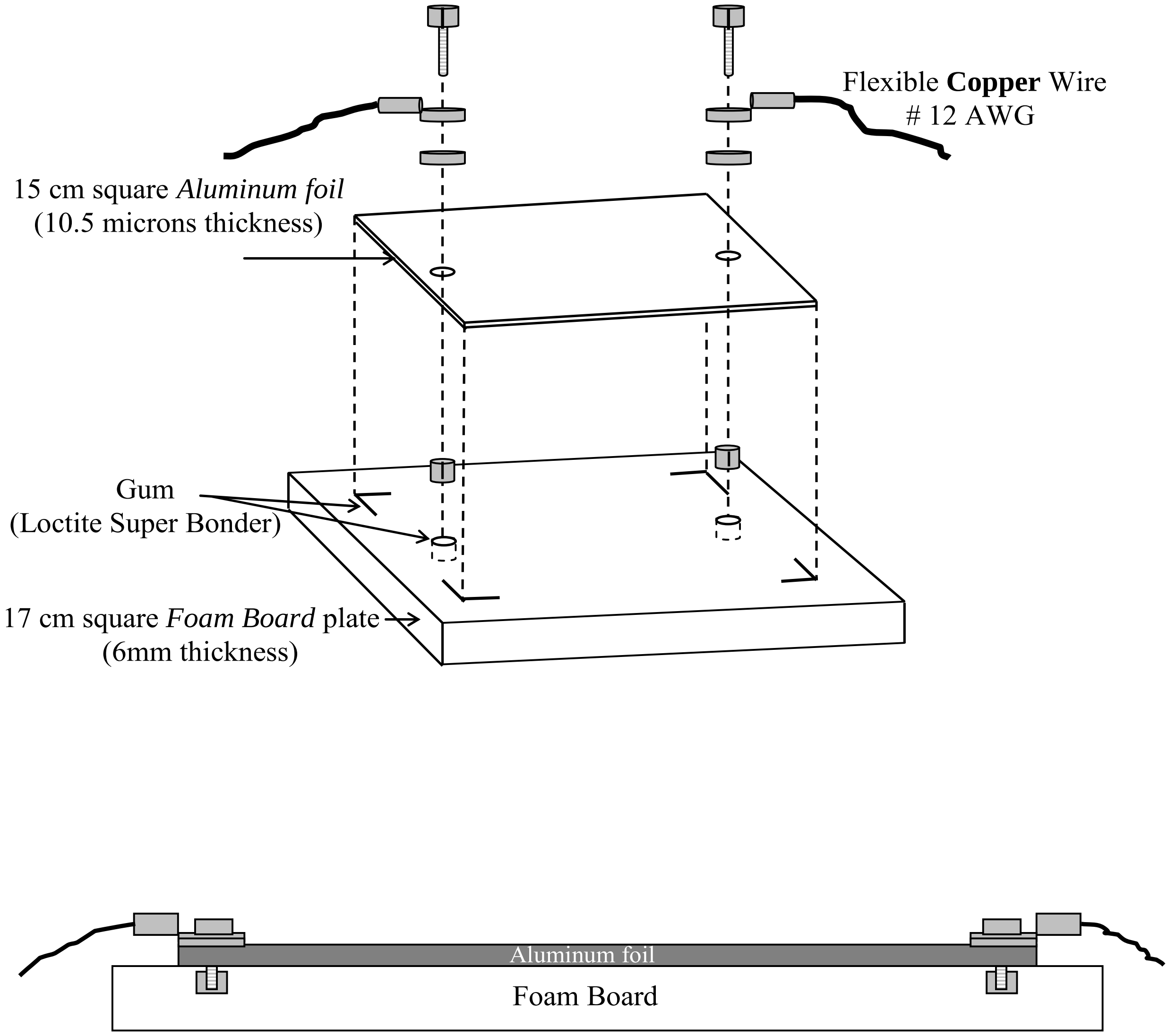

Fig. A3 – The Simplest *Gravity Control Cell* (GCC).

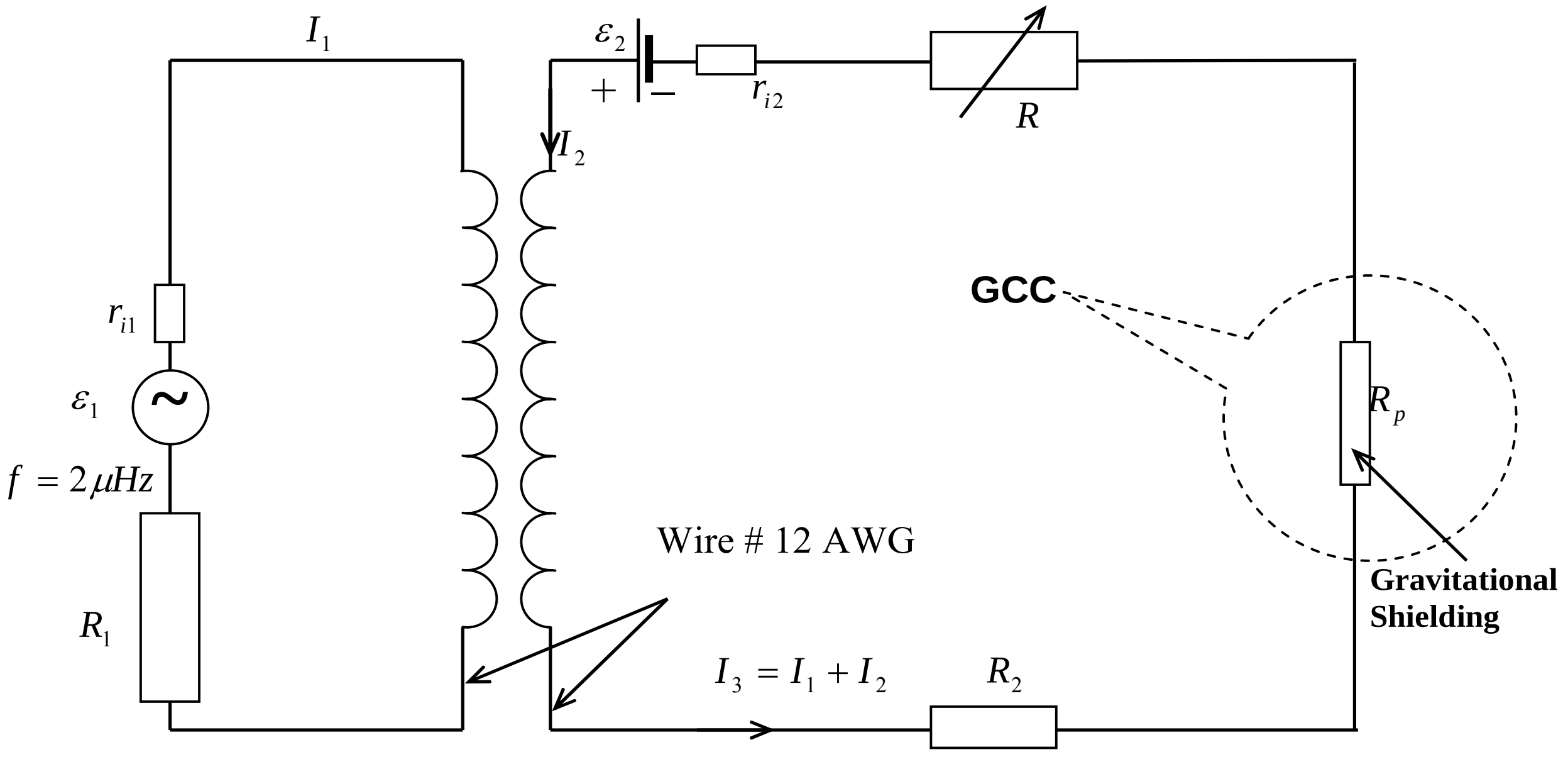

(a)

$\varepsilon_1$ = Function Generator HP3325A(*Option* 002 High Voltage Output)

$r_{i1} < 2\Omega$; $R_1 = 500\Omega - 2\ W$; $\varepsilon_2 = 12V\ DC$; $r_{i2} < 0.1\Omega\ (Battery)$;

$R_2 = 4\Omega - 40W$; $R_p = 2.5\times10^{-3}\Omega$; $Reostat = 0 \le R \le 10\Omega - 90W$

$I_1^{max} = 56mA\ (rms)$; $I_2^{max} = 3A$ ; $I_3^{max} \cong 3A\ (rms)$

*Coupling Transformer* to isolate the *Function Generator* from the Battery

- Air core 10 - mm diameter; wire #12 AWG; $N_1 = N_2 = 20; l = 42mm$

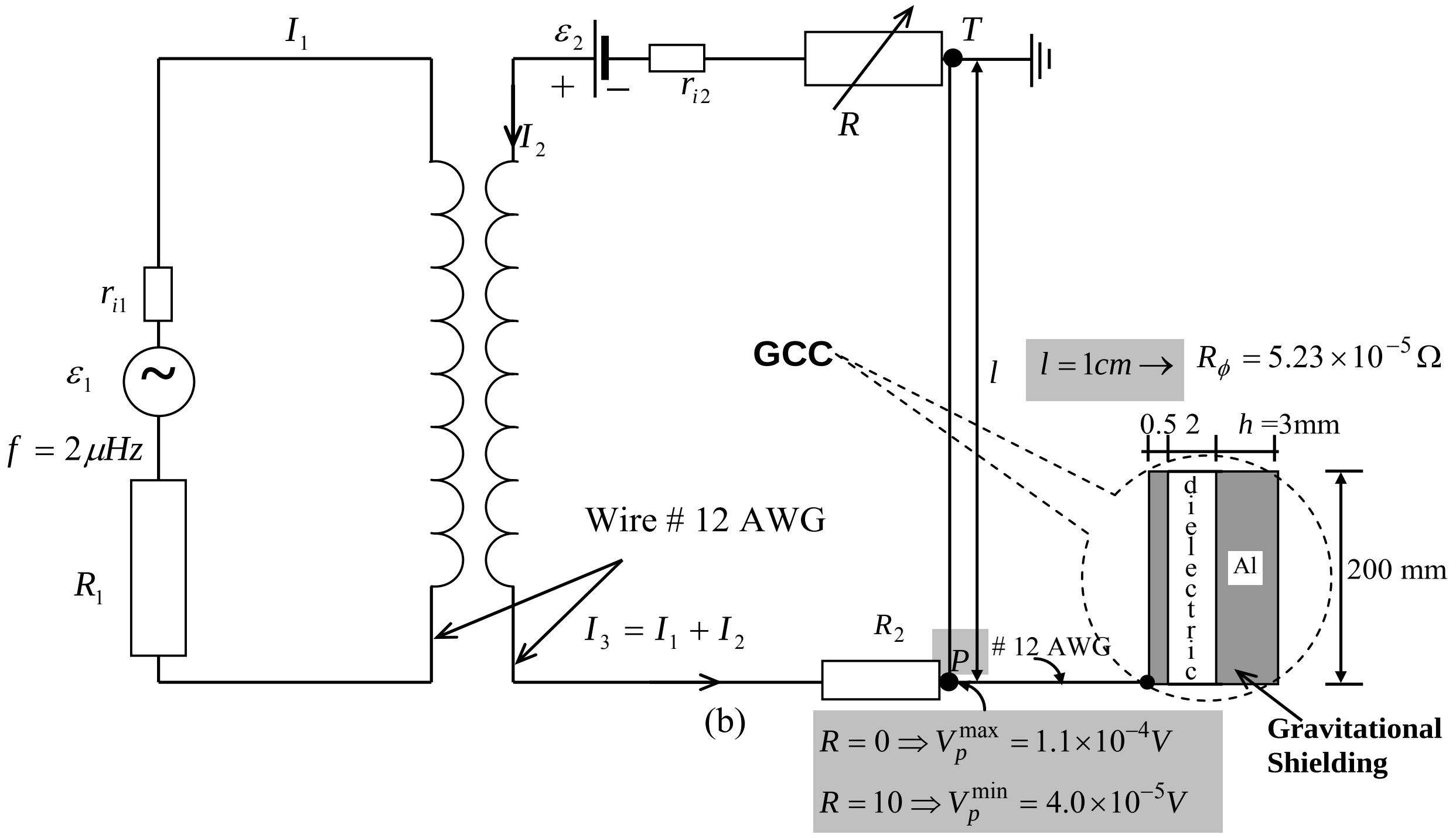

Fig. A4 – Equivalent Electric Circuits

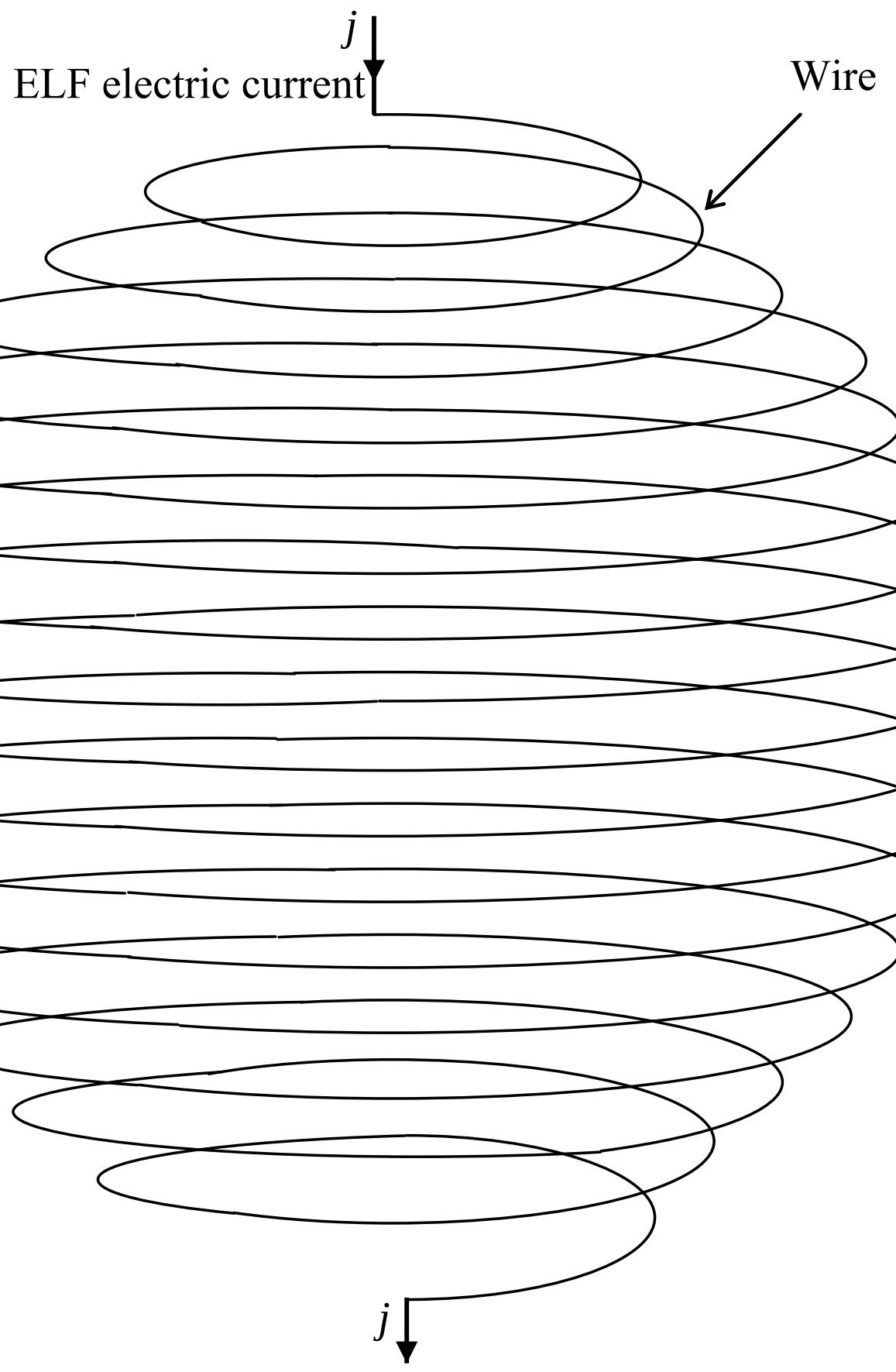

$$m_g = \left\{1 - 2\left[\sqrt{1 + 1.758\times10^{-27}\frac{\mu_r j^4}{\sigma\rho^2 f^3}} - 1\right]\right\} m_{i0}$$

Fig. A5 – An ELF electric current through a wire, that makes a spherical form as shown above, reduces the gravitational mass of the wire and the gravity inside sphere at the same proportion $\chi = m_g / m_{i0}$ (Gravitational Shielding Effect). Note that this spherical form can be transformed into an ellipsoidal form or a disc in order to coat, for example, a Gravitational Spacecraft. It is also possible to coat with a wire several forms, such as cylinders, cones, cubes, etc. The characteristics of the wire are expressed by: $\mu_r$, $\sigma$, $\rho$; $j$ is the electric current density and $f$ is the frequency.

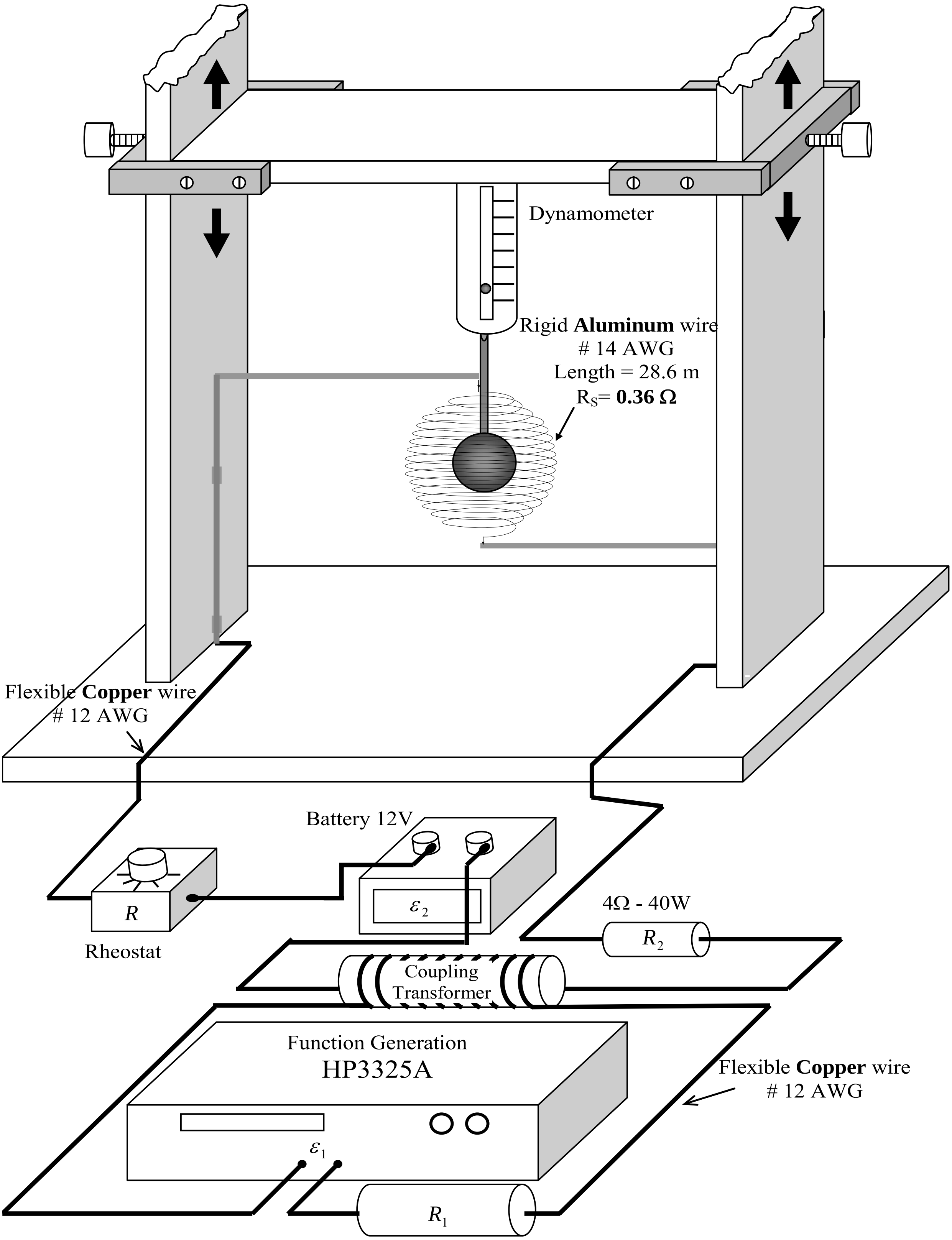

Fig. A6 – Experimental set-up 2.

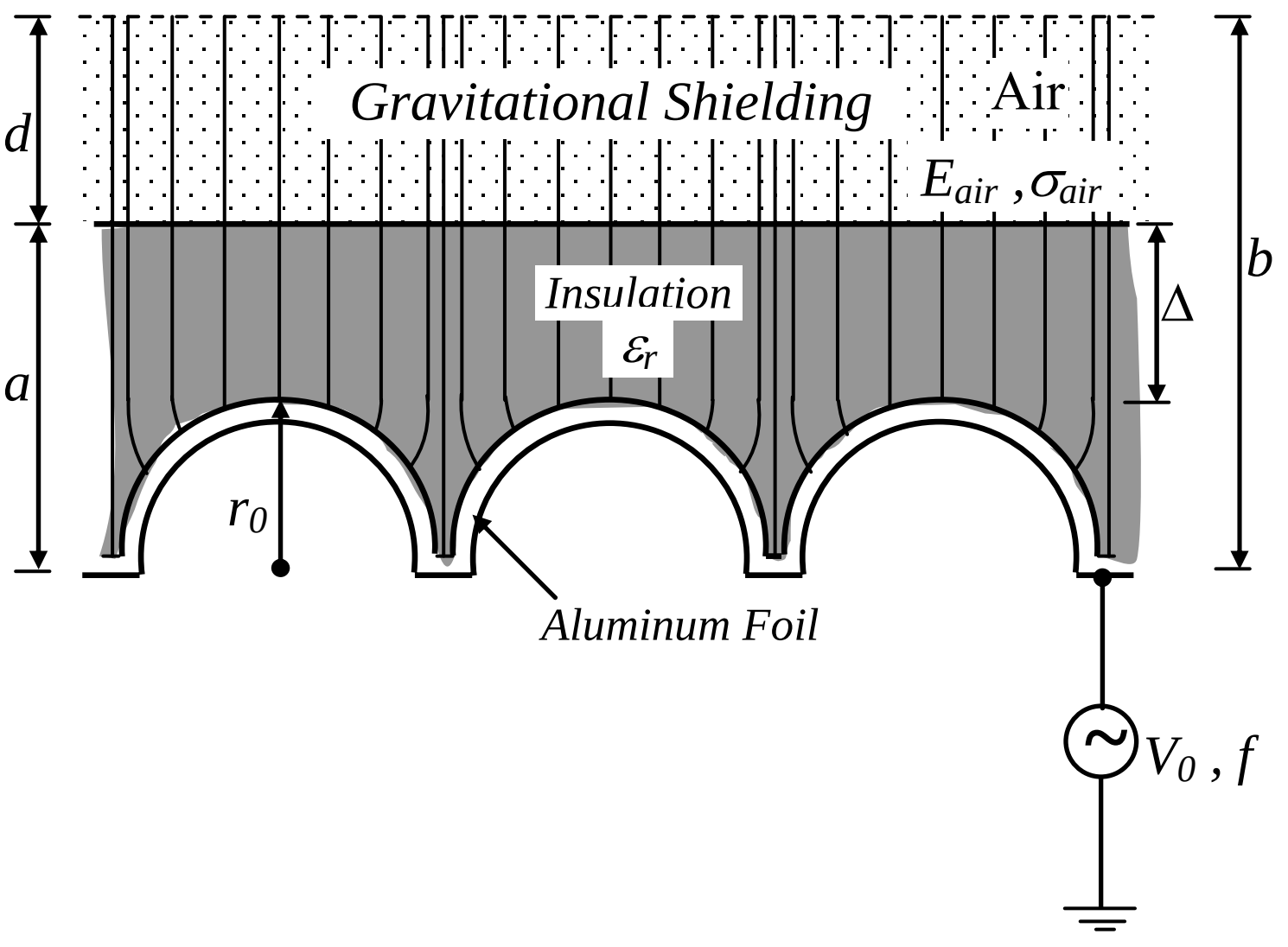

Fig A7 – *Gravitational shielding produced by semi-spheres stamped on the Aluminum foil* - By simply changing the geometry of the surface *of* the Aluminum foil it is possible to increase the working frequency $f$ up to more than *1Hz*.

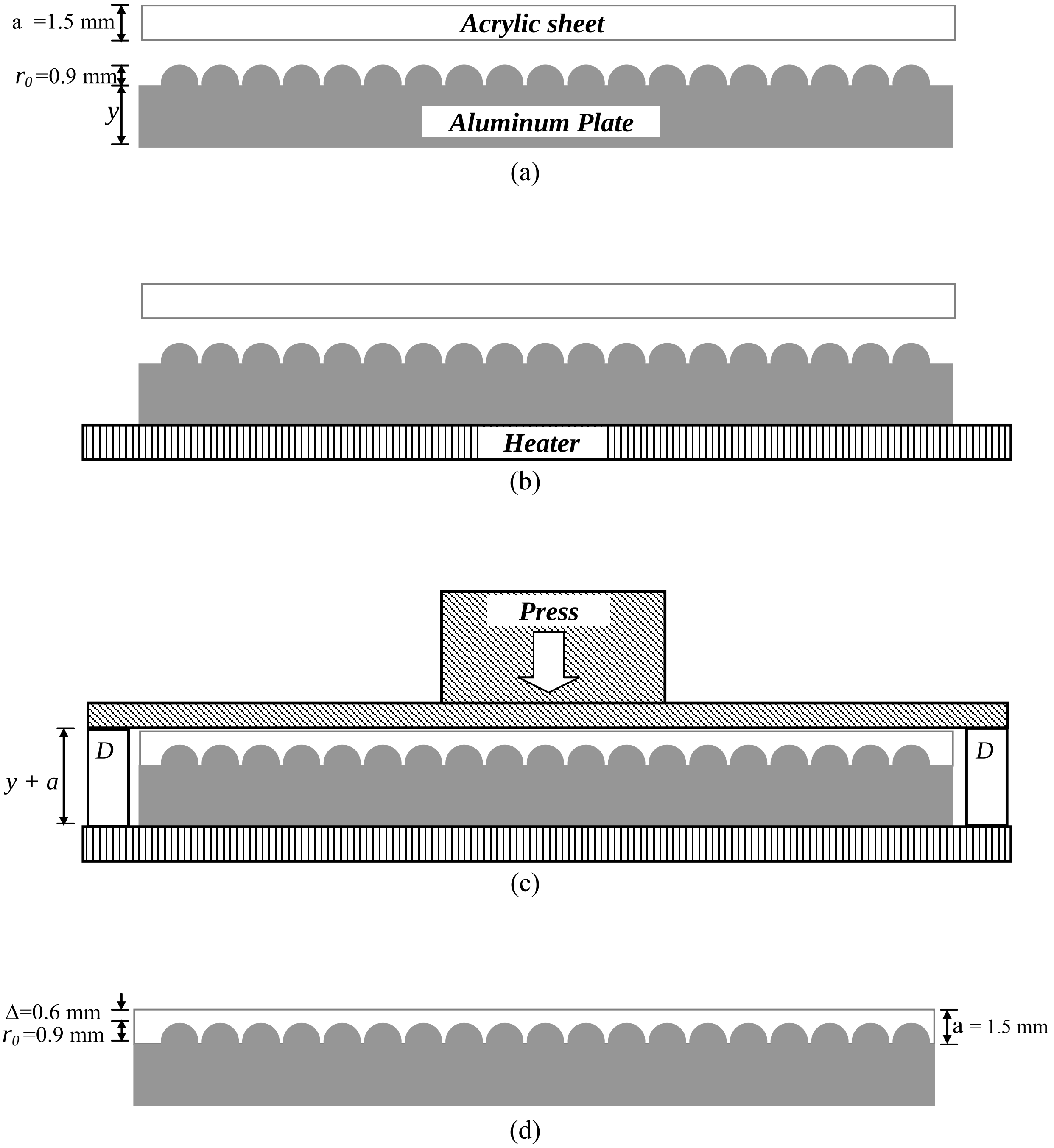

Fig A8 – *Method to coat the Aluminum semi-spheres with acrylic* $(\Delta = a - r_0 = 0.6mm)$. (a)Acrylic sheet (A4 format) with 1.5mm thickness and an Aluminum plate (A4) with several semi-spheres (radius $r_0 = 0.9\ mm$) stamped on its surface. (b)A heater is placed below the Aluminum plate in order to heat the Aluminum. (c)When the Aluminum is sufficiently heated up, the acrylic sheet and the Aluminum plate are pressed, one against the other (The two D devices shown in this figure are used in order to impede that the press compresses the acrylic and the aluminum besides distance $y + a$). (d)After some seconds, the press and the heater are removed, and the device is ready to be used.

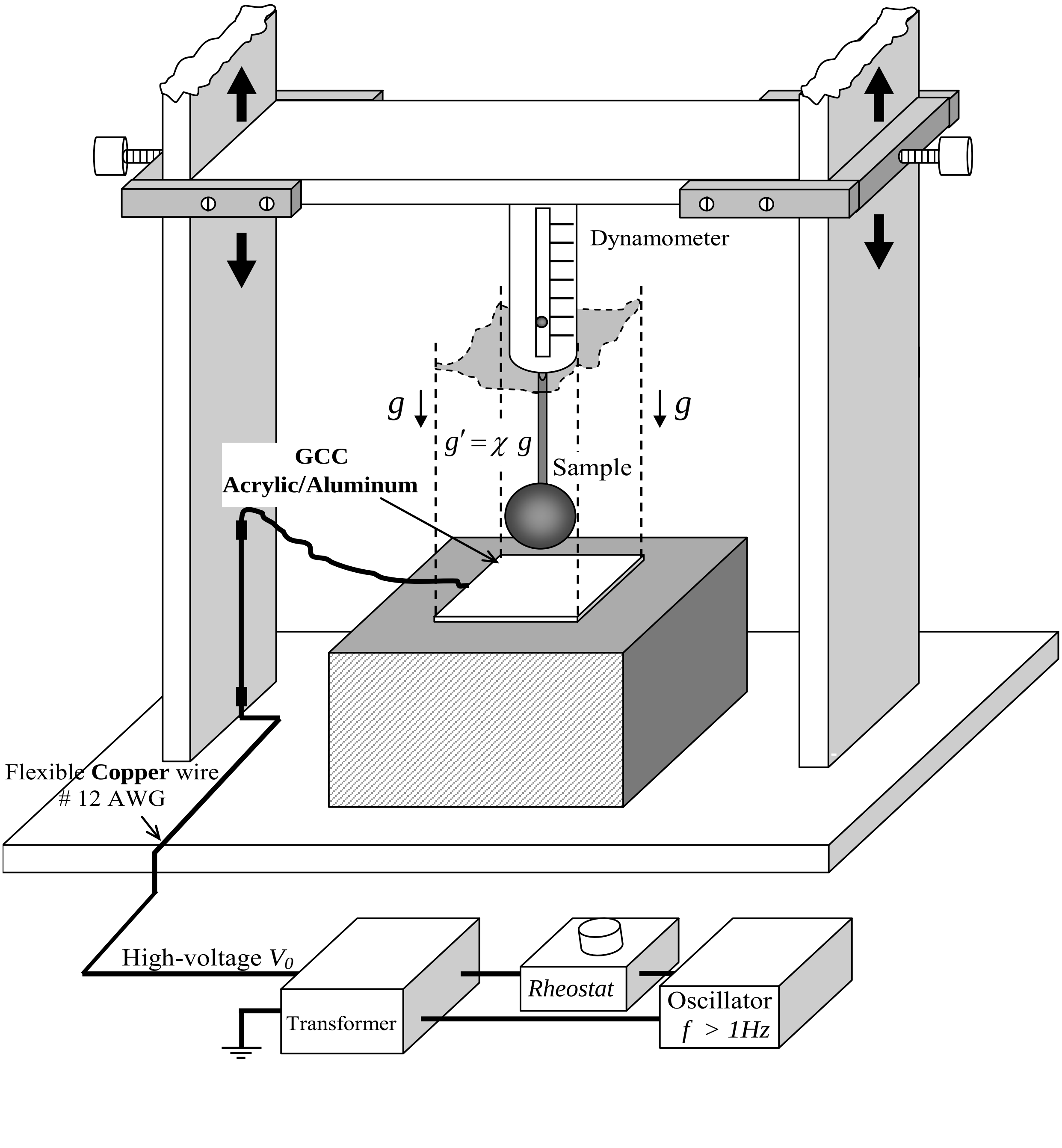

Fig. A9 – *Experimental Set-up using a GCC subjected to high-voltage* $V_0$ *with frequency* $f > 1Hz$. Note that in this case, the pan balance is not necessary because the substance of the Gravitational Shielding is an *air layer* with thickness $d$ *above* the acrylic sheet. This is therefore, more a type of Gravity Control Cell (GCC) with *external gravitational shielding*.

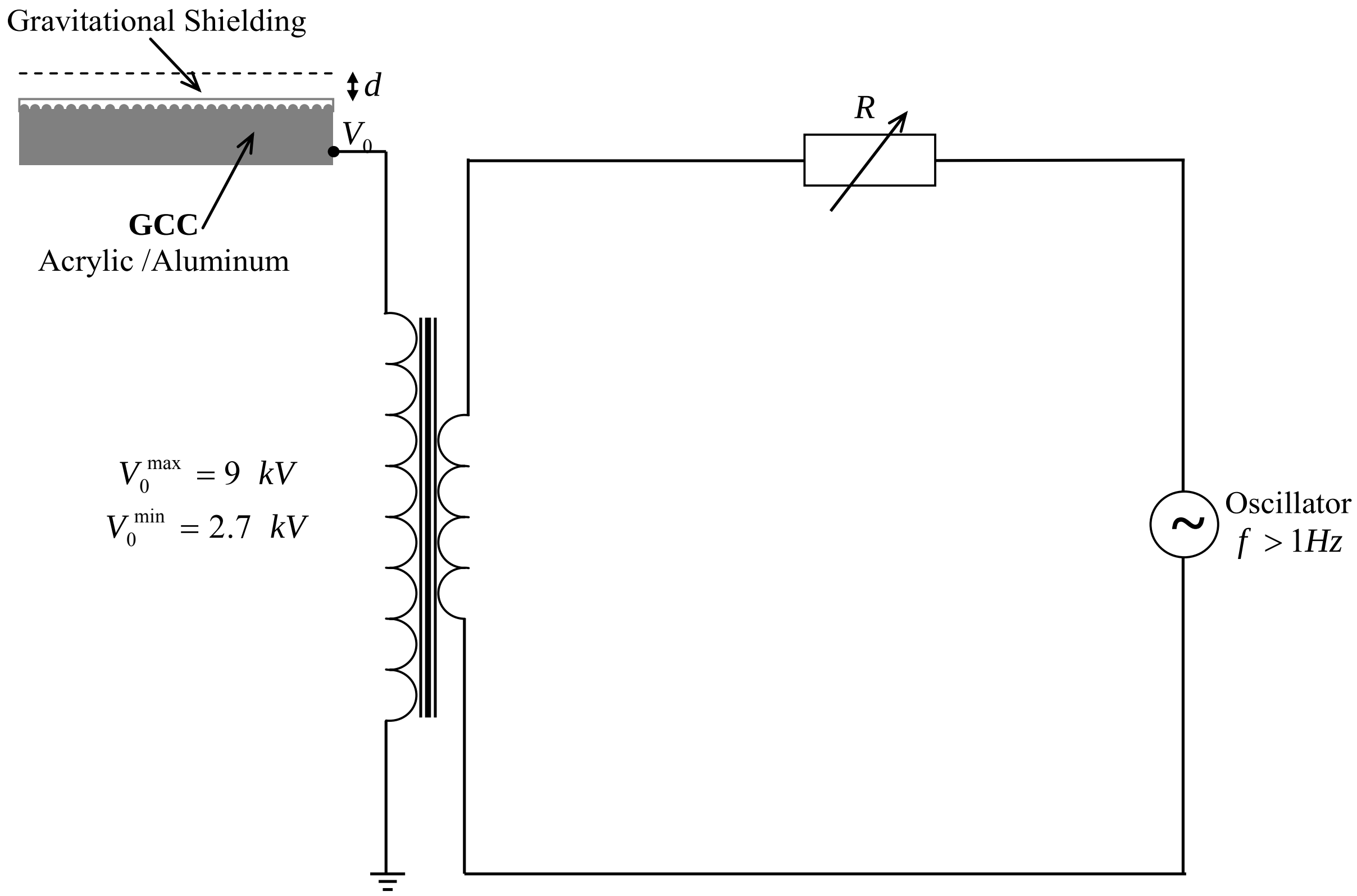

(a)

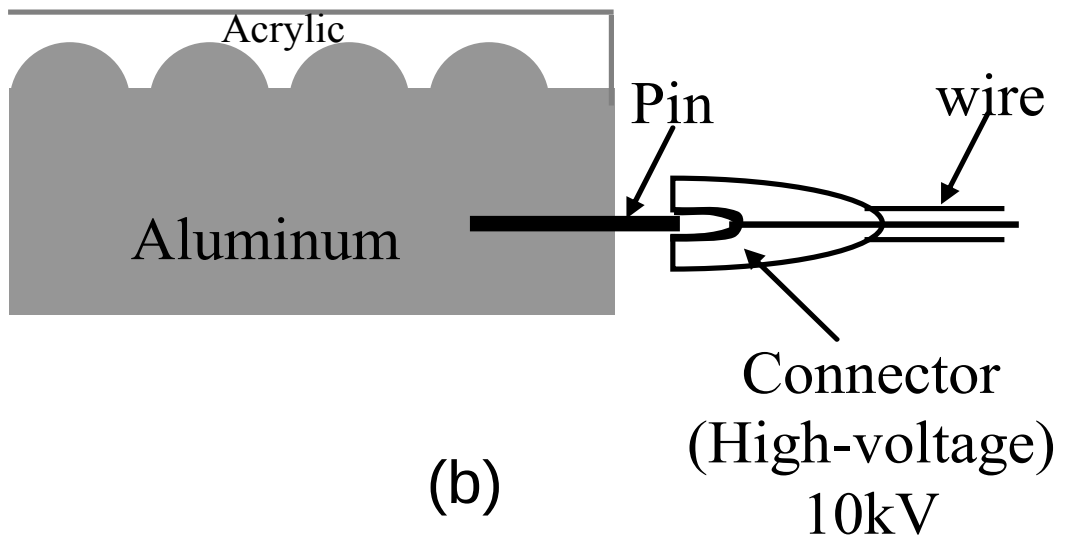

(b)

Fig. A10 – (a) *Equivalent Electric Circuit.* (b) Details of the electrical connection with the Aluminum plate. Note that others connection modes (by the top of the device) can produce destructible interference on the electric lines of the $E_{air}$ field.

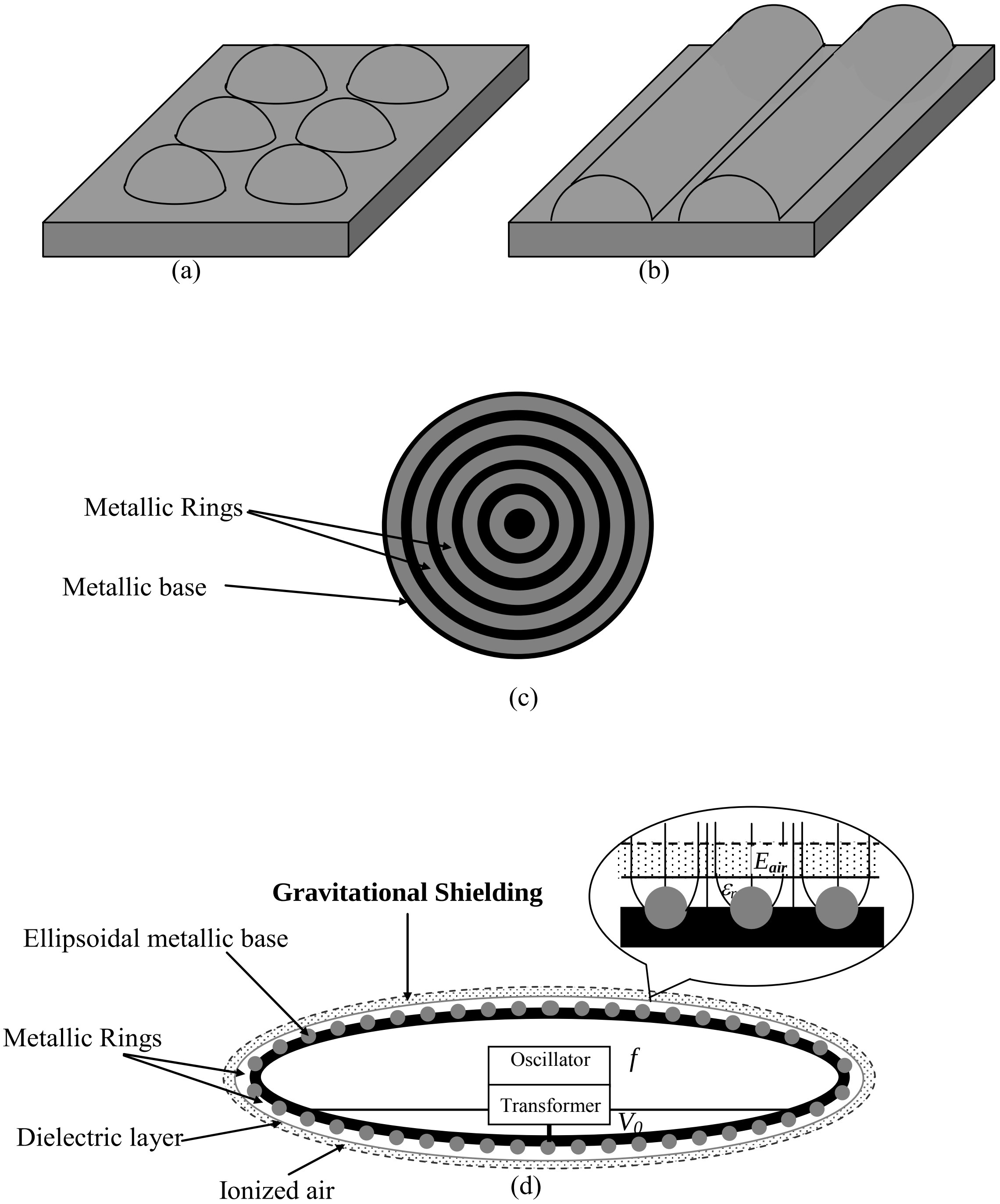

Fig. A11 – *Geometrical forms with similar effects as those produced by the semi-spherical form* – (a) shows the semi-spherical form stamped on the metallic surface; (b) shows the *semi-cylindrical* form (an obvious evolution from the semi-spherical form); (c) shows *concentric metallic rings* stamped on the metallic surface, an evolution from semi-cylindrical form. These geometrical forms produce the same effect as that of the semi-spherical form, shown in Fig.A11 (a). By using concentric metallic rings, it is possible to build *Gravitational Shieldings* around bodies or spacecrafts with several formats (spheres, ellipsoids, etc); (d) shows a Gravitational Shielding around a Spacecraft with *ellipsoidal form*.

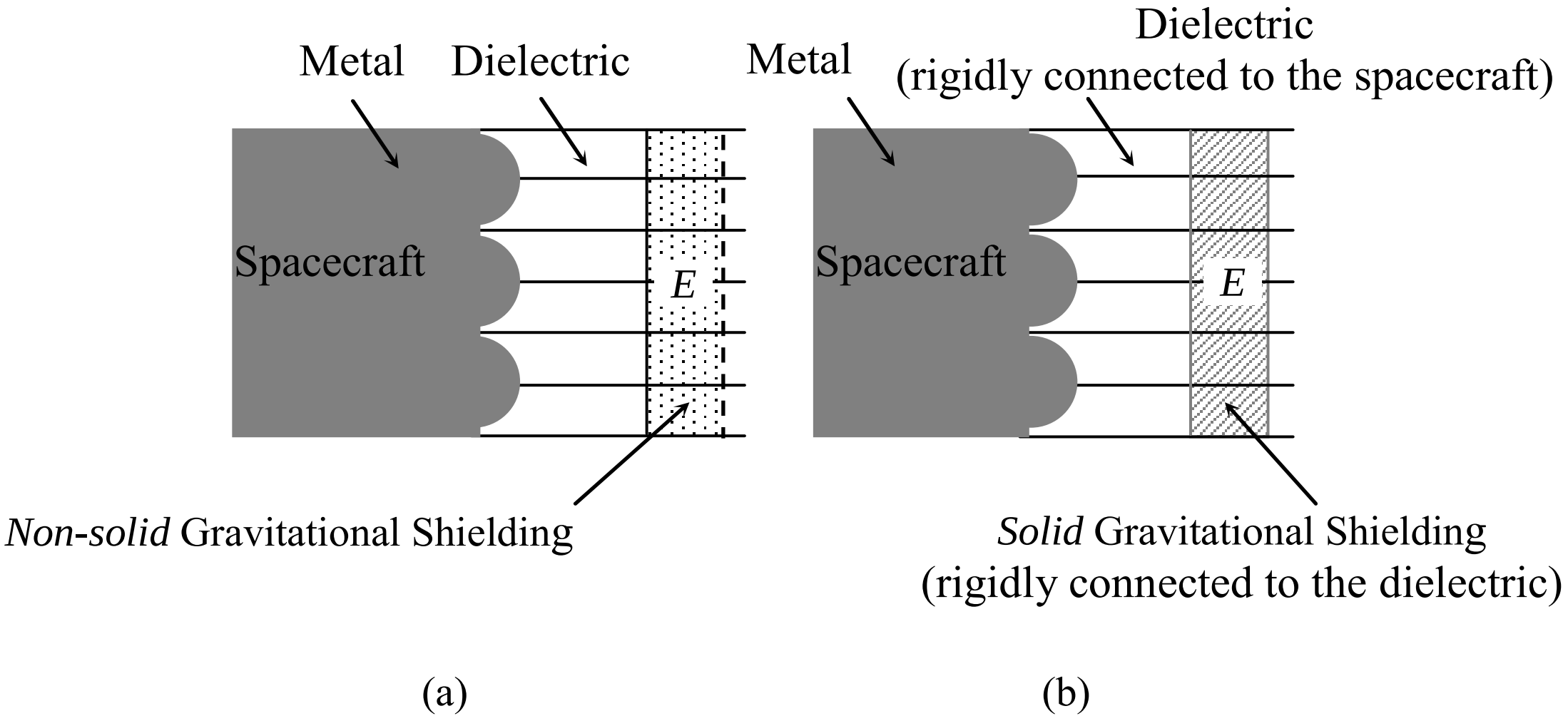

Fig. A12 – *Non-solid and Solid Gravitational Shieldings* - In the case of the Gravitational Shielding produced on a *solid substance* (b), when its molecules go to the *imaginary* space-time, *the electric field that produces the effect also goes to the imaginary space-time together with them*, because in this case, the substance of the Gravitational Shielding is *rigidly connected (by means of the dielectric) to the metal* that produces the electric field. This does not occur in the case of *Air* Gravitational Shielding.

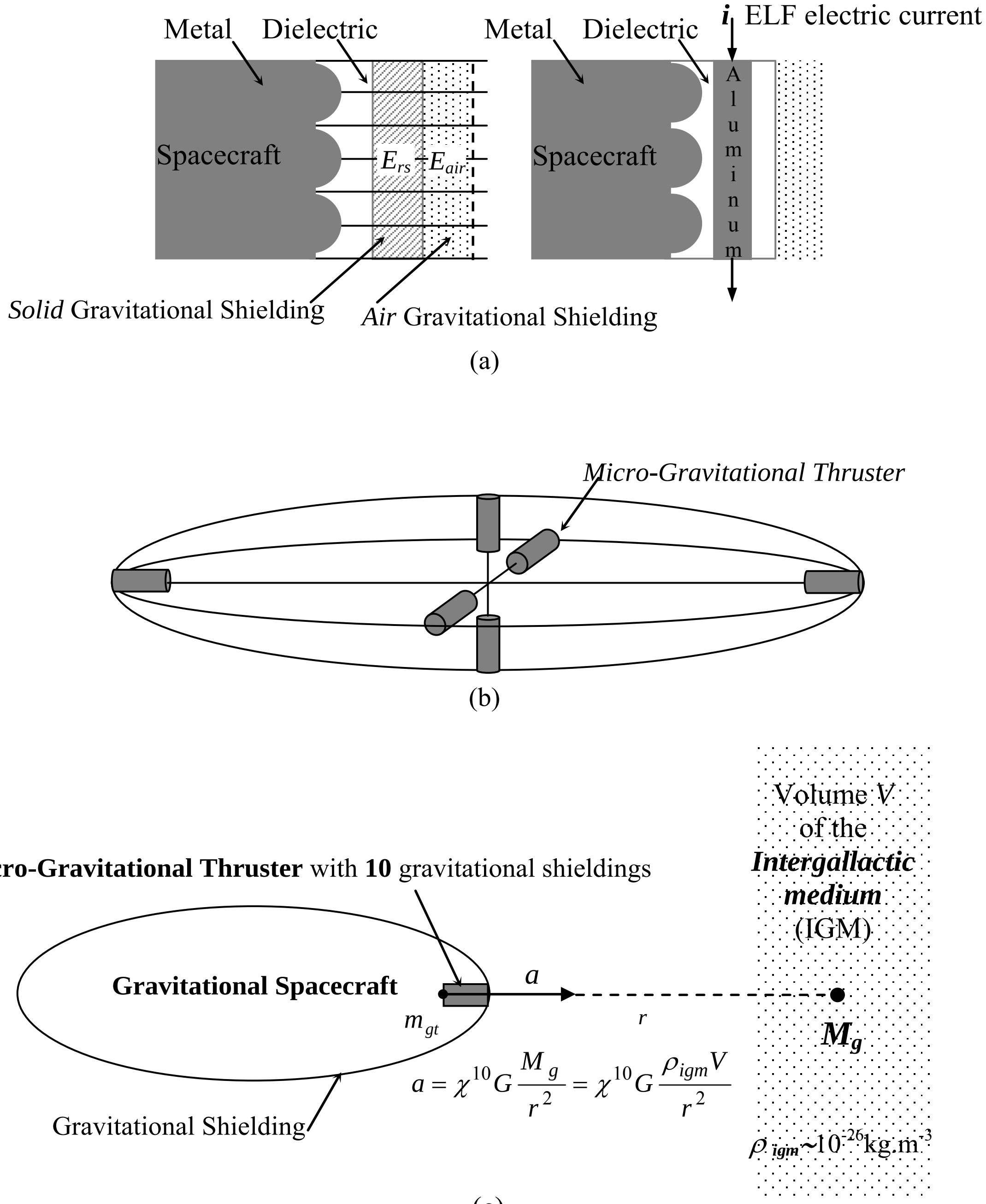

Fig. A13 – *Double Gravitational Shielding and Micro-thrusters* – (a) Shows a double gravitational shielding that makes possible to decrease the *inertial effects* upon the spacecraft when it is traveling both in the *imaginary* space-time and in the *real* space-time. The *solid* Gravitational Shielding also can be obtained by means of *an ELF electric current through a metallic lamina* placed *between the semi-spheres and the Gravitational Shielding of Air* as shown above. (b) Shows 6 *micro-thrusters* placed inside a Gravitational Spacecraft, in order to propel the spacecraft in the directions x, y and z. Note that the Gravitational Thrusters in the spacecraft must have a very small diameter (of the order of *millimeters*) because the hole through the Gravitational Shielding of the spacecraft cannot be large. Thus, these thrusters are in fact *Micro-thrusters*. (c) Shows a micro-thruster inside a spacecraft, and in front of a volume *V* of the intergalactic medium (IGM). Under these conditions, the spacecraft acquires an acceleration *a* in the direction of the volume *V*.

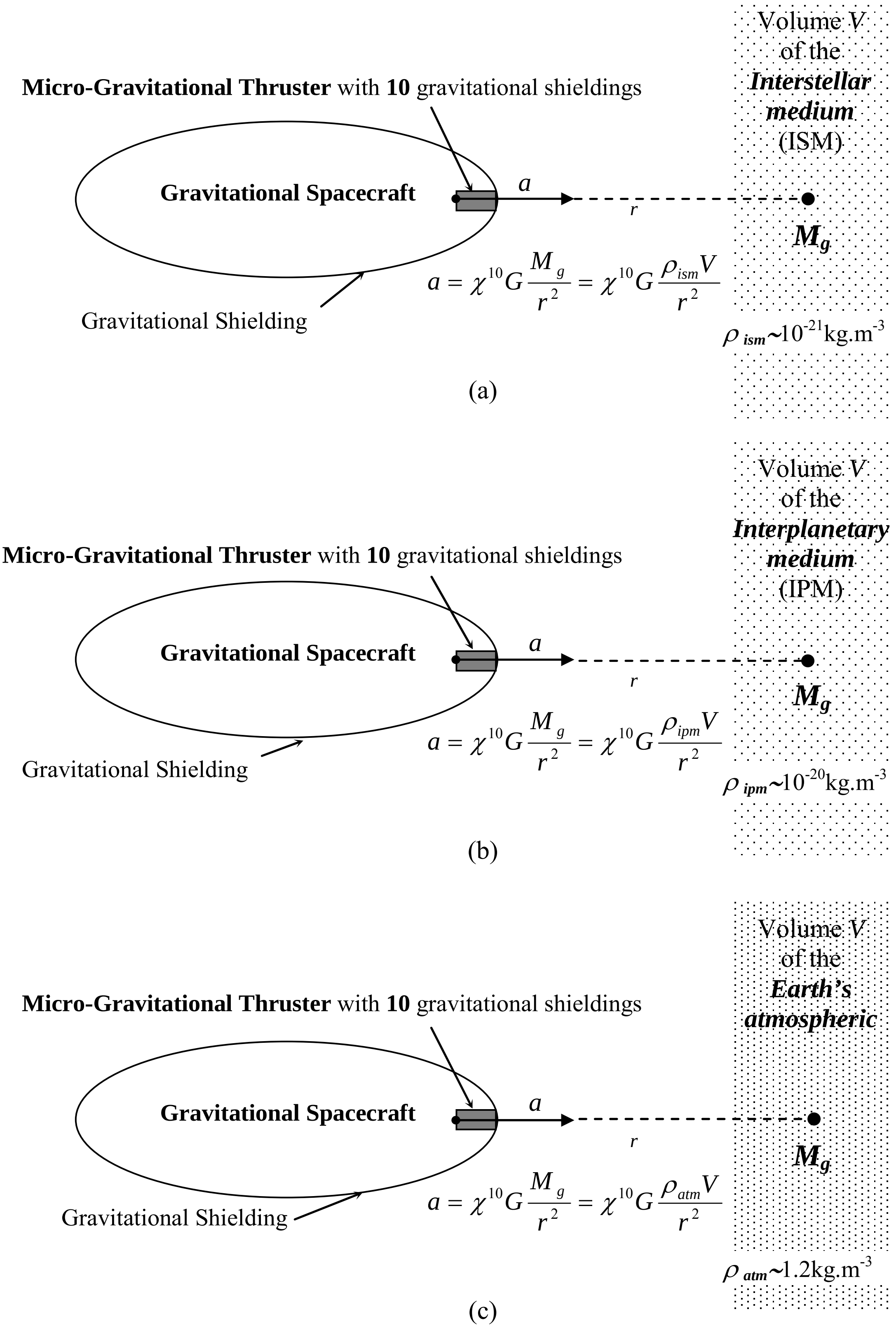

Fig. A14 – *Gravitational Propulsion using Micro-Gravitational Thruster* – (a) Gravitational acceleration produced by a gravitational mass $M_g$ of the *Interstellar Medium*. The density of the Interstellar Medium is about $10^5$ times greater than the density of the *Intergalactic Medium* (b) Gravitational acceleration produced in the *Interplanetary Medium*. (c) Gravitational acceleration produced in the *Earth's atmosphere*. Note that, in this case, $\rho_{atm}$ (*near to the Earth's surface*)is about $10^{26}$ times greater than the density of the *Intergalactic Medium*.

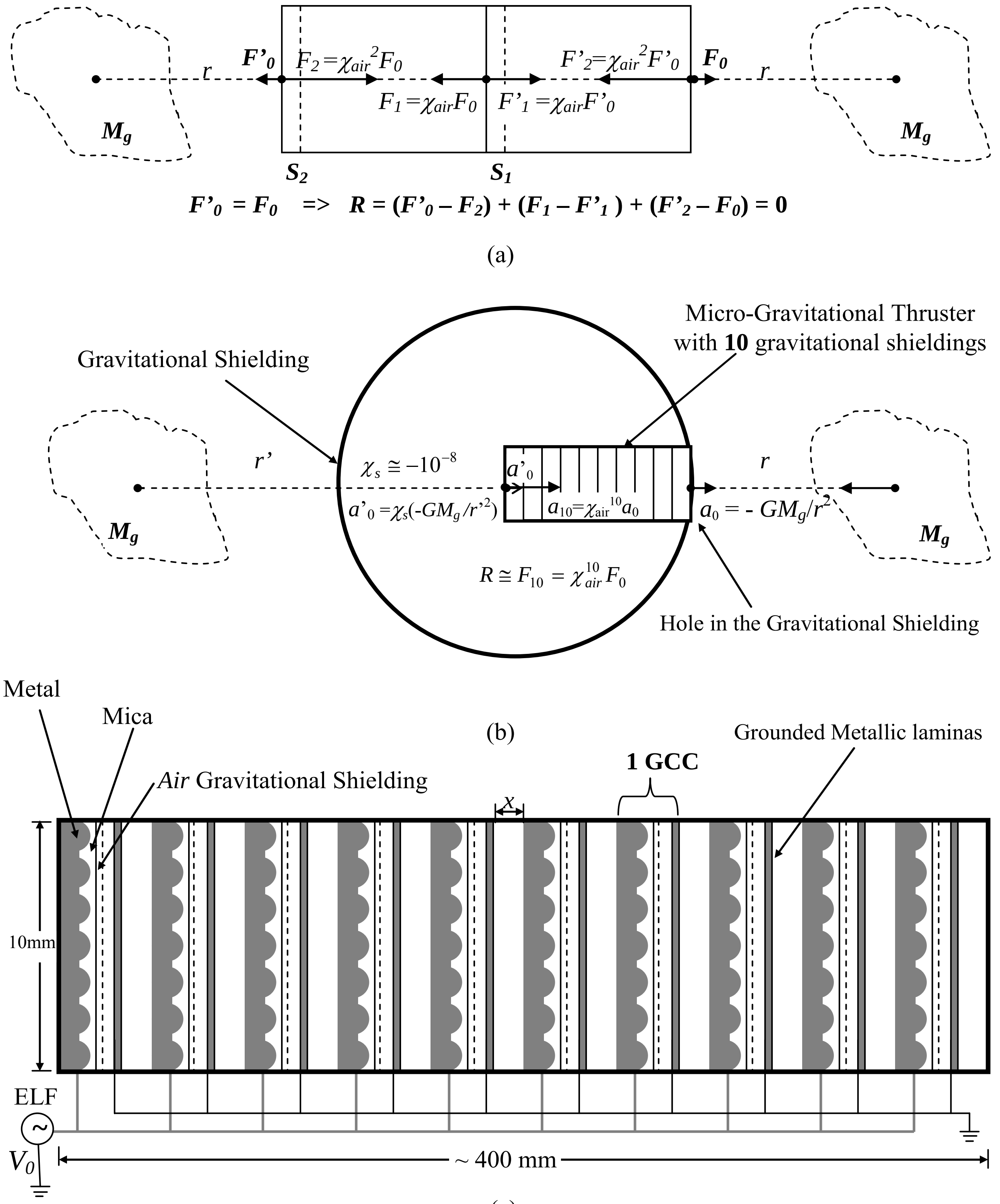

(c)

Fig. A15 – *Dynamics and Structure of the Micro-Gravitational Thrusters* - (a) The Micro-Gravitational Thrusters do not work *outside* the Gravitational Shielding, because, in this case, *the resultant upon the thruster is null* due to the symmetry. (b) The Gravitational Shielding $\left(\chi_s \cong 10^{-8}\right)$ reduces strongly the intensities of the gravitational forces acting on the micro-gravitational thruster, except obviously, through the hole in the gravitational shielding. (c) Micro-Gravitational Thruster with *10 Air Gravitational Shieldings* (10GCCs). The grounded metallic laminas are placed so as to retain the electric field produced by metallic surface behind the semi-spheres.

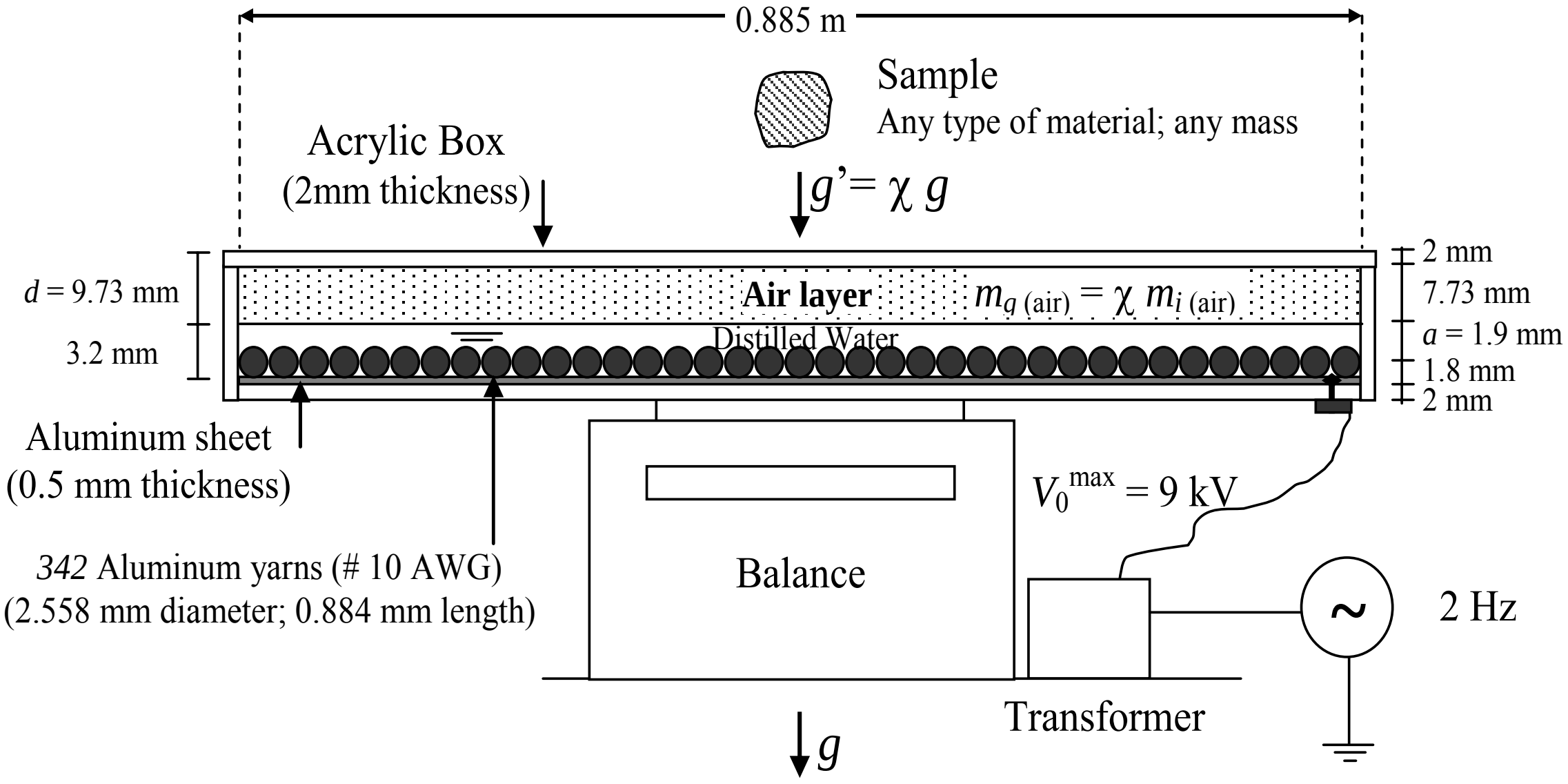

**GCC Cross-section Front view**

(a)

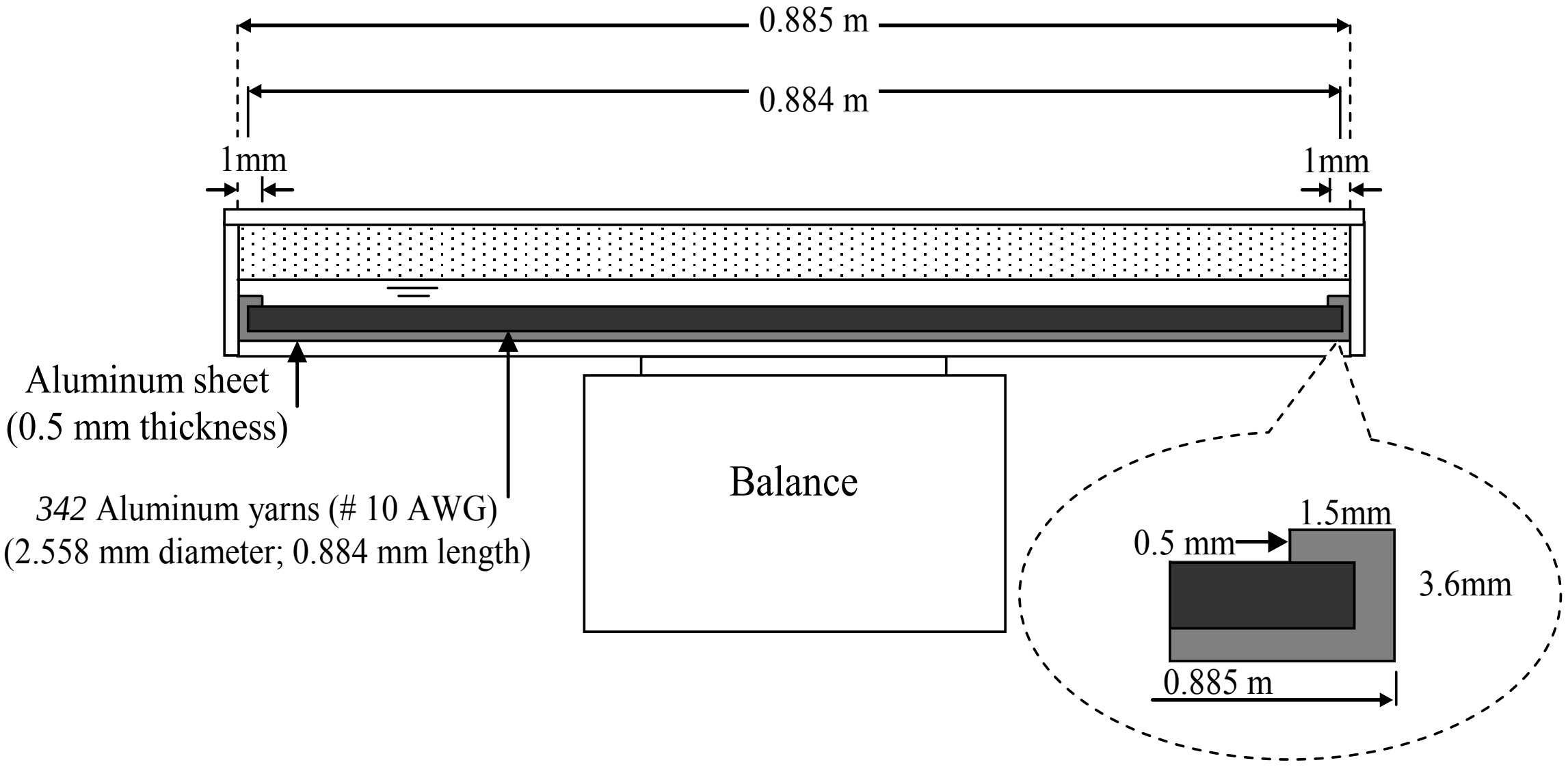

**GCC Cross-section Side View**

(b)

Fig. A16 – *A GCC using distilled Water.*

In total this GCC weighs about 6kg; the air layer 7.3 grams. The balance has the following characteristics: Range 0 – 6kg; readability 0.1g. The yarns are inserted side by side on the Aluminum sheet. Note the detail of fixing of the yarns on the Aluminum sheet.

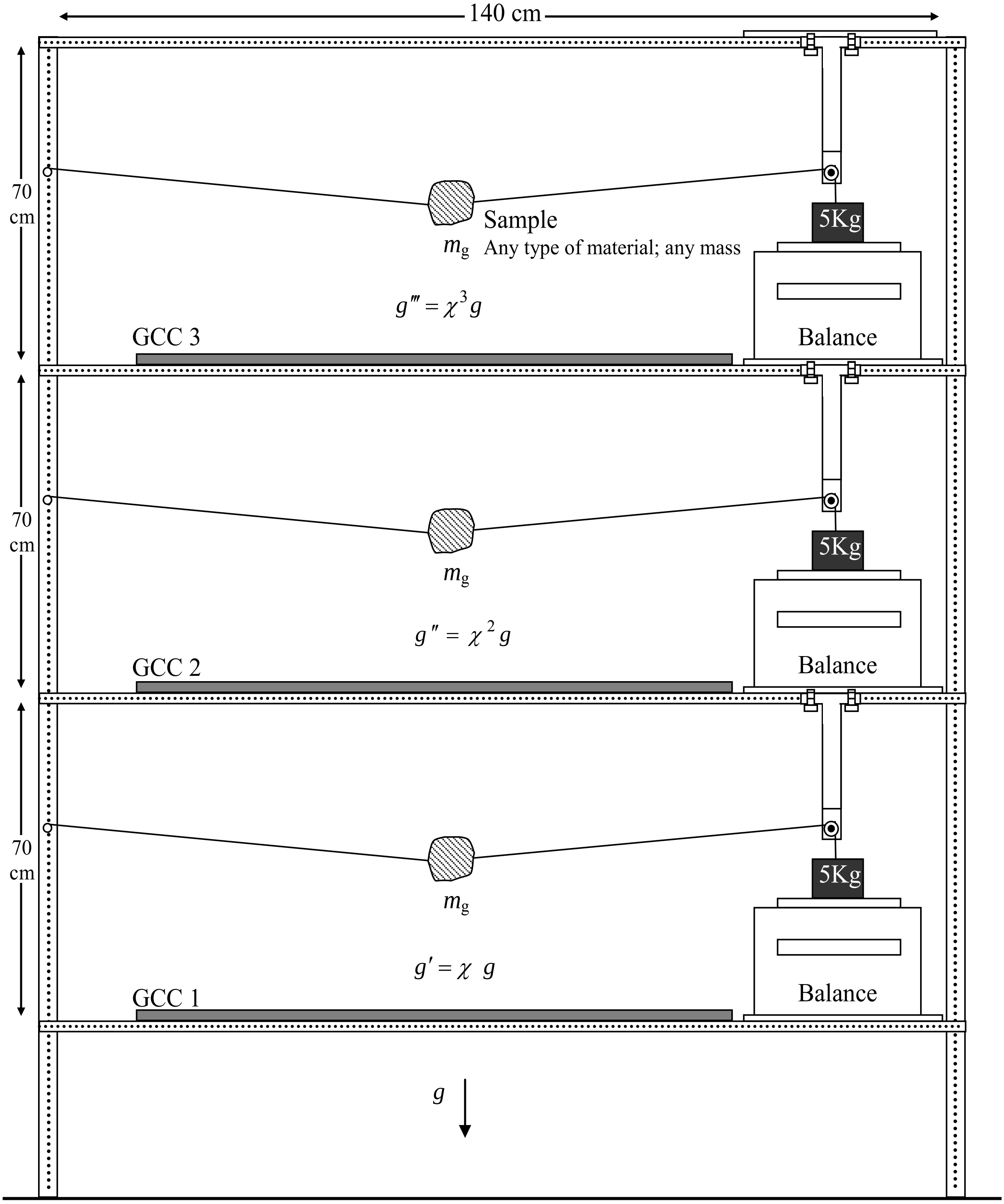

Fig. A17 – *Experimental set-up.* In order to prove *the exponential effect* produced by the superposition of the Gravitational Shieldings, we can take three similar GCCs and put them one above the other, in such way that above the GCC 1 the gravity acceleration will be $g' = \chi g$; above the GCC2 $g'' = \chi^2 g$, and above the GCC3 $g''' = \chi^3 g$. Where $\chi$ is given by Eq. (A47). The arrangement above has been designed for values of $m_g < 13g$ and $\chi$ up to -9 or $m_g < 1kg$ and $\chi$ up to -2 .

# APPENDIX B: Gravity Control Cells (GCCs) made from *Semiconductor Compounds*.

There are some semiconductors compounds with electrical conductivity between $10^4$S/m to 1 S/m, which can have their gravitational mass strongly decreased when subjected to ELF electromagnetic fields.

For instance, the polyvinyl chloride (PVC) compound, called Duracap™ 86103.

It has the following characteristics:

$$\mu_r = 1;\ \varepsilon_r = 3$$
$$\sigma = 3333.3 S/m$$
$$\rho = 1400 kg.m^{-3}$$
$$dieletric \quad strength = 98 KV / mm$$

Then, according to the following equation below (derived from Eq.A14)

$$m_g = \left\{1 - 2\left[\sqrt{1 + 1.758 \times 10^{-27}\left(\frac{\mu_r \sigma^3}{\rho^2 f^3}\right)E_{rms}^4} - 1\right]\right\} m_{i0} \qquad (B1)$$

the *gravitational mass*, $m_g$, of the Duracap™ 86103, when subjected to an electromagnetic field of frequency $f$, is given by

$$m_g = \left\{1 - 2\left[\sqrt{1 + 3.3 \times 10^{-23}\frac{E_{rms}^4}{f^3}} - 1\right]\right\} m_{i0} \qquad (B2)$$

Note that, if the electromagnetic field through the Duracap has *extremely-low frequency*, for example, if $f = 2Hz$, and

$$E_{rms} = 9.4 \times 10^5 V/m \qquad (0.94 kV/mm)$$

Then, its *gravitational mass* will be reduced down to $m_g \cong -1.1 m_{i0}$, reducing in this way, the initial *weight* $(P_0 = m_g g = m_{i0} g)$ of the Duracap down to $-1.1 P_0$.

BACKGROUND FOR EXPERIMENTAL

The Duracap™ 86103 is sold under the form of small cubes. Its melting temperature varies from 177ºC to 188ºC. Thus, a 15cm square Duracap plate with 1 mm thickness can be shaped by using a suitable mold, as the shown in Fig.B1.

Figure B2(a) shows the Duracap plate between the Aluminum plates of a parallel plate capacitor. The plates have the following dimensions: 19cm x 15cm x 1mm. They are painted with an insulating varnish spray of high dielectric strength (ISOFILM). They are connected to the secondary of a transformer, which is connected to a Function Generator. The distance between the Aluminum plates is $d = 1mm$. Thus, the electric field through the Duracap is given by

$$E_{rms} = \frac{E_m}{\sqrt{2}} = \frac{V_0}{\varepsilon_r d\sqrt{2}} \qquad (B3)$$

where $\varepsilon_r$ is the relative permittivity of the dielectric (Duracap), and $V_0$ is the amplitude of the wave voltage applied on the capacitor.

In order to generate *ELF wave voltage* of $f = 2Hz$, we can use the widely-known Function Generator HP3325A (Op.002 High Voltage Output) that can generate sinusoidal voltages with *extremely-low* frequencies and amplitude up to *20V* ($40V_{pp}$ into $500\Omega$ load). The maximum output current is $0.08 A_{pp}$; output impedance <2Ω at ELF.

The turns ratio of the transformer (Bosch red coil) is $200:1$. Thus, since the maximum value of the amplitude of the voltage produced by the Function Generator is $V_p^{max} = 20\ V$, then the maximum secondary voltage will be $V_s^{max} = V_0^{max} = 4kV$. Consequently, Eq. (B3) gives

$$E_{rms}^{max} = 2.8 \times 10^6 V/m$$

Thus, for $f = 2Hz$, Eq. (B2) gives

$$m_g = -29.5 m_{i0}$$

The *variations on the gravitational mass* of the Duracap plate can be measured by a pan balance with the following characteristics: range 0 – 1.5kg ; readability 0.01g, using the set-up shown in Fig. B2(a).

Figure B2(b) shows the set-up to measure *the gravity acceleration variations above* the Duracap plate (Gravitational Shielding effect). The samples used in this case, can be of several types of material.

Since voltage waves with frequencies very below 1Hz have a very long period, we cannot consider, in practice, their *rms* values. However, we can add a sinusoidal voltage $V_{osc} = V_0 \sin \omega t$ with a DC voltage $V_{DC}$, by means of the circuit shown in Fig.B3. Thus, we obtain $V = V_{DC} + V_0 \sin \omega t$ ; $\omega = 2\pi f$. If $V_0 << V_{DC}$ then $V \cong V_{DC}$. Thus, the voltage $V$ varies with the frequency $f$, but its intensity is approximately equal to $V_{DC}$, i.e., $V$ will be practically constant. This is of fundamental importance for maintaining the value of the gravitational mass of the body, $m_g$, sufficiently stable during all the time, in the case of $f << Hz$.

We have shown in this paper that it is possible to control the gravitational mass of a spacecraft, simply by controlling the gravitational mass of a body *inside* the spacecraft (Eq.(10)). This body can be, for example, the *dielectric* between the plates of a capacitor, whose gravitational mass can be easily controlled by means of an ELF electromagnetic field produced between the plates of the capacitor. We will call this type of capacitor of *Capacitor of Gravitational Mass Control* (CGMC).

Figure B 4(a) shows a CGMC placed in the center of the spacecraft. Thus, the gravitational mass of the spacecraft can be controlled simply by varying the gravitational mass of the dielectric of the capacitor by means of an ELF electromagnetic field produced between the plates of the capacitor. Note that the Capacitor of Gravitational Mass Control can have the *spacecraft's own form* as shown in Fig. B 4(b). The dielectric can be, for example, a Duracap plate, as shown in this appendix. In this case, the gravitational mass of the dielectric is expressed by Eq. (B2). Under these circumstances, the *total* gravitational mass of the spacecraft will be given by Eq.(10):

$$M_{g(spacecraf)} = M_{i0} + \chi_{dielectric} m_{i0}$$

where $M_{i0}$ is the rest inertial mass of the spacecraft(without the dielectric) and $m_{i0}$ is the rest inertial mass of the dielectric; $\chi_{dielectric} = m_g / m_{i0}$, where $m_g$ is the gravitational mass of the dielectric. By decreasing the value of $\chi_{dielectric}$, the gravitational mass of the spacecraft decreases. It was shown, that the value of $\chi$ can be negative. Thus, for example, when $\chi_{dielectric} \cong -M_{i0}/m_{i0}$, the *gravitational mass of the spacecraft gets very close to zero.* When $\chi_{dielectric} < -M_{i0}/m_{i0}$, the gravitational mass of the spacecraft becomes negative.

Therefore, *for an observer out of the spacecraft* the gravitational mass of the spacecraft is $M_{g(spacecraft)} = M_{i0} + \chi_{dielectric} m_{i0}$, and not $M_{i0} + m_{i0}$.

Since the dielectric strength of the Duracap is $98kV/mm$, a Duracap plate with *1mm* thickness can withstand up to $98kV$. In this case, the value of $\chi_{dielectric}$ for $f = 2Hz$, according to Eq. (B2), is

$$\chi_{dielectric} = m_g / m_{i0} \cong -10^4$$

Thus, for example, if the inertial mass of the spacecraft is $M_{i0} \cong 10021.0014kg$ and, the inertial mass of the *dielectric* of the *Capacitor of Gravitational Mass Control* is $m_{i0} \cong 1.0021kg$, then the gravitational mass of the spacecraft becomes

$$M_{g(spacecraf)} = M_{i0} + \chi_{dielectric} m_{i0} \cong 10^{-3} kg$$

This value is much smaller than $+0.159 M_{i0}$.

It was shown [1] that, when the gravitational mass of a particle is reduced to values between $+0.159 M_i$ and $-0.159 M_i$, it becomes *imaginary*, i.e., the gravitational and the inertial masses of the particle become *imaginary*. Consequently, the particle *disappears* from our ordinary space-time.

This means that we cannot reduce the gravitational mass of *the spacecraft* below $+0.159 M_i$, *unless we want to make it imaginary*.

Obviously this limits the minimum value of $\chi_{dielectric}$, i.e. $\chi_{dielctric}^{min} = 0.159$. Consequently, if the gravity acceleration *out of the spacecraft* (in a given direction) is $g$, then, according to the Gravitational Shielding Principle, the corresponding gravity acceleration upon the crew of the spacecraft can be reduced just down to $0.159g$. In addition, since the Mach's principle says that *the local inertial forces are produced by the gravitational interaction of the local system with the distribution of cosmic masses* then *the inertial effects* upon the crew would be reduced just by $\chi_{dielectric} = 0.159$.

However, there is a way to strongly reduce the inertial effects upon the crew of the spacecraft without making it *imaginary*. As shown in Fig. B4 (c), we can build an *inertial shielding*, with $n$ superimposed CGMCs. In this case, according to the Gravitational Shielding Principle, the gravity upon the crew will be given by $g_n = \chi_{dielectric}^n g$, where $g$ is the gravity acceleration out of the spacecraft (in a given direction) and $\chi_{dielectric} = m_g / m_{i0}$ ; $m_g$ and $m_{i0}$ are, respectively, the gravitational mass and the inertial mass of the dielectric. Under these conditions the *inertial effects upon the crew* will be reduced by $\chi_{dielectric}^n$.

Thus, for $n = 10$ (*ten* superimposed CGMCs), and $\chi_{dielectric} \cong 0.2$, the *inertial effects upon the crew* will be reduced by $\chi_{dielectric}^{10} \cong 1 \times 10^{-7}$. Therefore, if the maximum thrust produced by the thrusters of the spacecraft is $F = 10^5 N$, then the intensities of the inertial forces upon the crew will not exceed $0.01N$, i.e. they will be practically negligible.

Under these circumstances, the gravitational mass of the spacecraft, for an observer out of the spacecraft, will be just approximately equal to the gravitational mass of the *inertial shielding*, i.e. $M_{g(spacecraft)} \cong M_{g(inertial.shield)}$.

If $M_{g(inertial.shield)} \cong 10^3 kg$, and the thrusters of the spacecraft are able to produces up to $F = 3\times10^5 N$, the spacecraft will acquires an acceleration given by

$$a_{spacecraft} = \frac{F}{M_{g(spacecraft)}} \cong 3\times10^2 m.s^{-2}$$

With this acceleration it can reach velocities close to *Mach 10* in some seconds.

The velocity that the spacecraft can reach in the *imaginary* spacetime is much greater than this value, since $M_{g(spacecraft)}$, as we have seen, can be reduced down to $\cong 10^{-3} kg$ or less.

Thus, if the thrusters of the spacecraft are able to produces up to $F = 3\times10^5 N$, and $M_{g(spacecraft)} \cong 10^{-3} kg$, the spacecraft will acquires an acceleration given by

$$a_{spacecraft} = \frac{F}{M_{g(spacecraft)}} \cong 3\times10^8 m.s^{-2}$$

With this acceleration it can reach velocities close to the light speed in less than 1 second. After 1 month, the velocity of the spacecraft would be about $10^{15} m/s$ (remember that in the *imaginary* spacetime the maximum velocity of propagation of the interactions is *infinity* [1]).

OTHER SEMICONDUCTOR COMPOUNDS

A semiconductor compound, which can have its gravitational mass strongly decreased when subjected to ELF electromagnetic fields is the CoorsTek Pure SiC™ LR CVD *Silicon Carbide*, 99.9995% ‡‡‡‡. This Low-resistivity (LR) pure Silicon Carbide has electrical conductivity of *5000S/m* at room temperature; $\varepsilon_r = 10.8$ ; $\rho = 3210 kg.m^{-3}$; dielectric strength >10 KV/mm; maximum working temperature of 1600°C.

Another material is the *Alumina-CNT*, recently discovered §§§§. It has electrical conductivity of *3375 S/m* at 77°C in samples that were 15% nanotubes by volume [17]; $\varepsilon_r = 9.8$; $\rho = 3980 kg.m^{-3}$ ; dielectric strength 10-20KV/mm; maximum working temperature of 1750°C.

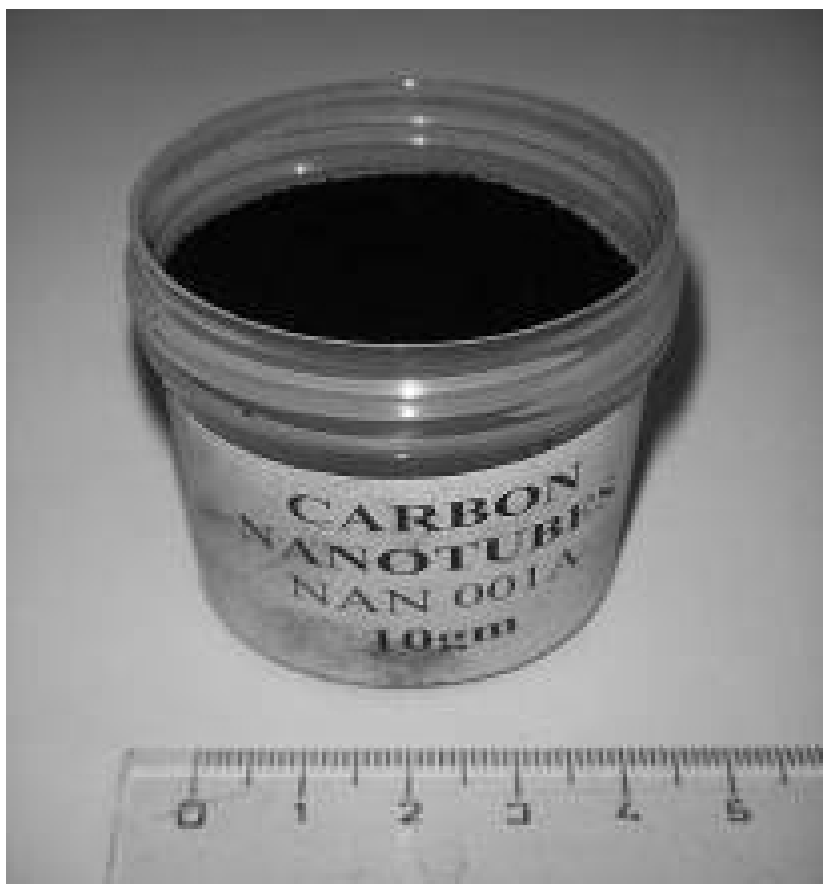

The novel *Carbon Nanotubes Aerogels******, called *CNT Aerogels* are also suitable to produce Gravitational Shieldings, mainly due to their very small densities. The electrical conductivity of the *CNT Aerogels* is *70.4S/m* for a density of $\rho = 7.5 kg.m^{-3}$ [18]; $\varepsilon_r \approx 10$. Recently (2010), it was announced the discovered of *Graphene Aerogel* with $\sigma =\sim 1\times10^2 S/m$ and $\rho = 10 kg.m^{-3}$ [19] (Aerogels exhibit higher dielectric strength than expected for porous materials).

‡‡‡‡ www.coorstek.com

§§§§ Recently, it was discovered that Carbon nanotubes (CNTs) can be added to *Alumina* ($Al_2O_3$) to convert it into a good electrical conductor.

***** In 2007, Mateusz Brying *et al.* working with Prof. Arjun Yodh at the University of Pennsylvania produced the first aerogels made entirely of carbon nanotubes (CNT Aerogels) [20] that, depending on the processing conditions, can have their electrical conductivity ranging as high as 100 S/m.

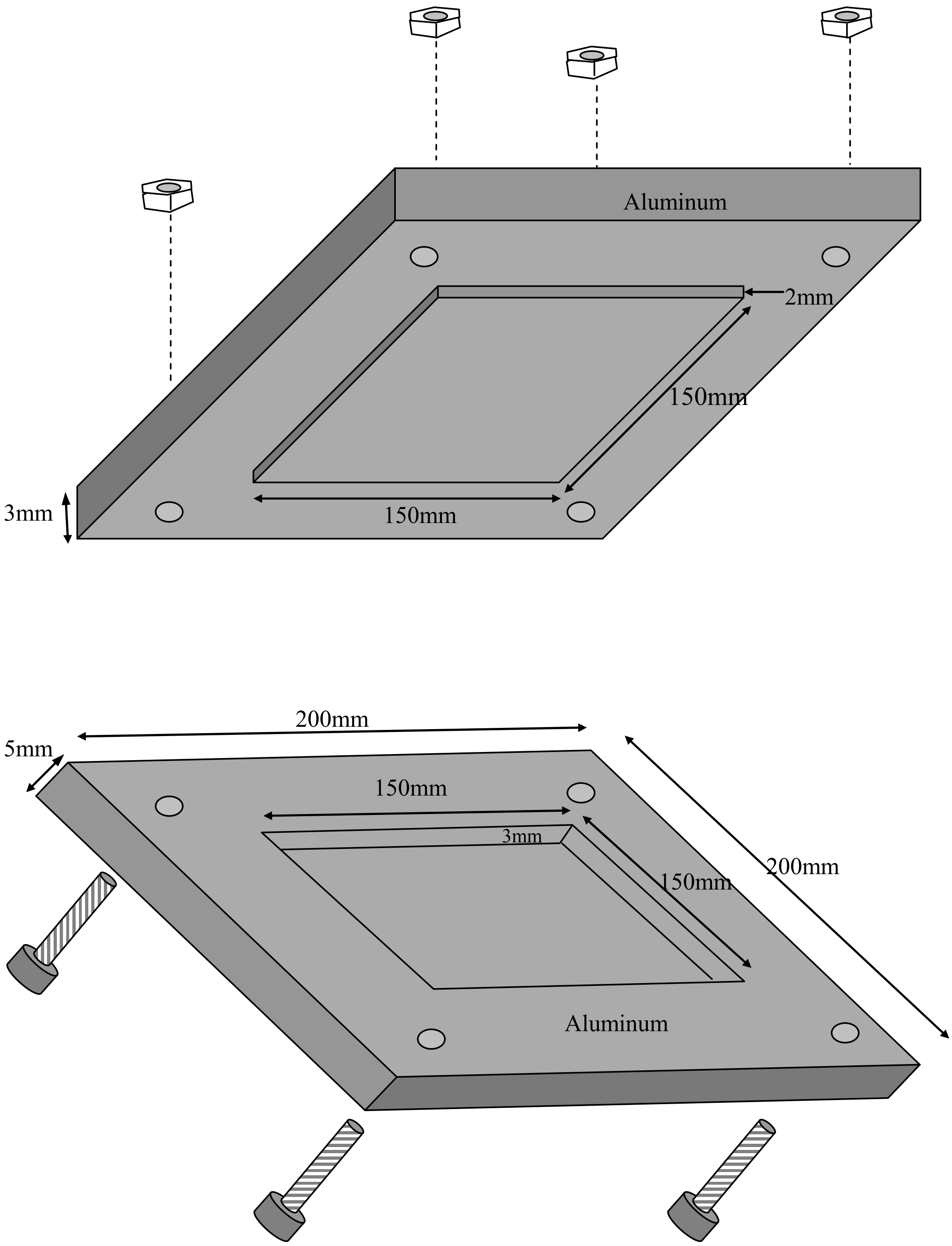

Fig.B1 – Mold design

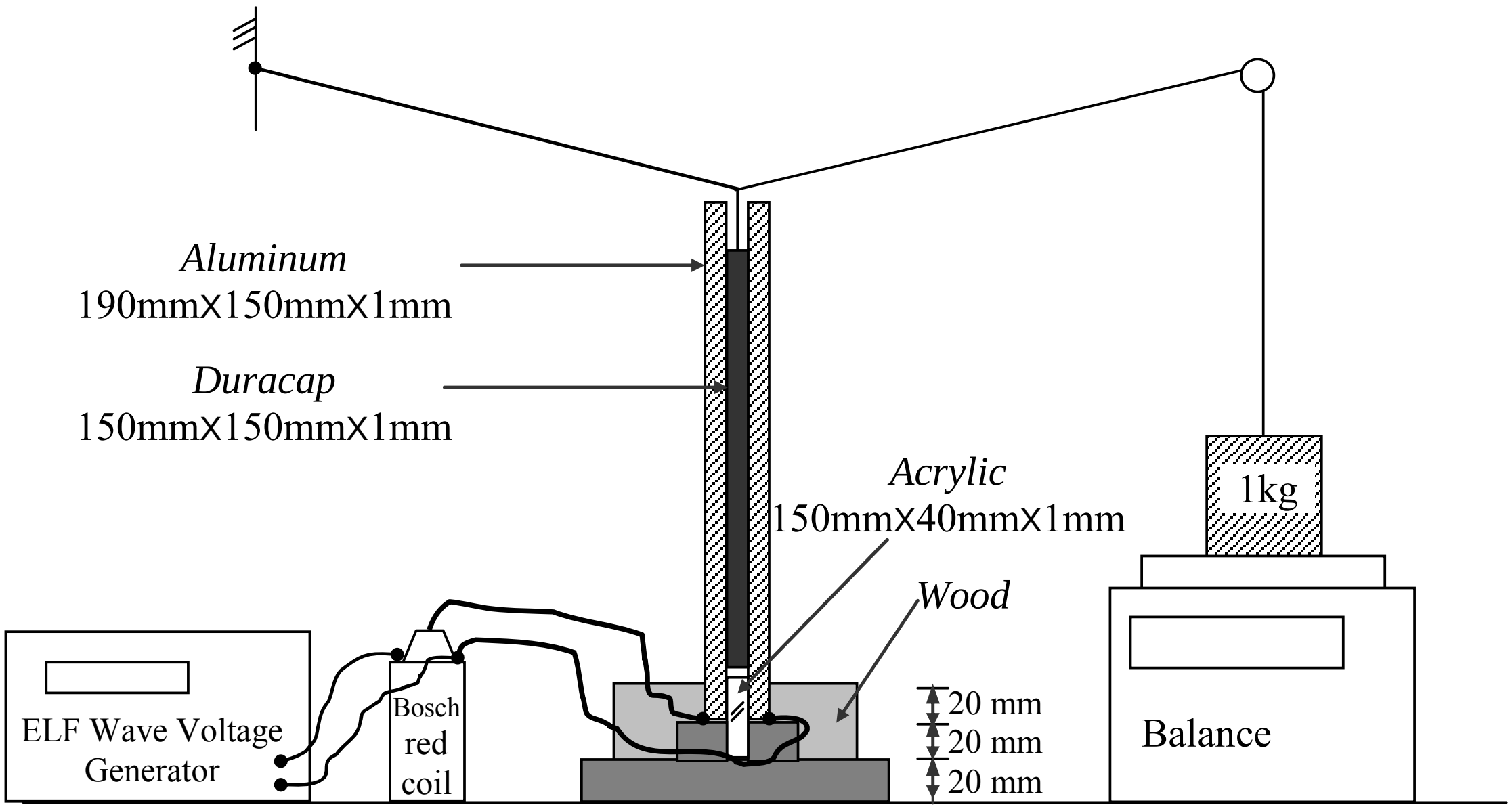

(a)

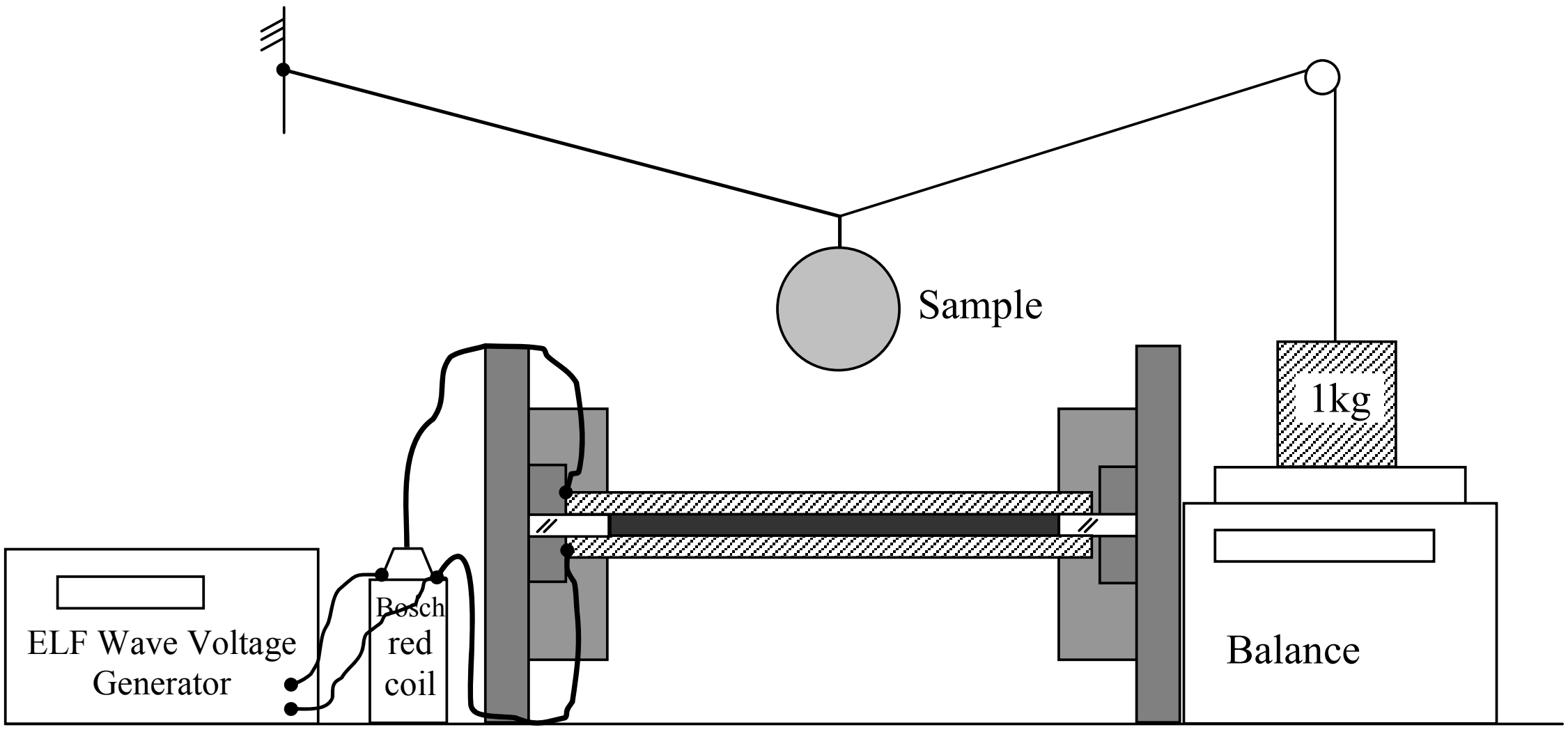

(b)

Fig.B2 – Schematic diagram of the experimental set-up

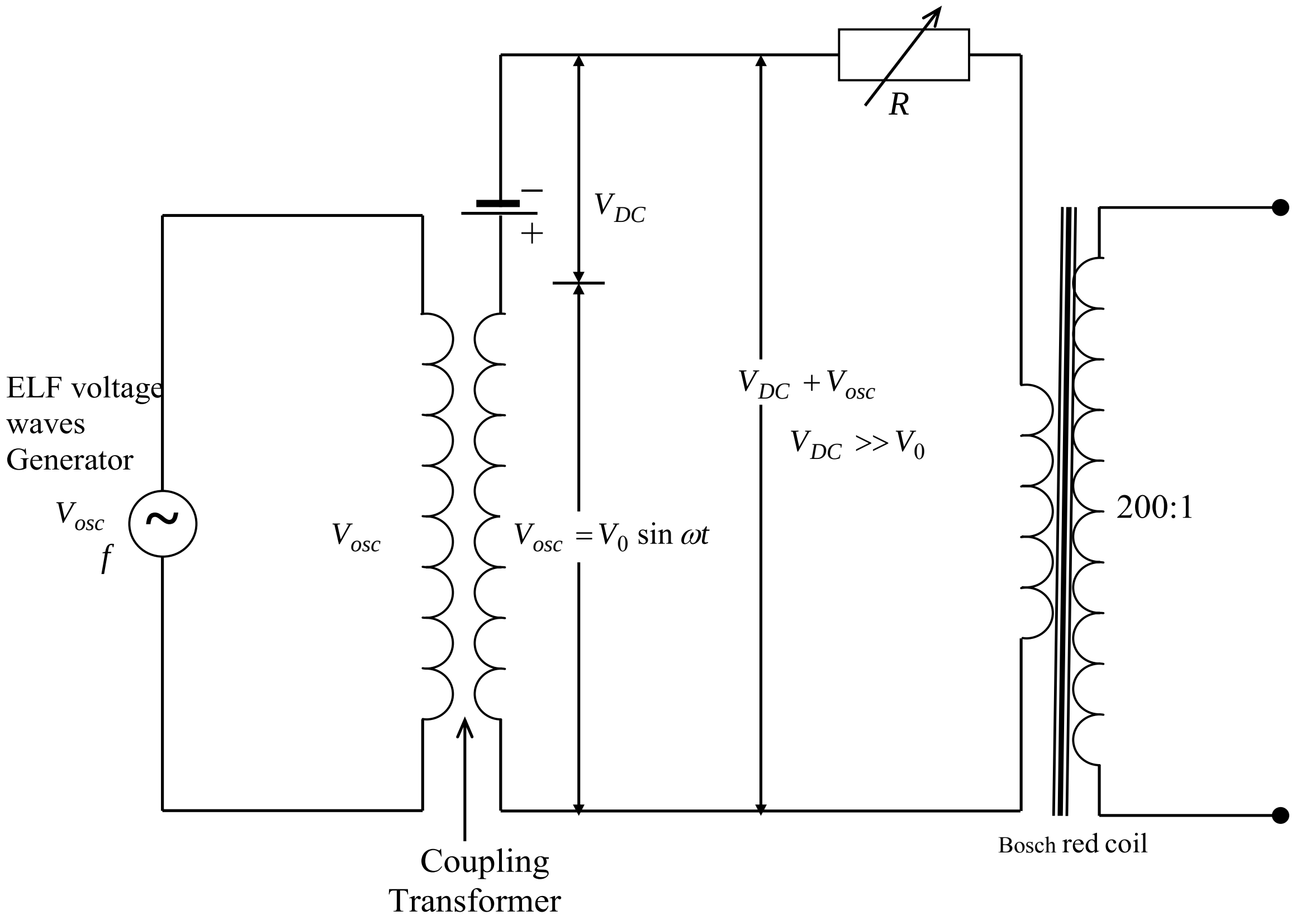

Fig. B3 – Equivalent Electric Circuit

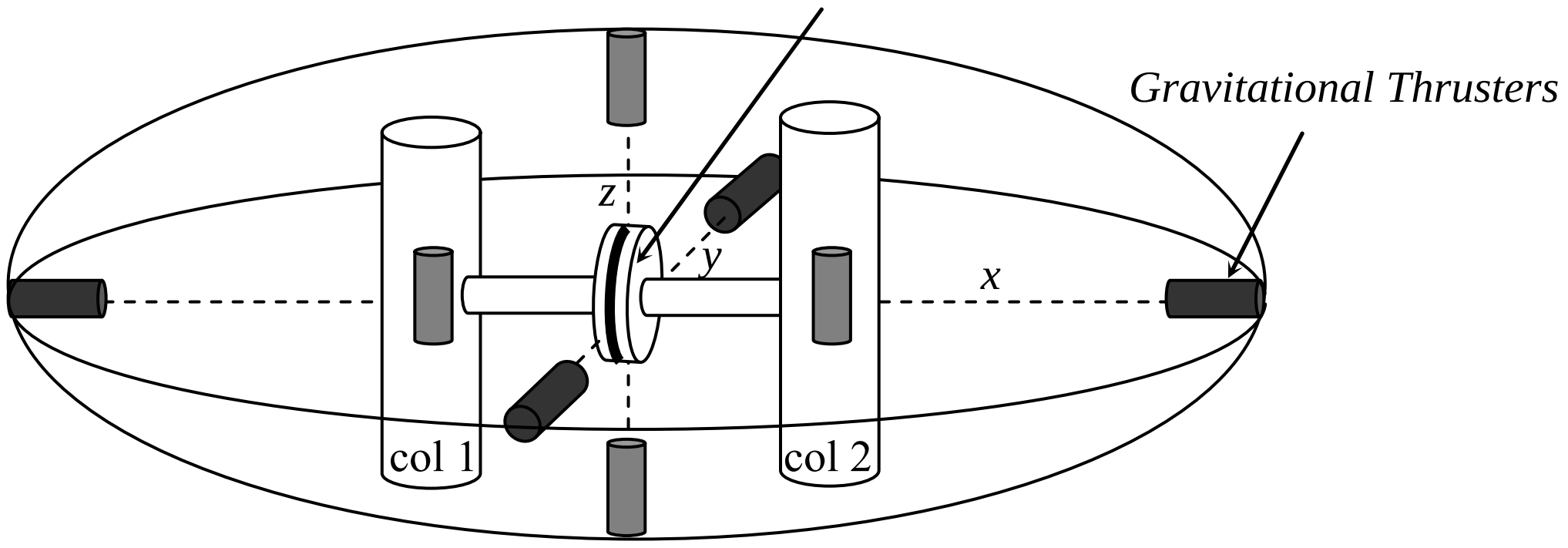

(a)

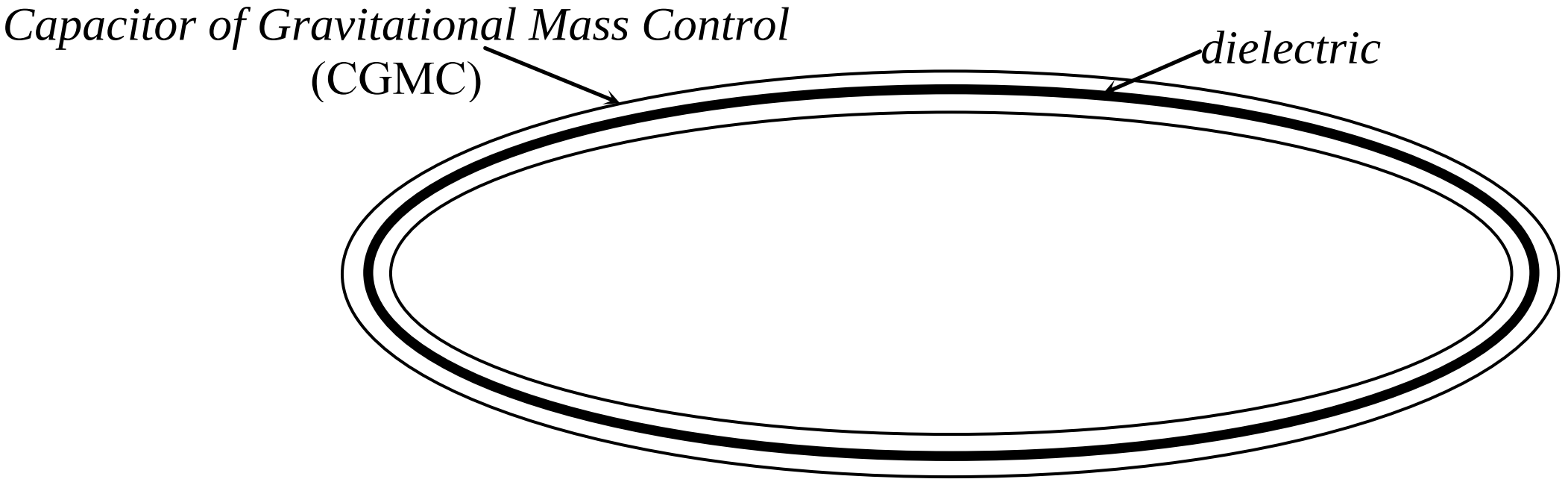

(b)

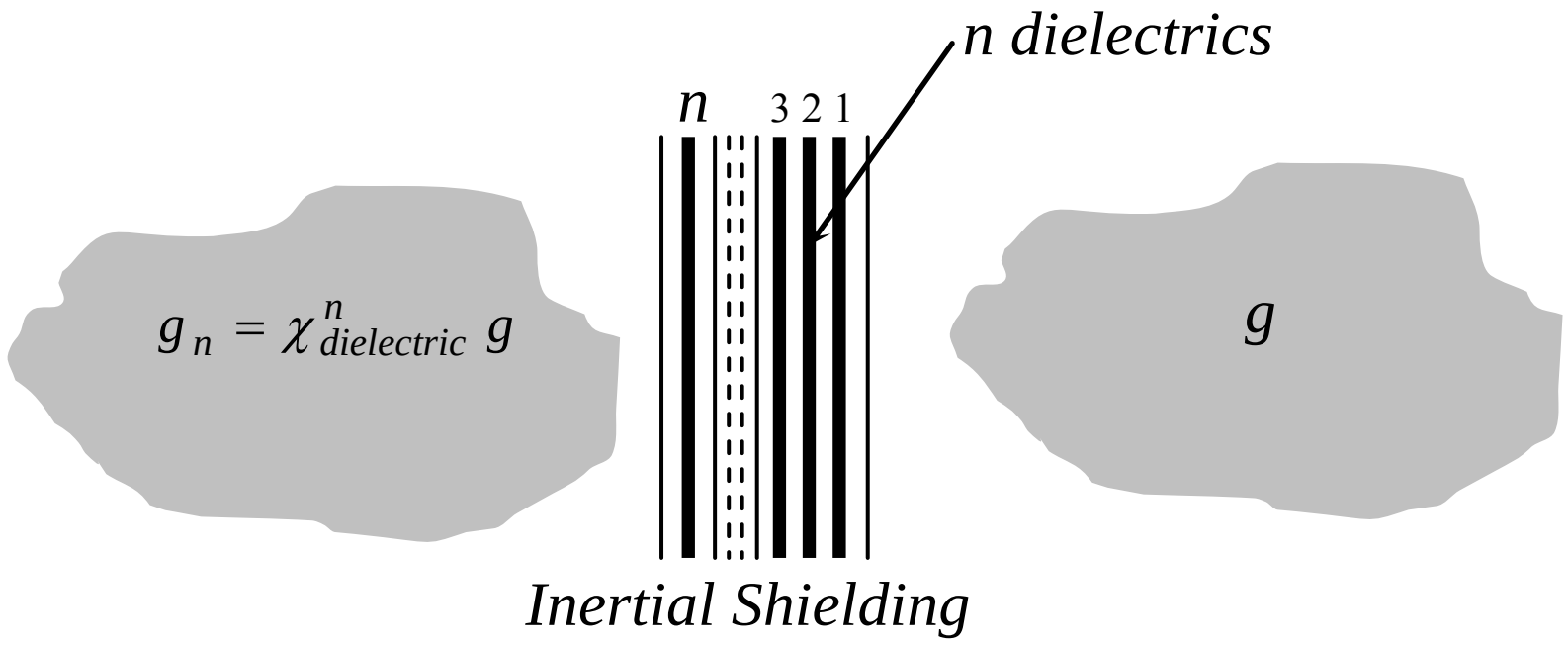

(c)

Fig.B4 – Gravitational Propulsion System and *Inertial Shielding* of the Gravitational Spacecraft – (a) eight *gravitational thrusters are* placed inside a Gravitational Spacecraft, in order to propel the spacecraft along the directions x, y and z. Two *gravitational thrusters* are inside the columns 1 and 2, in order to rotate the spacecraft around the y-axis. The functioning of the Gravitational Thrusters is shown in Fig.A14. *The gravitational mass of the spacecraft is controlled by the Capacitor of Gravitational Mass Control* (CGMC). Note that the CGMC can have the *spacecraft's own form*, as shown in (b). In order to strongly reduce the inertial effects upon the crew of the spacecraft, we can build an inertial shielding, with several CGMCs, as shown above (c). In this case, the gravity upon the crew will be given by $g_n = \chi^n_{dielectric}\, g$, where $g$ is the gravity acceleration out of the spacecraft (in a given direction) and $\chi_{dielectric} = m_g / m_{i0}$ ; $m_g$ and $m_{i0}$ are, respectively, the gravitational mass and the inertial mass of the dielectric. Under these conditions the *inertial effects upon the crew* will be reduced by $\chi^n_{dielectric}$. Thus, for example, if $n = 10$ and $\chi_{dielectric} \cong 0.2$, the inertial effects will be reduced by $\chi^{10}_{dielectric} \cong 1\times 10^{-7}$. If the maximum thrust produced by the thrusters is $F = 10^5\,N$, then the intensities of the inertial forces upon the crew will not exceed $0.01N$.

# APPENDIX C: Longer-Duration Microgravity Environment Produced by *Gravity Control Cells* (GCCs).

The acceleration experienced by an object in a *microgravity* environment, by definition, is one-millionth ($10^{-6}$) of that experienced at Earth's surface (1g). Consequently, a *microgravity* environment is one where the acceleration induced by gravity has little or no measurable effect. The term *zero-gravity* is, obviously inappropriate since the *quantization of gravity* [1] shows that the gravity can have only discrete values *different of zero* [1, Appendix B].

Only three methods of creating a microgravity environment are currently known: to travel far enough into deep space so as to reduce the effect of gravity by *attenuation*, by *falling*, and by *orbiting* a planet.

The first method is the simplest in conception, but requires traveling an enormous distance, rendering it most impractical with the conventional spacecrafts. The second method, *falling*, is very common but approaches microgravity only when the fall is in a vacuum, as air resistance will provide some resistance to free fall acceleration. Also it is difficult to fall for long enough periods of time. There are also problems which involve avoiding too sudden of a stop at the end. The NASA Lewis Research Center has several drop facilities. One provides a 132 meter drop into a hole in the ground similar to a mine shaft. This drop creates a reduced gravity environment for *5.2 seconds*. The longest drop time currently available (about *10 seconds*) is at a 490 meter deep vertical mine shaft in Japan that has been converted to a drop facility.

Drop towers are used for experiments that only need a *short duration of microgravity*, or for an initial validation for experiments that will be carried out in longer duration of microgravity.

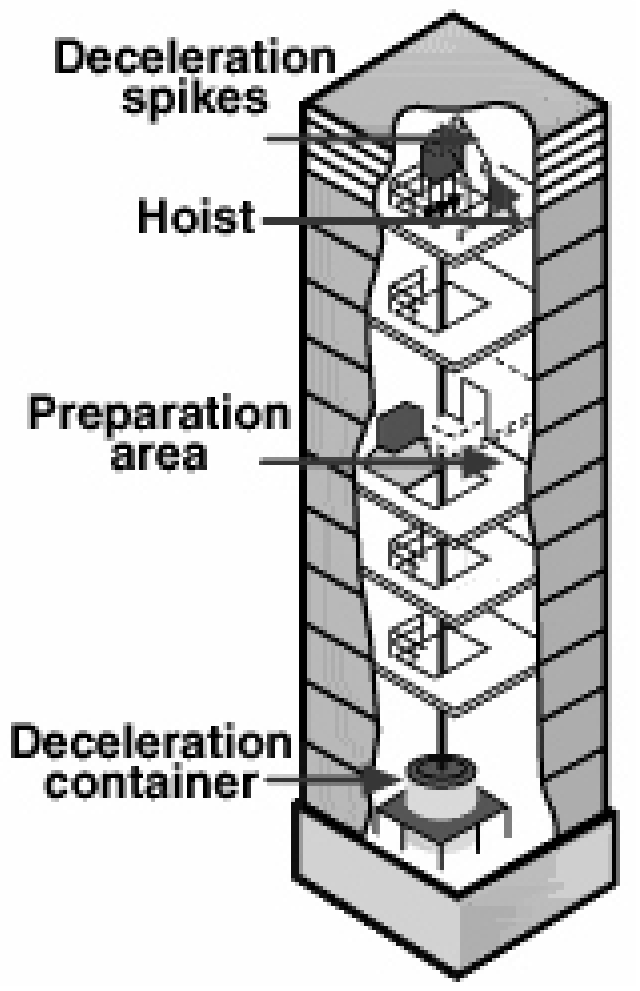

Aircraft can fly in parabolic arcs to achieve period of microgravity of 20 to 25 seconds with g-level of approximately 0.02 g. The airplane climbs rapidly until its nose is about 45-degree angle to the horizon then the engines are cut back. The airplane slows; the plane remains in free fall over the top of the parabola, then it nose-dives to complete the parabola, creating microgravity conditions.

Aircraft parabolic flights give the opportunity to perform medical experiments on human subjects in real microgravity environment. They also offer the possibility of direct intervention by investigators on board the aircraft during and between parabolas. In the mid-1980s, NASA KC-135, a modified Boeing 707, provided access to microgravity environment. A parabolic flight provided 15 to 20 seconds of 0.01 g or less, followed by a 2-g pull out. On a typical flight, up to 40 parabolic trajectories can be performed. The KC-135 can accommodate up to 21 passengers performing 12 different experiments. In 1993, the Falcon-20 performed its first parabolic flight with microgravity experiment on board. This jet can carry two experimenters and perform up to 3 experiments. Each flight can make up to 4 parabolic trajectories, with each parabola lasting 75 seconds, with 15 to 20 seconds of microgravity at 0.01g or less.

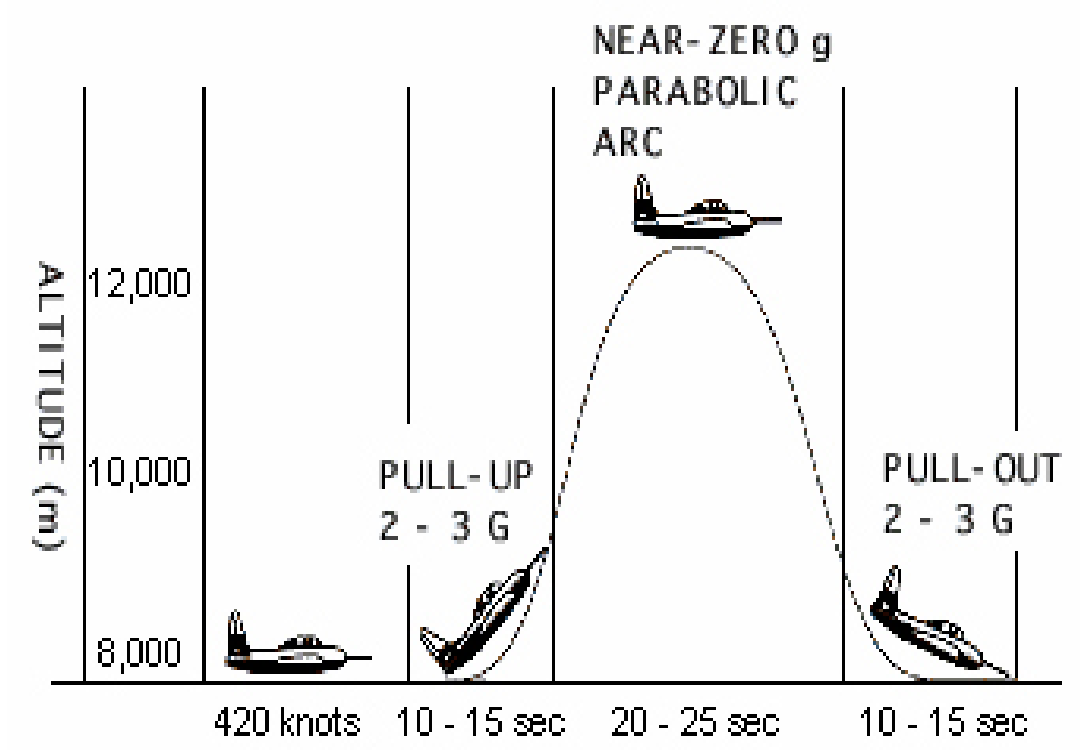

The third method of creating a microgravity environment is *orbiting* a planet. This is the environment commonly experienced in the space shuttle, International Space Station, Mir (no longer in orbit), etc. While this scenario is the most suitable for scientific experimentation and commercial exploitation, it is still quite expensive to operate in, mostly due to launch costs.

A space shuttle provides an ideal laboratory environment to conduct microgravity research. A large panoply of experiments can be carried out in microgravity conditions for up to *17 days*, and scientists can make adjustment to avoid experiment failure and potential loss of data. Unmanned capsules, platforms or satellites, such as the European retrievable carrier Eureka, DLR's retrievable carrier SPAS, or the Russian Photon capsules, the US Space Shuttle (in connection with the European Spacelab laboratory or the US Spacelab module), provide *weeks* or *months* of microgravity.

A space station, maintaining a low earth orbit for several decades, greatly improves access to microgravity environment for up to *several months*.

Thus, microgravity environment can be obtained via different means, providing different duration of microgravity. While *short-duration microgravity environments* can be achieved on Earth with relative easiness, *longer-duration microgravity environments* are too expensive to be obtained.

Here, we propose to use the Gravity Control Cells (GCCs), shown in this work, in order to create *longer-duration microgravity environments*. As we have seen, just above a GCC the gravity can be strongly reduced (*down to* $1\mu$g or less). In this way, the gravity above a GCC can remain at the microgravity ranging during a very long time (*several years*). Thus, GCCs can be used in order to create longer-duration microgravity environments on Earth. In addition, due to the cost of the GCCs to be relatively low, also the longer-duration microgravity environments will be produced with low costs.

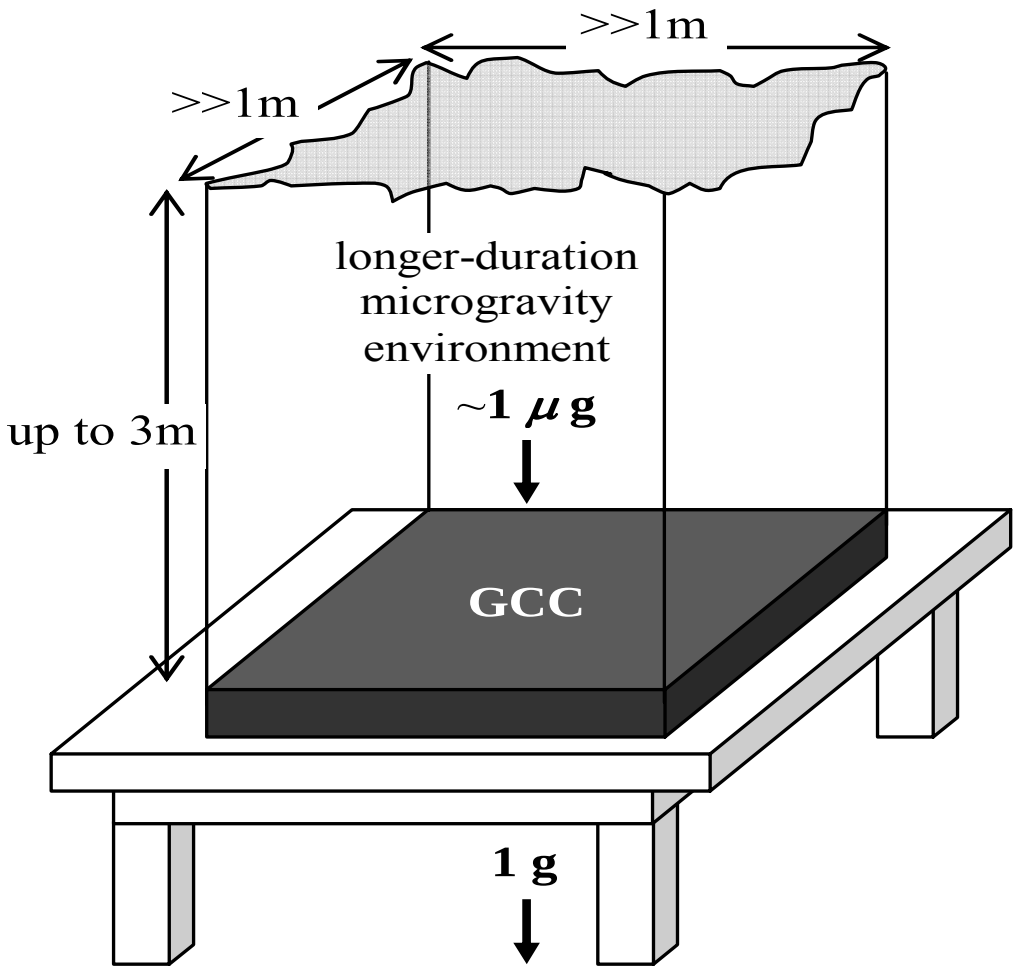

This possibility appears to be absolutely new and unprecedented in the literature since longer-duration microgravity environments are usually obtained via airplanes, sounding rockets, spacecraft and space station.

It is easy to see that the GCCs can be built with *width* and *length* of until some meters. On the other hand, as the effect of gravity reduction above the GCC can reach up to 3m, we can then conclude that the *longer-duration microgravity environments* produced above the GCCs can have sufficiently large volumes to perform any microgravity experiment on Earth.

The longer-duration microgravity environment produced by a GCC will be a special tool for microgravity research. It will allow to improve and to optimize physical, chemical and biological processes on Earth that are important in science, engineering and also medicine. The reduction of gravitational effects in a microgravity environment shows, for example, that temperature differences in a fluid do not produce convection, buoyancy or sedimentation. The changes in fluid behavior in microgravity lie at the heart of the studies in materials science, combustion and many aspects of space biology and life sciences. Microgravity research holds the promise to develop new materials which can not be made on Earth due to gravity. These new materials shall have properties that are superior to those made on Earth and may be used to:

-increase the speed of future computers,
-improve fiber optics,
-make feasible Room Temperature Superconductors,
-enable medical breakthroughs to cure several diseases (e.g., diabetes).

In a microgravity environment protein crystals can be grown larger and with a purity that is impossible to obtain under gravity of 1g. By analyzing the space-grown crystals it is possible to determine the structure and function of the thousands of proteins used in the human body and in valuable plants and animals. The determination of protein structure represents a huge opportunity for pharmaceutical companies to develop new drugs to fight diseases.

Crystal of HIV protease inhibitor grown in microgravity are significantly larger and of higher quality than any specimens grown under gravity of 1g. This will help in defining the structure of the protein crucial in fighting the AIDS virus.

Protein Crystal Isocitrate Lysase is an enzyme for fungicides. The isocitrate lysase crystals grown in microgravity environments are of larger sizes and fewer structural defects than crystals grown under gravity of 1g. They will lead to more powerful fungicides to treat serious crop diseases such as rice blast, and increase crop output.

Improved crystals of human insulin will help improve treatment for diabetes and *potentially create a cure*.

*Anchorage dependent cells* attached to a polymer and grown in a bioreactor in microgravity will lead to the production of a protein that is closer in structure and function to the three-dimensional protein living in the body.

Anchorage Dependant Cells
Attached to a Polymer

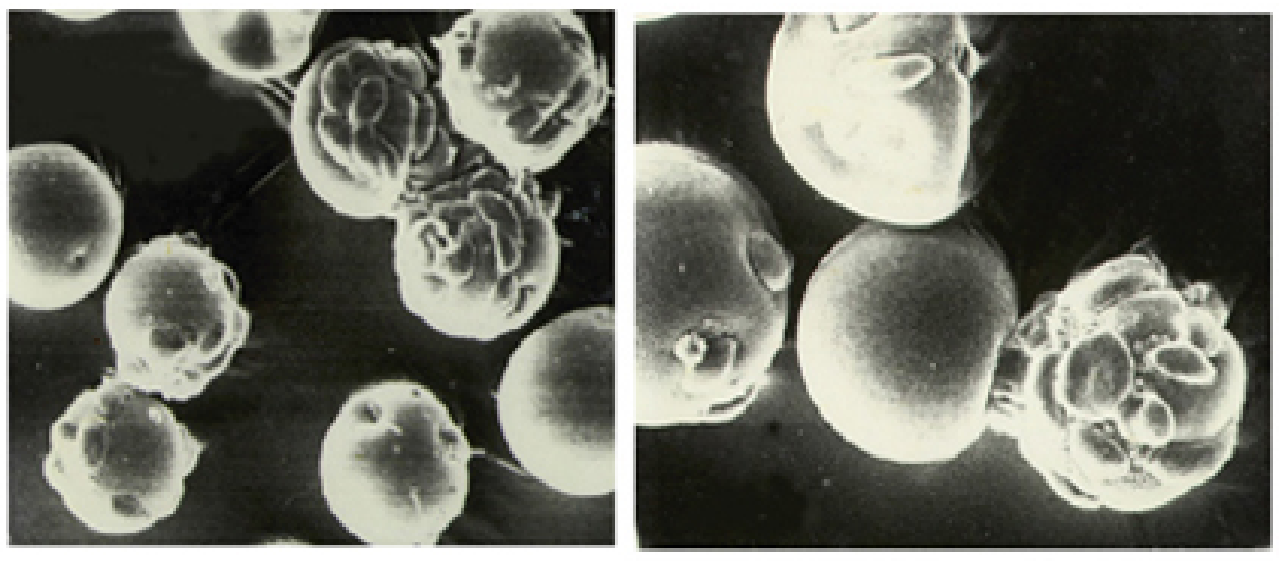

1g μg

This should help reduce or eliminate *transplant rejection* and is therefore critical for organ transplant and for the replacement of damaged bone and tissues. Cells grown on Earth are far from being three-dimensional due to the effect of 1g gravity.

The ZBLAN is a new substance with the potential to revolutionize fiber optics communications. A member of the heavy metal fluoride family of glasses, ZBLAN has promising applications in fiber optics. It can be used in a large array of industries, including manufacture of ultra high purity fiber optics, optical switches for computing, telecommunications, medical surgery and cauterization, temperature monitoring, infrared imaging, fiber-optic lasers, and optical power transmission. A ZBLAN fiber optic cable manufactured in a *microgravity environment* has the potential to carry 100 times the amount of data conveyed by conventional silica-based fibers.

ZBLAN Fiber

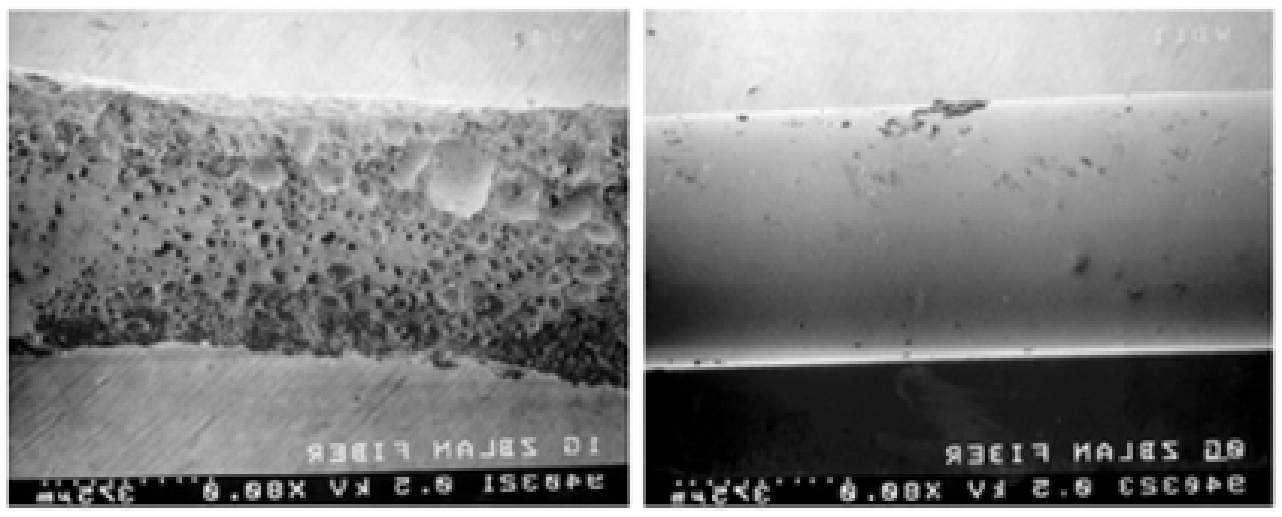

1g μg

In microgravity environment where complications of gravity-driven convection flows are eliminated, we can explore the fundamental processes in fluids of several types more easily and test fundamental theories of three-dimensional laminar, oscillatory and turbulent flow generated by various other forces.

By improving the basics for predicting and controlling the behavior of fluids, we open up possibilities for improving a whole range of industrial processes:

- Civil engineers can design safe buildings in earthquake-prone areas thanks to a better understanding of the fluid-like behavior of soils under stress.

- Materials engineers can benefit from a deeper knowledge of the determination of the structure and properties of a solid metal during its formation and can improve product quality and yield, and, in some cases, lead to the introduction of new products.

- Architects and engineers can design more stable and performing power

plants with the knowledge of the flow characteristics of vapor-liquid mixture.

- Combustion scientists can improve fire safety and fuel efficiency with the knowledge of fluid flow in microgravity.

In microgravity environment, medical researchers can observe the functional changes in cells when the effect of gravity is practically removed. It becomes possible to study fundamental life processes down to the cellular level.

Access to microgravity will provide better opportunities for research, offer repeated testing procedures, and enormously improve the test facilities available for life sciences investigations. This will provide valuable information for medical research and lead to improvements in the health and welfare of the six billion people, which live under the influence of 1g gravity on the Earth's surface.

The utilization of microgravity to develop new and innovative materials, pharmaceuticals and other products is waiting to be explored. Access to microgravity environments currently is limited. Better access, as the produced by GCCs, will help researchers accelerate the experimentation into these new products.

*Terrafoam* is a rigid, silicate based inorganic foam. It is *nonflammable* and does not five off noxious fumes when in the presence of fire. *It does not conduct heat to any measurable degree* and thus is an outstanding and possible unsurpassed thermal insulator. In addition, it appears to have unique *radiation shielding* capabilities, including an ability to block *alpha*, *beta*, gamma rays). Terrafoam can be constructed to be extremely lightweight. Altering the manufacturing process and the inclusion of other materials can vary the properties of Terrafoam. Properties such as cell structure, tensile strength, bulk density and temperature resistance can be varied to suit specific applications. It self-welds to concrete, aluminum and other metals. The useful variations on the base product are potentially in the thousands. Perhaps the most exciting potential applications for Terrafoam stem from its extraordinary capability as an ultra-lightweight thermal and radioactive shield.

Also, the formation of nanoscale *carbon structures* by electrical arc discharge plasma synthesis has already been investigated in microgravity experiments by NASA. Furthermore, complex plasmas are relevant for processes in which a particle formation is to be prevented, if possible, as, for example, within plasma etching processes for microchip production.

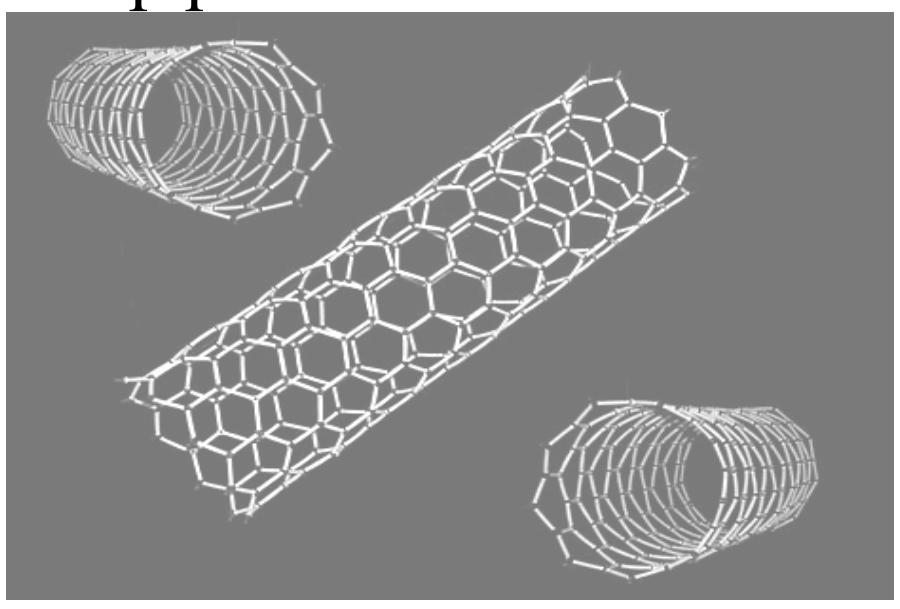

People will benefit from numerous microgravity experiments that can be conducted in Longer-Duration Microgravity Environment Produced by *Gravity Control Cells* (GCCs) on Earth.

## APPENDIX D: Antenna with Gravitational Transducer for *Instantaneous Communications* at any distance

It was previously shown in this article that *Quantum Gravitational Antennas* (GCC antennas, Fig.8) can emit and detect *virtual gravitational* radiation. The velocity of this radiation is *infinite*, as we have seen. This means that these quantum antennas can transmit and receive communications *instantaneously* to and from anywhere in the Universe. Here, it is shown how to transmit and receive communications *instantaneously* from any distance in the Universe by utilizing *virtual electromagnetic* (EM) radiation instead of *virtual gravitational* radiation. Starting from the principle that the antennas of usual transceivers (*real* antennas) radiate *real* EM radiation, then we can expect that *imaginary* antennas radiate *imaginary* EM radiation or *virtual* EM radiation. The velocity of this radiation is also *infinite*, in such a way that it can transmit communications *instantaneously* from any distance in the Universe.

It was shown [1] that when the *gravitational mass* of a body is decreased down to the range of $+0.159m_i$ to $-0.159m_i$ ($m_i$ is its inertial mass), the body becomes *imaginary* and goes to an *imaginary* Universe which contains our *real* Universe. Thus, we have the method to convert *real* antennas to *imaginary* antennas.

Now, consider a Gravitational Shielding $S$, whose gravitational mass is decreased down to the range of $+0.159m_{iS}$ to $-0.159m_{iS}$. By analogy, it becomes imaginary and goes to the imaginary Universe. It is easy to show that, in these circumstances, also a body inside the shielding $S$ becomes imaginary and goes to the imaginary Universe together with the gravitational shielding $S$. In order to prove it, consider, for example, Fig.D1 where we clearly see that the *Gravitational Shielding Effect* is equivalent to a decrease of $\chi = m_{gS}/m_{iS}$ in the gravitational masses of the bodies $A$ and $B$, since the *initial* gravitational masses: $m_{gA} \cong m_{iA}$ and $m_{gB} \cong m_{iB}$ become respectively $m_{gA} = \chi m_{iA}$ and $m_{gB} = \chi m_{iB}$, when the gravitational shielding is activated. Thus, when $\chi$ becomes less than $+0.159$, both the gravitational masses of $S$ and $A$ become respectively:

$$m_{gS} < +0.159 m_{iS}$$

and

$$m_{gA} < +0.159 m_{iA}$$

This proves, therefore, that when a Gravitational Shielding $S$ becomes *imaginary*, any particle (including *photons*‡‡‡‡‡) inside $S$, also becomes *imaginary* and goes to the *imaginary*

‡‡‡‡‡ As shown in the article "*Mathematical Foundations of the Relativistic Theory of Quantum Gravity*", *real* photons become *imaginary* photons or *virtual* photons.

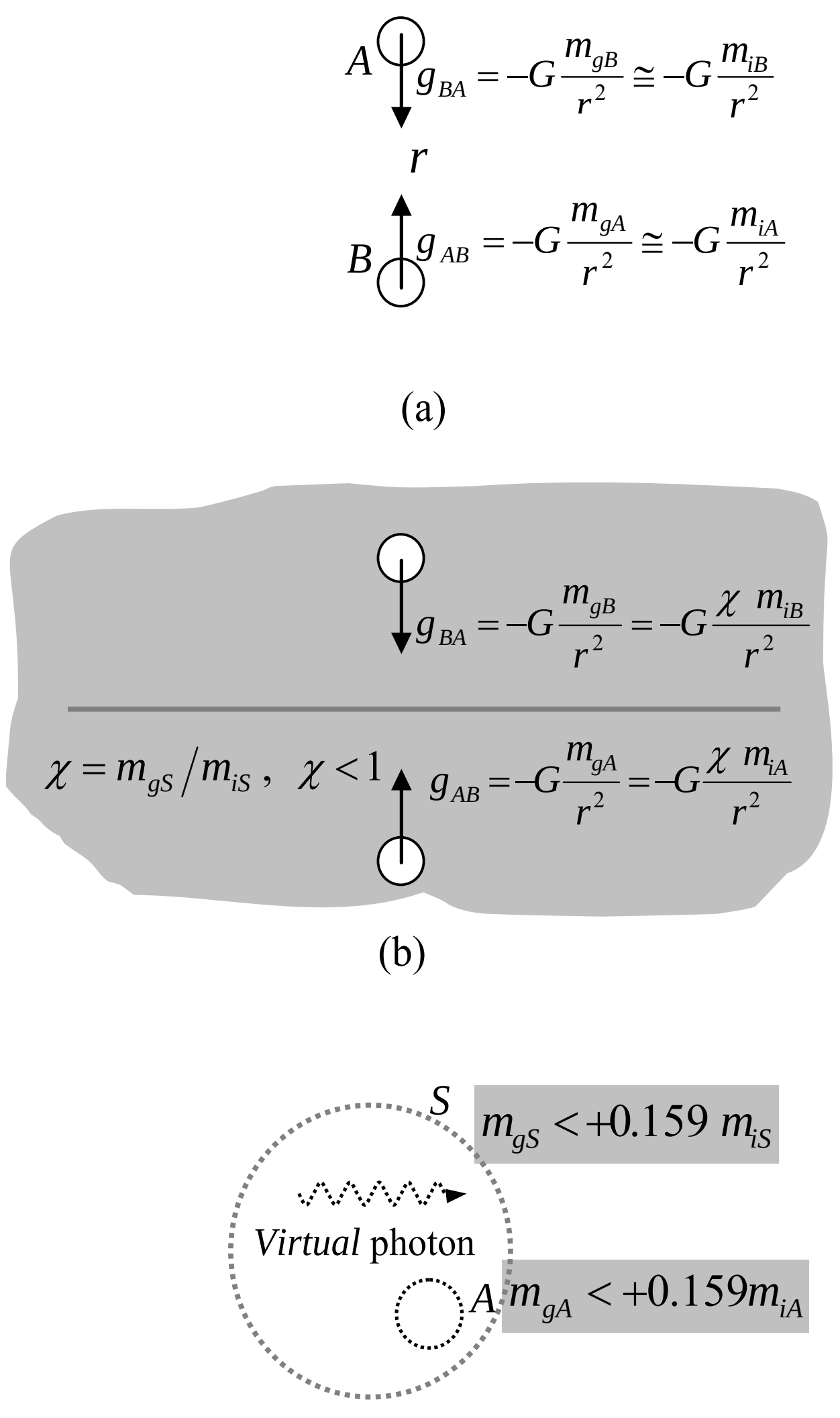

Fig.D1 – (a) (b) The *Gravitational Shielding Effect* is equivalent to a decrease of $\chi = m_{gS}/m_{iS}$ in the gravitational masses of the bodies $A$ and $B$. (c) When a Gravitational Shielding $S$ becomes *imaginary*, any particle (including *photons*) inside $S$, also becomes *imaginary*.

Universe together with the *Gravitational Shielding* $S$ §§§§§.

Now, consider a transceiver antenna inside a Gravitational Shielding $S$. When the gravitational mass of $S$, $m_{gS}$, is reduced down to the range of $+0.159 m_{iS}$ to $-0.159 m_{iS}$, the antenna becomes imaginary, and, together with $S$, it goes to the imaginary Universe. In these circumstances, the *real* photons radiated from the antenna also become imaginary photons or *virtual* photons. Since the velocity of these photons is infinite, they can reach instantaneously the receiving antenna, *if it is also an imaginary antenna in the imaginary Universe*.

Therefore, we can say that the Gravitational Shielding around the antenna works as a *Gravitational* Transducer ****** converting *real* EM energy into *virtual* EM energy.

In practice, we can encapsulate antennas of transceivers with Aluminum cylinders, as shown in Fig.D2(a). By applying an appropriate ELF electric current through the Al cylinders, in order to put the gravitational masses of the cylinders within the range of $+0.159 m_{iCyl}$ to $-0.159 m_{iCyl}$, we can transform *real* antennas into *imaginary* antennas, making possible *instantaneously communications* at any distance, including astronomical distances.

Figure D2 (b) shows usual transceivers operating with *imaginary* antennas, i.e., real antennas turned into imaginary antennas. It is important to note that the communications between them occur through the *imaginary* space-time. At the end of transmissions, when the Gravitational Transducers are turned off, the antennas reappear in the *real* space-time, i.e., they become *real* antennas again.

§§§§§ Similarly, the bodies inside a Gravitational Spacecraft become also imaginaries when the Gravitational Spacecraft becomes imaginary.

****** A *Transducer* is substance or device that converts input energy of one form into output energy of another.

Imagine now cell phones using antennas with gravitational transducers. There will not be any more need of *cell phone signal transmission stations* because the reach of the *virtual* EM radiation is *infinite* (without *scattering*). The new cell phones will transmit and receive communications directly to and from one another. In addition, since the *virtual* EM radiation does not interact with matter, then there will not be any biological effects, as it happens in the case of usual cell phones.

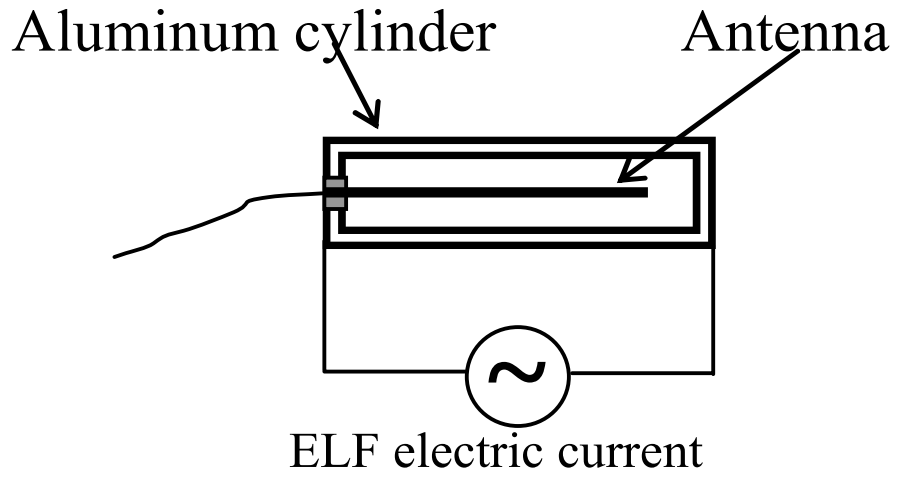

(a)

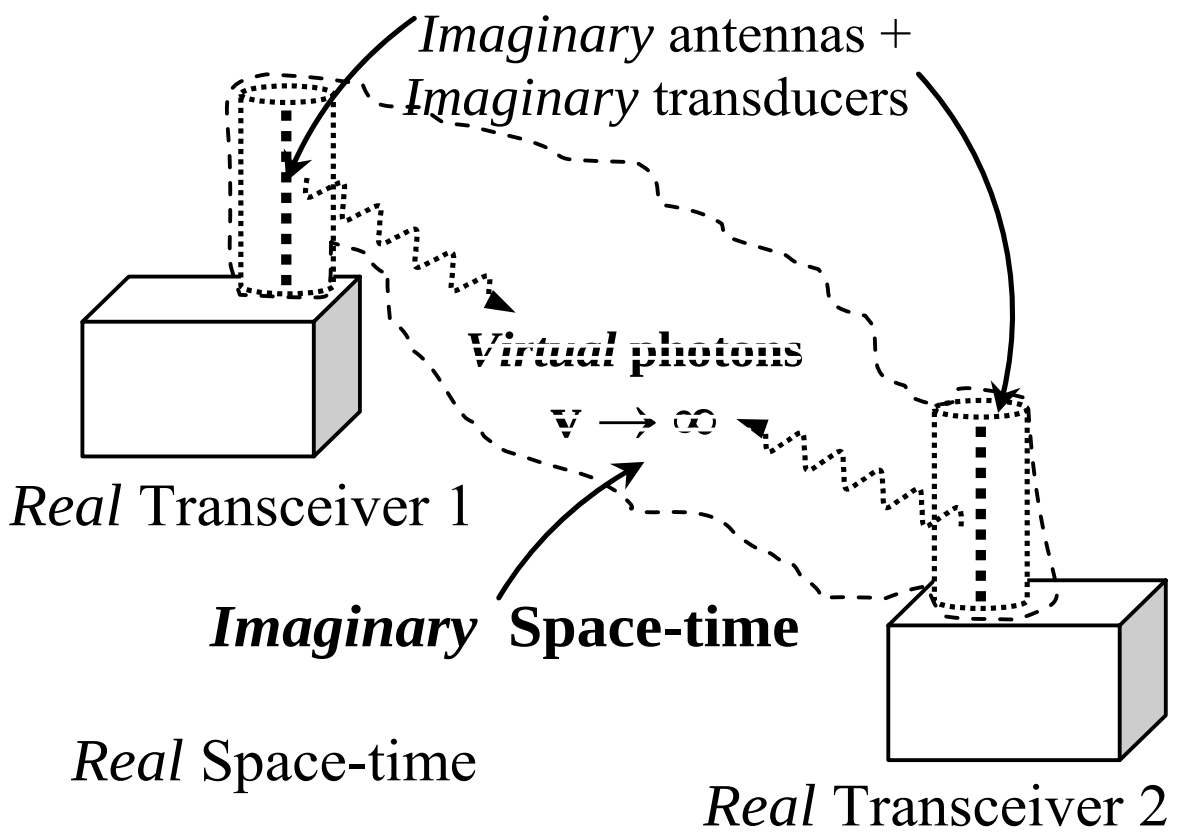

(b)

Fig. D2 – (a) Antenna with *Gravitational* Transducer. (b) Transceivers operating with *imaginary* antennas (*instantaneous* communications *at any distance*, including astronomical distances).

Let us now consider the case where a transceiver is *totally* turned into *imaginary* (Fig.D3). In order to convert real antennas into imaginary antennas, we have used the gravitational shielding effect, as we have already seen. Now, it is necessary to put the transceiver *totally* inside a Gravitational Shielding. Then, consider a transceiver $X$ inside the gravitational shielding of a Gravitational Spacecraft. When the spacecraft becomes imaginary, so does the transceiver $X$. Imagine then, another real transceiver $Y$ with *imaginary* antenna. With their antennas in the imaginary space-time, both transceivers $X$ and $Y$ are able to transmit and receive communications instantaneously between them, by means of *virtual* EM radiation (See Fig. D3(a)). Figure D3(b) shows another possibility: instantaneous communications between two transceivers at *virtual* state.

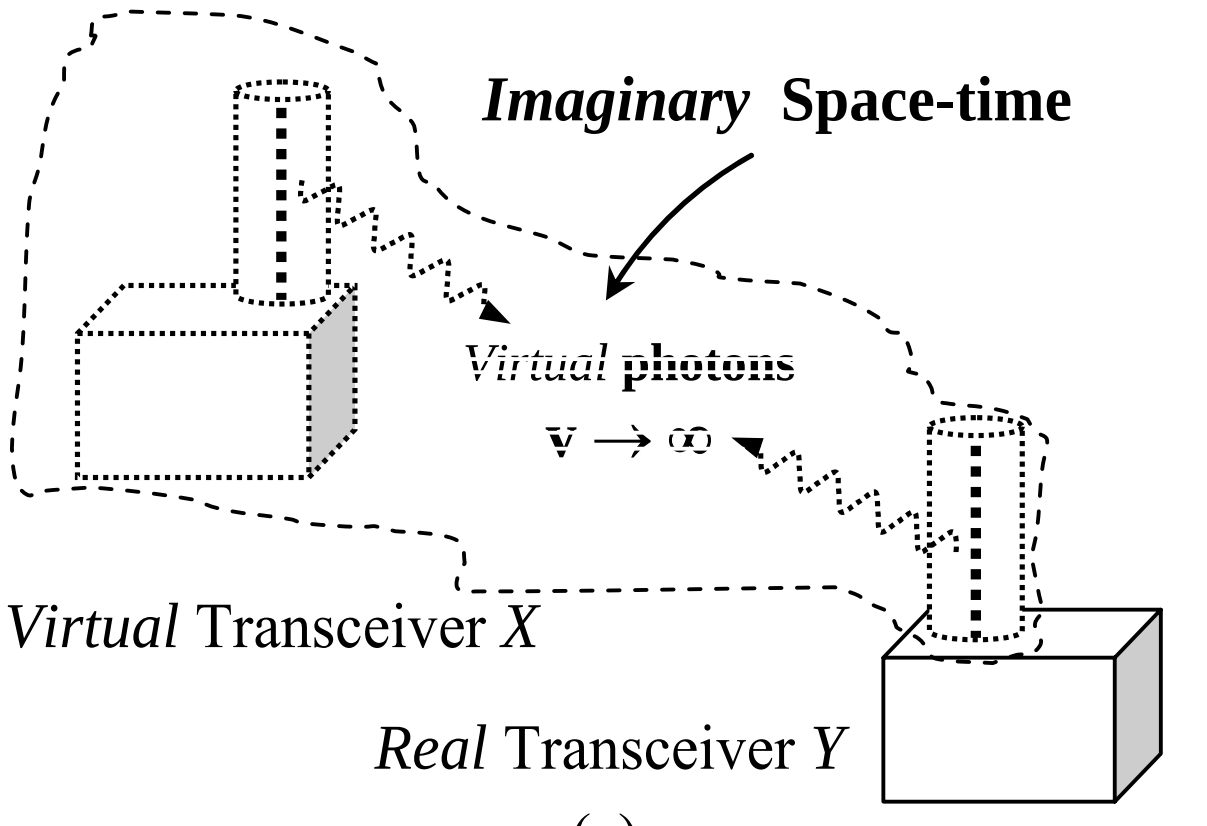

(a)

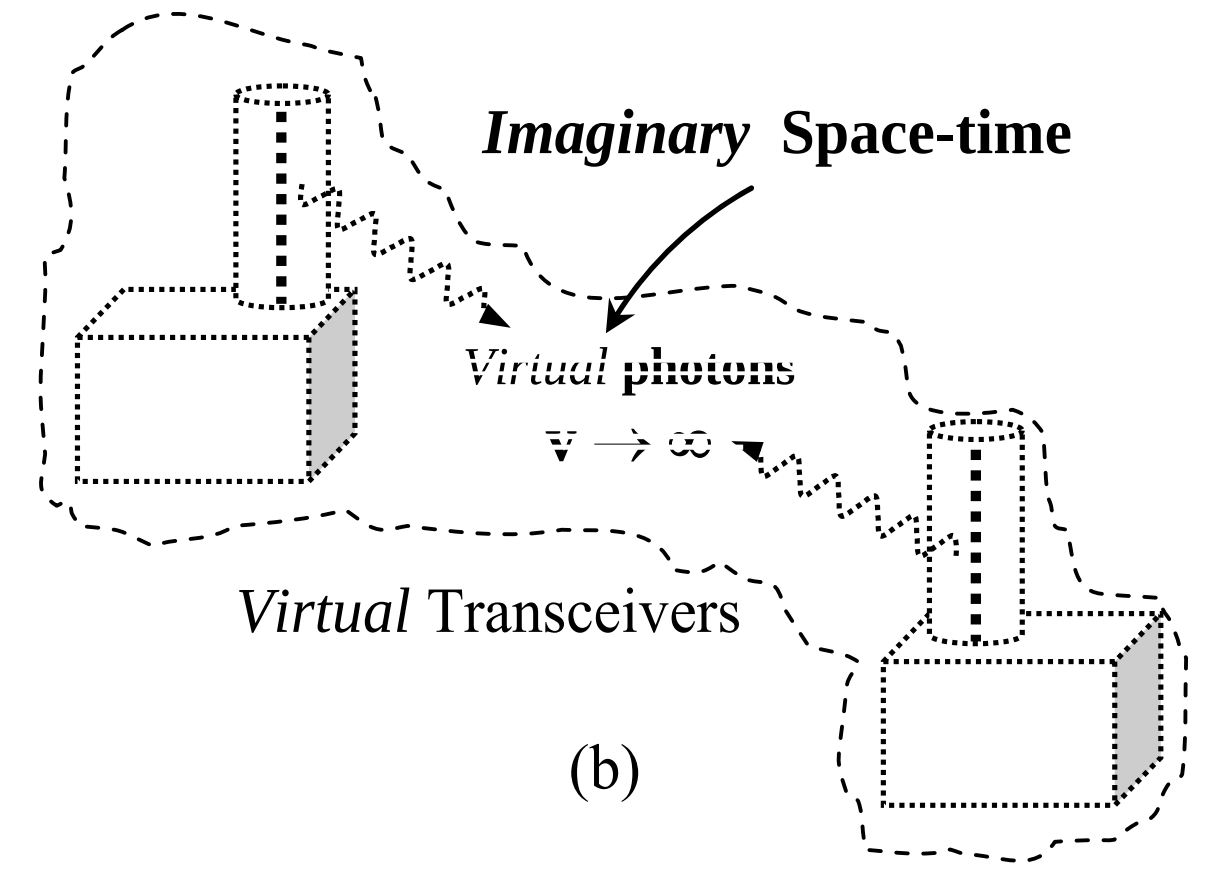

(b)

Fig. D3 – (a) Instantaneous communications between the *real* Universe and the *imaginary* Universe.(b) Instantaneous communications between two Virtual Transceivers in the *imaginary* Universe.